\documentclass[twocolumn,iop]{emulateapj}
\pdfoutput=1
\usepackage{graphics,color}
\usepackage{amssymb,amsmath}

\newcommand{\degsq}{\,deg$^{2}$}
\newcommand{\perdegsq}{\,deg$^{-2}$}
\newcommand\geqsim{\lower.73ex\hbox{$\sim$}\llap{\raise.4ex\hbox{$>$}}$\,$}
\newcommand\leqsim{\lower.73ex\hbox{$\sim$}\llap{\raise.4ex\hbox{$<$}}$\,$}
\newcommand{\project}[1]{\emph{#1}}
\newcommand{\sdss}{\project{SDSS}}

\newcommand{\drsev}{\project{DR7}}
\newcommand{\dre}{\project{DR8}}
\newcommand{\drn}{\project{DR9}}
\newcommand{\drt}{\project{DR10}}
\newcommand{\drtw}{\project{DR12}}
\newcommand{\drtwq}{\project{DR12Q}}
\newcommand{\sdssi}{\project{SDSS-I}}
\newcommand{\sdssii}{\project{SDSS-II}}
\newcommand{\sdssiii}{\project{SDSS-III}}
\newcommand{\sdssiv}{\project{SDSS-IV}}
\newcommand{\sequels}{\project{SEQUELS}}

\newcommand{\panstarrs}{\project{Pan-STARRS}}

\newcommand{\ptf}{\project{PTF}}
\newcommand{\iptf}{\project{iPTF}}
\newcommand{\wise}{\project{WISE}}

\newcommand{\xd}{\project{XD}}
\newcommand{\xdqso}{\project{XDQSO}}
\newcommand{\xdqsoz}{\project{XDQSOz}}

\newcommand{\boss}{\project{BOSS}}
\newcommand{\eboss}{\project{eBOSS}}
\newcommand{\tdss}{\project{TDSS}}
\newcommand{\spiders}{\project{SPIDERS}}

\newcommand{\first}{\project{FIRST}}

\newcommand{\kms}{\,km\,s$^{-1}$}
\newcommand{\LyA}{Lyman-$\alpha$}
\newcommand{\perMpc}{\,Mpc$^{-1}$}

\begin{document}

\title{The SDSS-IV extended Baryon Oscillation Spectroscopic Survey: Quasar Target Selection}
\slugcomment{Accepted to ApJS}
\author{
Adam~D.~Myers\altaffilmark{1,2},
Nathalie~{Palanque-Delabrouille}\altaffilmark{3},
Abhishek~Prakash\altaffilmark{4},
Isabelle~P\^aris\altaffilmark{5},
Christophe~Yeche\altaffilmark{3},
Kyle~S.~Dawson\altaffilmark{6},
Jo~Bovy\altaffilmark{7},
Dustin~Lang\altaffilmark{8,9},
David~J.~Schlegel\altaffilmark{10},
Jeffrey~A.~Newman\altaffilmark{4},
Patrick~Petitjean\altaffilmark{11},
Jean-Paul~Kneib\altaffilmark{12,13},
Pierre~Laurent\altaffilmark{3},
Will~J.~Percival\altaffilmark{14},
Ashley~J.~Ross\altaffilmark{14,15},
Hee-Jong~Seo\altaffilmark{16},
Jeremy~L.~Tinker\altaffilmark{17},
Eric~Armengaud\altaffilmark{3},
Joel~Brownstein\altaffilmark{6},
Etienne~Burtin\altaffilmark{3},
Zheng~Cai\altaffilmark{18},
Johan~Comparat\altaffilmark{19}
Mansi~Kasliwal\altaffilmark{20,21,22},
Shrinivas~R.~Kulkarni\altaffilmark{23},
Russ~Laher\altaffilmark{24},
David~Levitan\altaffilmark{25},
Cameron~K.~McBride\altaffilmark{26},
Ian~D.~McGreer\altaffilmark{18}
Adam~A.~Miller\altaffilmark{22,23,27},
Peter~Nugent\altaffilmark{10,28},
Eran~Ofek\altaffilmark{29},
Graziano~Rossi\altaffilmark{30},
John~Ruan\altaffilmark{31},
Donald~P.~Schneider\altaffilmark{32,33},
Branimir~Sesar\altaffilmark{34},
Alina~Streblyanska\altaffilmark{35,36},
Jason~Surace\altaffilmark{24}
}

\altaffiltext{1}{amyers14@uwyo.edu}

\altaffiltext{2}{
Department of Physics and Astronomy, 
University of Wyoming, 
Laramie, WY 82071, USA.
}

\altaffiltext{3}{
CEA, Centre de Saclay, Irfu/SPP,  F-91191 Gif-sur-Yvette, France.
}

\altaffiltext{4}{
Department of Physics and Astronomy and PITT PACC, 
University of Pittsburgh, Pittsburgh, PA 15260, USA.
}

\altaffiltext{5}{
INAF - Osservatorio Astronomico di Trieste, Via G. B. Tiepolo 11, I-34131 Trieste, IT.
}

\altaffiltext{6}{
Department of Physics and Astronomy, 
University of Utah, Salt Lake City, UT 84112, USA.
}

\altaffiltext{7}{
Department of Astronomy and Astrophysics, University of Toronto, 50 St. George Street, Toronto, ON, M5S 3H4, Canada
}

\altaffiltext{8}{
Bruce and Astrid McWilliams Center for Cosmology,
Department of Physics, 
Carnegie Mellon University, 5000 Forbes Ave, Pittsburgh, PA 15213, USA.
}

\altaffiltext{9}{
McWilliams fellow.
}

\altaffiltext{10}{
Lawrence Berkeley National Laboratory, One Cyclotron Road,
Berkeley, CA 94720, USA.
}

\altaffiltext{11}{
UPMC-CNRS, UMR 7095, 
Institut d'Astrophysique de Paris, 
98 bis Boulevard Arago, 75014, Paris, France.
}

\altaffiltext{12}{
Laboratoire d'Astrophysique, Ecole Polytechnique F\'ed\'erale de Lausanne
Observatoire de Sauverny, 1290 Versoix, Switzerland.
}

\altaffiltext{13}{
Aix Marseille Universit\'e, CNRS, LAM
(Laboratoire d'Astrophysique de Marseille),
UMR 7326, 13388, Marseille, France
}

\altaffiltext{14}{
Institute of Cosmology \& Gravitation, Dennis Sciama Building, University of Portsmouth, Portsmouth, PO1 3FX, UK.
}

\altaffiltext{15}{
Center for Cosmology and AstroParticle Physics, The Ohio State University, Columbus, OH 43210, USA.}

\altaffiltext{16}{
Department of Physics and Astronomy,
Ohio University,
251B Clippinger Labs, Athens, OH 45701, USA.
}

\altaffiltext{17}{
Center for Cosmology and Particle Physics,
Department of Physics, New York University,
4 Washington Place, New York, NY 10003, USA.
}

\altaffiltext{18}{
Steward Observatory, University of Arizona, 933 N Cherry Ave, Tucson, AZ 85721, USA.
}

\altaffiltext{19}{
Instituto de F\'{\i}sica Te\'orica, (UAM/CSIC), 
Universidad Aut\'onoma de Madrid, Cantoblanco, E-28049 Madrid, Spain.
}

\altaffiltext{20}{
The Observatories, Carnegie Institution for Science, 813 Santa Barbara Street, Pasadena, CA 91101, USA.
}

\altaffiltext{21}{
Carnegie-Princeton Fellow.
}

\altaffiltext{22}{
Hubble fellow.
}

\altaffiltext{23}{
California Institute of Technology, Pasadena, CA 91125, USA.
}

\altaffiltext{24}{
Spitzer Science Center, California Institute of Technology, M/S 314-6, 
Pasadena, CA 91125, U.S.A.
}

\altaffiltext{25}{
Division of Physics, Mathematics and Astronomy, California Institute of Technology, Pasadena,
CA 91125, USA.
}

\altaffiltext{26}{
Harvard-Smithsonian Center for Astrophysics,
Harvard University,
60 Garden St.,
Cambridge MA 02138, USA.
}

\altaffiltext{27}{
Jet Propulsion Laboratory, California Institute of Technology, 4800 Oak Grove Drive, MS 169-506, Pasadena, CA 91109, USA.
}

\altaffiltext{28}{
Department of Astronomy, 
University of California, Berkeley, CA 94720, USA.
}

\altaffiltext{29}{
Department of Particle Physics and Astrophysics, The Weizmann Institute of Science, Rehovot 76100, Israel.
}

\altaffiltext{30}{
Department of Astronomy and Space Science, Sejong University, Seoul, 143-747, Korea.
}

\altaffiltext{31}{
Department of Astronomy, University of Washington, Box 351580, Seattle, WA 98195, USA.
}

\altaffiltext{32}{
Department of Astronomy and Astrophysics, 525 Davey Laboratory, 
The Pennsylvania State University, University Park, PA 16802, USA.
}

\altaffiltext{33}{
Institute for Gravitation and the Cosmos, 
The Pennsylvania State University, University Park, PA 16802, USA.
}

\altaffiltext{34}{
Max-Planck-Institut f\"ur Astronomie, K\"onigstuhl 17, D-69117
Heidelberg,
Germany.
}

\altaffiltext{35}{
Instituto de Astrofisica de Canarias (IAC), E-38200 La Laguna, Tenerife, Spain.
}

\altaffiltext{36}{
Universidad de La Laguna (ULL), Dept. Astrofisica, E-38206 La Laguna, Tenerife, Spain.
}

\shorttitle{eBOSS quasars}

\begin{abstract}

As part of the {\em Sloan Digital Sky Survey IV} the
{\em extended Baryon Oscillation Spectroscopic Survey} (\eboss) will
improve measurements of the cosmological distance scale by 
applying the Baryon Acoustic Oscillation (BAO) method to quasar samples.
\eboss\ will adopt two approaches to target quasars over 7500\degsq.
First, a ``CORE'' quasar sample
will combine optical selection in $ugriz$ using a 
likelihood-based routine called \xdqsoz, with a mid-IR-optical color-cut. 
\eboss\ CORE selection (to $g < 22$ OR 
$r < 22$) should return $\sim 70$\perdegsq\ quasars at redshifts $0.9 < z < 2.2$
and $\sim 7$\perdegsq\ $z > 2.1$ quasars. Second, a selection based on variability
in multi-epoch imaging from the {\em Palomar Transient Factory} should recover an additional
$\sim 3$--4\perdegsq\ $z > 2.1$ quasars to $g < 22.5$. A linear model of how imaging systematics affect target density 
recovers the angular distribution of \eboss\ CORE quasars over 96.7\% (76.7\%) of the \sdss\ North (South)
Galactic Cap area. 
The \eboss\ CORE quasar sample
should thus be sufficiently dense and homogeneous over $0.9 < z < 2.2$
to yield the first few-percent-level BAO constraint near 
$\bar{z}\sim1.5$. 
\eboss\ quasars at $z > 2.1$ will be used to
improve BAO measurements in the \LyA\ Forest.
Beyond its key cosmological goals, \eboss\ should be {\em the} next-generation quasar survey, 
comprising $> 500{,}000$ {\em new} quasars 
and $> 500{,}000$ uniformly selected spectroscopically confirmed $0.9 < z < 2.2$ quasars.
At the conclusion of \eboss, the \sdss\ will have provided unique spectra of over $800{,}000$ quasars. 

\end{abstract}


\section{Introduction}\label{sec:intro}

\setcounter{footnote}{0}

Over 50 years have elapsed since the discoveries that quasars are bright, blue, extragalactic 
sources in optical imaging \citep{Sch63} and that the vast majority of unresolved,
extragalactic objects that are bluer than the stellar main sequence are quasars 
\citep{San65}. Since this time, many imaging surveys used a UV-excess (UVX) criterion, as 
manifested in simple optical color cuts, to provide a mechanism for targeting quasars 
\citep[e.g.][]{San69,Bra70,For80,Gre86,Boy90}. The UVX approach, which mainly 
targets quasars at redshifts around $0.5 < z < 2.5$, precipitated increasingly extensive 
spectroscopically confirmed quasar samples as the capabilities of imaging surveys improved, 
such as 
the Large Bright Quasar Survey \citep{LBQS}, the 2dF QSO Redshift Survey 
\citep{2QZ}, and the 2dF-SDSS LRG and QSO Survey \citep{2SLAQ}.

Modifications of the UVX approach to target all of color space beyond the stellar locus, 
rather than just the blue side \citep[e.g.][]{War87,Ken95,New97}, extended the selection of 
large numbers of quasars to $z > 2.5$. The {\em Sloan Digital Sky Survey} \citep[\sdss;][]{Yor00}
applied this methodology to imaging taken 
using a new $ugriz$ filter system \citep{Fuk96}. 
\sdss\ eventually spectroscopically confirmed an 
unprecedentedly large sample of over one-hundred-thousand quasars \citep{Ric02,DR7QSO} as
part of the \sdssi\ and \project{II} surveys. 

In addition to optical color space, \sdssi\ and \project{II} selected about
10\% of their quasar sample via radio matches to the FIRST survey \citep{FIRST,Hel15}, or 
X-ray matches to the ROSAT All Sky Survey \citep{ROSAT}. The proliferation of 
such large, multi-wavelength surveys, as well as multi-epoch surveys, has made
quasar classification approaches that do not rely on optical colors (but 
still may use optical imaging to constrain morphology or brightness) increasingly attractive. Such approaches include: the use of the radio \citep[e.g.][]{Whi00,Mcg09},
near-infrared \citep[e.g.][]{Ban12}, or both \citep[e.g.][]{Gli12}; the lack of an observed proper motion 
\citep[e.g.][]{Kro81}, the use of the mid-infrared \citep[e.g.][]{Lac04,Ste05,Ric09,Ste12}, X-rays
\citep[e.g.][]{Tri12}, or both \citep[e.g.][]{Lac07,Hic07,Hic09}; the use of 
slitless spectroscopy \citep[e.g.][]{Osm82,Sch86} and the use of variability 
\citep[e.g.][]{Ush78,Ren04,Sch10,But11,Mac11,Pal11}. 

Even after the first iterations of the \sdss, the selection of quasars at 
$z~\geqsim\,2.5$ remained relatively incomplete. This problem arose 
partially because \sdssi\ and \project{II} 
targeted quasars a magnitude or more brighter than the limits of \sdss\ imaging, thus sampling only
the high luminosity regime at these redshifts, 
and partially because the stellar and quasar loci intersect in $ugriz$ color space 
around the ``quasar redshift desert" near $z\sim2.7$ \citep{Fan99}. In order to target
quasars at $z > 2.1$ for cosmological studies of the Lyman-$\alpha$ Forest, the
\sdssiii \citep{Eis11} {\em Baryon Oscillation Spectroscopic Survey}
\citep[\boss;][]{Daw13} attempted to circumvent these problems of quasar selection 
near $z \sim 3$ by  
applying sophisticated, multi-wavelength, multi-epoch star-quasar separation techniques
to the full depth of \sdss\ imaging. \boss\ 
spectroscopically identified $\sim170{,}000$ new quasars of redshift $2.1 \leq z < 3.5$
to a depth of $g < 22$ (I.\ P{\^a}ris et al.\ 2016, in preparation; henceforth \drtwq), a sample about ten times larger than
for the same redshift range in \sdssi\ and \project{II}. \boss\ may only be $\sim60$\%
complete \citep[e.g.][]{Ros13}, raising the possibility that there are additional $g < 22$ quasars to be
discovered in this redshift regime.

In combination, \sdssi/\project{II}/\project{III} targeted quasars
at $2.1~\leqsim\,z~\leqsim\,4$ to a magnitude limit of $g < 22$ or $r < 21.85$ \citep{Ros12} and
quasars at all redshifts to $i < 19.1$\footnote{In addition, smaller dedicated
programs affiliated with \sdss\ have targeted higher redshift quasars to fainter
limits} \citep{Ric02}. There remains an obvious, highly populated discovery 
space using \sdss\ imaging data---namely, $z < 2.1$ quasars fainter
than $i = 19.1$. In addition, since the advent of \boss, new and extensive
multi-wavelength and multi-epoch imaging has become available, allowing
$z > 2.1$ quasars to be targeted that may have been missed by \boss. In particular,
mid-IR colors provide a powerful mechanism for separating quasars and stars 
and Wide-field Infrared Survey Explorer 
\citep[\wise;][]{WISE} data therefore provide additional information for targeting quasars
that otherwise resemble stars in optical color space \citep[e.g.][]{Ste12,Ass13,Yan13}.

The remaining potential of \sdss\ and other imaging for targeting new quasars 
has obvious synergy with the now mature field
of using Baryon Acoustic Oscillation features (BAOs) to measure the expansion
of the Universe \citep{Eis98,Seo03,Lin03}. No strong BAO constraint currently exists 
in the redshift range $1~\leqsim\,z~\leqsim\,2$, and BAO measurements 
at yet higher redshift remain a particularly potent constraint on the 
evolution of the angular diameter distance, $D_A(z)$ and of the Hubble Parameter $H(z)$ \citep{Aub14}.
These factors led to the conception of a new survey---the 
{\em extended Baryon Oscillation Spectroscopic Survey}
\citep[\eboss;][]{ebosspaper???} as part of \sdssiv.

It has been difficult to detect BAO features using quasars as direct tracers
due to their low space density. \eboss\ will circumvent this issue by surveying quasars
over a huge volume, corresponding to 7{,}500\degsq\ of sky. The quasar component of \eboss\ will attempt to statistically target and measure redshifts for
$\sim500{,}000$ quasars at $0.9 < z < 2.2$ (including spectroscopically confirmed
quasars from \sdssi/\project{II}, which will not need to be retargeted). We will refer
to this homogeneous tracer sample as the \eboss\ {\em CORE quasar target selection}.
\boss\ targeted quasars at $z > 2.2$ with the main goal
of using them as indirect tracers to study cosmology in the \LyA\ Forest. In contrast,
\eboss\ will open up the $i > 19.1$, $z < 2.2$ parameter space 
to {\em directly use quasars themselves} as cosmological tracers.

In addition, analyses of the \LyA\ Forest with \boss\ have provided substantial new insights into
cosmological constraints \citep[e.g.][]{Slo11,Slo13,Not12,Bus13,Kir13,Pal13,Fon14,Del14}.
\eboss\ will therefore also (heterogeneously) observe over $\sim60{,}000$ new $z > 2.1$
quasars and will reobserve low signal-to-noise ratio $z > 2.1$ quasars from \boss.
The main goals of this targeting campaign are to produce measurements of
the BAO scale (in both $d_A(z)$ and $H(z)$) in 
the Ly$\alpha$ Forest that approach $\sim1.5\%$ at 
$z\sim2.5$ and that probe an entirely new
redshift regime via quasar clustering 
at $z\sim1.5$ with $\sim2\%$ precision (see \S\ref{sec:goals}). 

In total, at the conclusion
of \eboss, the \sdss\ surveys will have spectroscopically confirmed more than 800{,}000
quasars. The scope of the science that can be conducted with a large sample of quasars across
a range of redshifts has been shown to be vast. Beyond \LyA\ Forest science, 
\boss\ also facilitated additional, diverse quasar science, from measurements of quasar
clustering and the quasar luminosity function to studies of Broad Absorption Line quasars.
\citep[e.g][]{Fil12,Fil13,Fil14,Whi12,Ale13,Fin13,McG13,Ros13,Vik13,Gre14,Eft15}.
\eboss\ will seek to augment many of these measurements. In addition to higher-redshift
studies, \sdssiv/\eboss\ will produce a $z < 2.2$ sample of quasars about
six times larger than the final \sdssii\ quasar catalog \citep{DR7QSO} and will further
benefit from upgrades conducted for \sdssiii\ \citep[such as larger wavelength
coverage for spectra; see][for extensive details of upgrades]{Sme13}. Many high-impact 
projects that used the original \sdssi/\project{II} quasar
samples can therefore potentially be revisited using much larger samples 
with \eboss, such as composite quasar spectra, rare types of quasars, and precision
studies of the quasar luminosity function \citep[e.g.][]{Van01,Ina03,McL04,Hen06,Ric06,Yor06,Net07,Kas07,She08,Bor09}.

In this paper, we describe quasar target selection for the \sdssiv/\eboss\ survey.
Further technical details about \eboss\ can be found in  
our companion papers which include an overview of \eboss\ \citep{ebosspaper???}
and discussions of targeting for Luminous Red Galaxies (\citealt{Prakash15inprep???}; see also \citealt{Pra15}),
and Emission Line Galaxies \citep{Comparat15inprep???}.~\eboss\ will run 
concurrently with two surveys; the 
SPectroscopic IDentification of ERosita Sources survey (\spiders)
and the Time Domain Spectroscopic Survey \citep[\tdss;][]{Mor15}.
These associated surveys are further outlined in our companion overview paper \citep{ebosspaper???}.

In \S2 we discuss how forecasts for BAO constraints at different redshifts drive
targeting goals for \eboss\ quasars.
The parent imaging used for \eboss\ quasar target selection 
is outlined in \S3. Those interested
in the main quasar targeting details for \eboss\ (targeting algorithms, the meaning
of targeting bits, the criteria for re-targeting of previously known
quasars) should read \S4 of this paper. In \S5, we use the results from an 
extensive pilot survey (\sequels; {\em The Sloan Extended QUasar, ELG and LRG Survey}, 
undertaken as part of \sdssiii)
to detail our expected efficiency and distribution of
quasars for \eboss. An important criterion for any large-scale structure survey
is sufficient homogeneity to facilitate modeling of the distribution of the tracer 
population---the ``mask'' of the survey. In \S6 we use the full \eboss\ target sample 
to characterize the homogeneity of \eboss\ quasar selection.
In \S7, we provide our overall conclusions regarding \eboss\ quasar targeting, and
provide a bulleted summary of the final \eboss\ CORE quasar selection algorithm.

Unless we state otherwise, all magnitudes and fluxes in this paper are corrected for Galactic extinction 
using the dust maps of \citet{Sch98}. Specifically, we use the correction based upon the recalibration
of the \sdss\ reddening coefficients measured by \citet{Sch11}. For
\wise\ we adopt the reddening coefficients from \citet{Fit99}.
The SDSS photometry has been demonstrated to have colors
that are within 3\% \citep{Sch11} of being on the AB system \citep{Oke83}. \wise\ is calibrated to be on the
Vega system.
We use a cosmology of
($\Omega_{\rm m}$, $\Omega_\Lambda$, $h\equiv H_0/100\,{\rm km\,s^{-1}\,Mpc^{-1}})= (0.315,0.685,0.67)$ 
consistent with recent results from {\em Planck} \citep[][]{Planck14}.

\section{Cosmological Goals of \eboss\ and Implications for Quasar Target Selection}
\label{sec:goals}

\subsection{CORE and Lyman-$\alpha$ quasars}

The goal of the \eboss\ quasar survey is to study the scale of the BAO in two distinct 
redshift regimes---$z\sim1.5$ using the clustering of quasars, and $z\sim2.5$ using 
high redshift quasars as backlights to illuminate the Lyman-$\alpha$ Forest. Broadly, 
this approach requires a sample of statistically selected quasars in the redshift range 
$0.9 < z < 2.2$---which we will refer to as ``CORE quasars''---and quasars selected at 
$z > 2.1$---which we will refer to as  ``Lyman-$\alpha$ quasars''. 

A major difference between the two
samples is the homogeneity of the target selection technique. The selection of CORE quasars 
must be statistically uniform. Lyman-$\alpha$ quasars, however, can be selected heterogeneously, 
as a clustering measurement using the Lyman-$\alpha$ Forest does not require the background 
quasars to have a uniform (or even a {\em reproducible}) selection. In fact, the full redshift range of the CORE
sample will extend well beyond $0.9 < z < 2.2$, and many CORE quasars can thus be utilized as
Lyman-$\alpha$ quasars. The terminology ``CORE quasars'' therefore refers to {\em how the
quasars were targeted} whereas the terminology ``Lyman-$\alpha$ quasars'' refers to 
the {\em redshift of the quasar}.



\subsection{Target Requirements for CORE and Lyman-$\alpha$ quasars}
\label{sec:SRD}

Full details of the techniques used to forecast requirements for \eboss\ quasars are provided in our companion 
overview paper \citep{ebosspaper???}. Those forecasts imply the following broad
requirements for quasar target selection, driven by instrument capabilities and a 2\% measurement of the 
BAO distance scale (G.\ Zhao et al.\ 2016, in preparation). For the CORE quasars:

\begin{enumerate}

\item Survey area $> 7500$\degsq
\item Total number of $0.9 < z < 2.2$ quasars $> 435{,}000$ (this corresponds to 58\perdegsq\ over exactly 7500\degsq)
\item A total density of {\em assigned fibers} of $< 90$\perdegsq\ (effectively a {\em target} density of $\leqsim 115$\perdegsq\ for reasons noted at the end of this section)
\item Redshift precision $< 300$\kms\ RMS for $z < 1.5$ and $(300 + 400(z - 1.5))$\kms\ for $z > 1.5$\footnote{see the \eboss\ overview paper \citep{ebosspaper???} for a discussion of this requirement and \citet{Hew10} for details of the precision of \sdss\ quasar redshifts.}
\item Catastrophic redshift errors (exceeding 3000\kms) $< 1$\%, where the redshifts are not known to be in error
\item Maximum absolute variation in expected target density as a function of imaging survey sensitivity, stellar density, and Galactic extinction of $< 15$\% within the survey footprint
\item Maximum fluctuations in target density due to imaging zero-point errors of $<15$\% in each individual band used for targeting

\end{enumerate}

Once these CORE requirements are met, remaining fibers not allocated to other \eboss\ target classes are assigned
to the \LyA\ target class. These \LyA\ quasars have the following {\em additional} constraints and requirements:

\begin{enumerate}

\item \boss\ quasars within the \eboss\ area with ${\rm SNR\,pixel^{-1}}$ = 0\footnote{SNR is defined as the mean S/N per \LyA\ Forest pixel measured over the rest-frame wavelength range of 
1040\,\AA$\,< \lambda < 1200$\,\AA. A ``pixel'' here refers to a single bin of wavelength in a \boss\ spectrum. The logic behind retargeting ${\rm SNR\,pixel^{-1}}$ = 0 spectra is that they are almost certainly
{\em bad}, whereas $0 \leq {\rm SNR\,pixel^{-1}}< 0.75$ spectra are ``good'' but are of irrecoverably low S/N (see \S\ref{sec:reobs}).
}, or $0.75 < {\rm SNR\,pixel^{-1}} < 3$ must be reobserved
\item Flux calibration at least as accurate as \boss
\item Recalibration of the \boss\ high-$z$ quasar sample using a spectroscopic pipeline that is consistent with that of \eboss\

\end{enumerate}

A subtlety arises for item (3) of the CORE requirements; targets with existing good spectroscopy from earlier iterations of the \sdss\ are not assigned fibers as part of \eboss\
(see \S\ref{sec:donotobs}). On average, this saves
25\,fibers\,\perdegsq. Typically, therefore, this paper will quote a total target density of 115\perdegsq\ but this corresponds to a density of assigned fibers of only 90\perdegsq\ for CORE quasars.

\begin{figure*}[t]
\centering
\includegraphics[width=0.99\textwidth,height=1.09\textwidth]{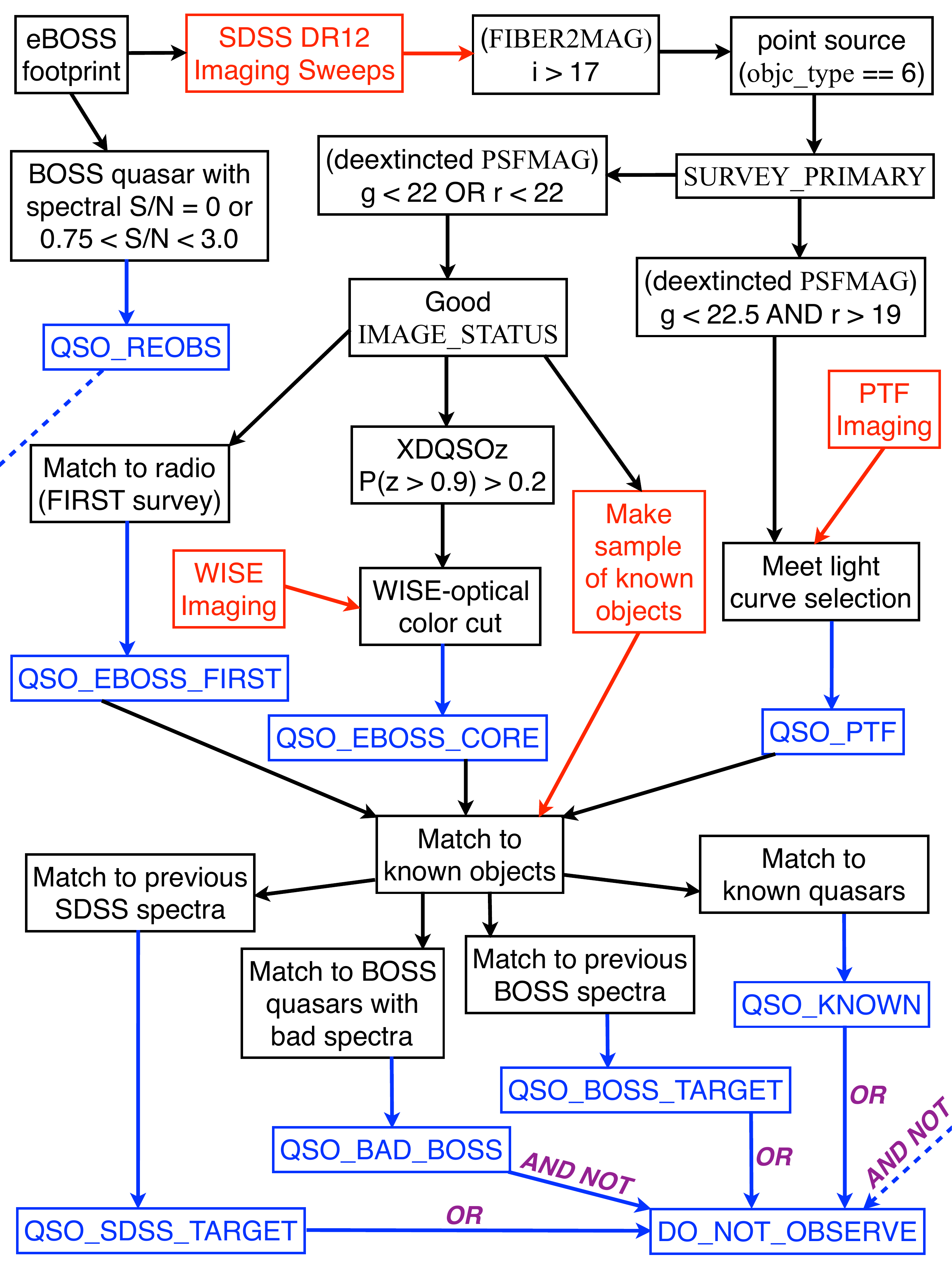}
\caption{\small 
Flowchart depicting \eboss\ quasar target selection. Red boxes represent sources of 
input information such as imaging (see \S\ref{sec:imaging})
or catalogs of known objects. Black boxes depict cuts that are made to the 
input sources as part of the target selection algorithm (see \S\ref{sec:qts}). Blue boxes 
depict output target selection bits (see \S\ref{sec:bits}). The Boolean
 terms in purple describe how the four bits produced by matching to previous spectra 
 are combined to set the {\tt DO\_NOT\_OBSERVE} bit 
(see \S\ref{sec:donotobs}).  The dashed blue arrow indicates that
{\tt QSO\_REOBS} targets are {\em always} reobserved, regardless of the value of {\tt DO\_NOT\_OBSERVE}.
The sample of known objects undergoes
the CORE flag and magnitude cuts rather than the \ptf\ magnitude cuts. Consequently, \ptf\
selection could re-target previously known objects with bad {\tt IMAGE\_STATUS} and/or with 
$22 < g < 22.5$.
}
\label{fig:flowchart}
\end{figure*}

\section{Parent Imaging for Target Selection}
\label{sec:imaging}

\subsection{Updated calibrations of \sdss\ imaging}
\label{sec:v5b}

All \eboss\ quasar targets are ultimately tied to the 
\sdss-\project{I}/\project{II}/\project{III}
images collected in the $ugriz$ system \citep{Fuk96} using the 
wide-field imager \citep{Gun98} on the \sdss\ telescope \citep{Gun06}. 
\sdss-\project{I}/\project{II} mostly derived imaging over the $\sim8400$\degsq\ ``Legacy'' area, 
$\sim90$\% of which was in the North Galactic Cap (NGC). This imaging was
released as part of \sdss\ Data Release 7 \citep[\drsev;][]{DR7}. The legacy imaging
area of the \sdss\ was expanded by $\sim2500$\degsq\ in the South Galactic Cap (SGC) as 
part of \dre\ \citep{DR8}. The \sdss-\project{III}/\boss\ survey used \dre\ imaging for target
selection over $\sim7600$\degsq\  in the NGC and $\sim3200$\degsq\  in the 
SGC \citep{Daw13}. Quasar targets are selected for \eboss\ over the same areas as \boss, and
ultimately \eboss\ will observe quasars over a subset of at least 7500\degsq\ of this area.

Although adopting the same {\em area} as \boss, \eboss\ target selection
takes advantage of updated calibrations of the \sdss\ imaging. 
\citet{Sch12} have applied the ``uber-calibration'' 
technique of \citet{Pad08} to \panstarrs\ imaging \citep{panstarrs},
achieving an improved global calibration compared to \sdss\dre. Targeting
for \eboss\ is conducted using \sdss\ imaging that is calibrated
to the \cite{Sch12} \panstarrs\ solution, as fully detailed in D.\ Finkbeiner et al.\ (2016, in preparation).
We will refer to this set of observations as the ``updated'' imaging.
 
The specific version of the updated \sdss\ imaging used in \eboss\ target selection is 
as stored in the {\tt calib\_obj} or ``data sweep'' files \citep{Bla05}. These data correspond to 
the native files used in the \sdss-\project{III} data model\footnote{e.g., \url{http://data.sdss3.org/datamodel/files/PHOTO\_SWEEP \\ /RERUN/calibObj.html}}
and the updated  \panstarrs-calibrated data sweeps will be made available in a future \sdss\ Data Release. 
The magnitudes derived
from these data sweeps are AB magnitudes {\em not}, e.g., asinh ``Luptitudes'' \citep{Lup99}. Note that
the \xdqsoz\ targeting technique \citep{Bov12} adopted by \eboss\ is designed to handle noisy data, so can rigorously
incorporate small (and even negative) fluxes when classifying quasars.

\subsection{WISE}
\label{sec:WISE}

The Wide-Field Infrared Survey Explorer (\wise; \citealt{WISE}) surveyed
the full sky in four mid-infrared bands centered on 3.4\,\micron, 4.6\,\micron, 
12\,\micron, and 22\,\micron, known as W1, W2, W3 and W4.  For \eboss\ we
use only the W1 and W2 bands, which are substantially deeper than W3 and W4.  Over the course of its primary mission and
``NEOWISE post-cryo" continuation, \wise\ completed two full scans of the sky
in W1 and W2.  Over 99\% of the sky has 23 or more exposures in W1 and W2; the median coverage is 33 exposures.
We investigate whether the non-uniform spatial distribution of WISE exposure depth 
presents a problem for modeling CORE quasar clustering in \S\ref{sec:homo}.

We use the ``unWISE'' coadded photometry from \citet{unwise} applied
to \sdss\ imaging sources \citep[as detailed in][]{Lan14}. This approach produces forced
photometry of custom coadds of the \wise\ imaging at the positions of all \sdss\
primary sources.  Using forced photometry rather than catalog-matching
avoids issues such as blended sources and non-detections.
Since the \wise\ scale is $2.75$\arcsec\,pixel$^{-1}$ (roughly seven times
as large as \sdss), and since many of our targets have \wise\ fluxes below
the ``official'' \wise\ catalog detection limits, using forced photometry is
of significant benefit.

\subsection{PTF}
\label{sec:PTF}

The Palomar Transient Factory\footnote{See \url{http://irsa.ipac.caltech.edu/Missions/ptf.html} for the public \ptf\ data} (\ptf)  is a wide-field photometric survey aimed at a systematic exploration of the optical transient sky via repeated imaging over 20,000\degsq\
in the Northern Hemisphere \citep{Rau09, Law09}. The PTF image processing is presented in \citet{Lah14}, while the photometric calibration, system and filters are discussed in \citet{Ofe12}.
In February 2013, the next phase of the program, \iptf\ (intermediate \ptf), began. Both surveys use the CFHT12K mosaic camera, mounted on the 1.2\,m Samuel Oschin Telescope at Palomar Observatory. The camera has an 8.1\degsq\  field of view and 1\arcsec\ sampling. Because one detector (CCD03) is non-functional, the usable field of view is reduced to 7.26\,\degsq. Observations are mostly performed in the Mould-$R$ broad-band filter, with some in the  SDSS $g$-filter. Under median seeing conditions, the images are obtained with 2.0\arcsec\ FWHM, and reach 5$\sigma$ limiting AB magnitudes of  $m_R \simeq 20.6$ and $m_{g'} \simeq 21.3$ in 60-second exposures. The cadence varies between fields, and can produce one measurement every five nights in regions of the sky dedicated to supernova searches. Four years of \ptf\ survey operations have yielded a coverage of $\sim90$\% of the 
\eboss\ footprint. 

 
Two automated data processing pipelines are used in parallel in the search for transients; a near-real-time image subtraction pipeline at Lawrence Berkeley National Laboratory (LBNL), and a database populated on timescales of a few days at the Infrared Processing and Analysis Center (IPAC). The \eboss\ analysis uses the individual calibrated frames available from IPAC~\citep{Lah14}. 

We have developed a customized pipeline
based on the SWarp \citep{Bertin2002} and SCAMP \citep{Bertin2006} public packages
to build coadded \ptf\ images on a timescale adapted to quasar targeting---i.e., typically 1 to 4 epochs per year depending on the cadence and total exposure time
within each field. Using the same algorithms, a full stack is also constructed by coadding all available images.
This full stack is complete at 3$\sigma$ to $g\sim22.0$, and has over 50\% completeness to quasars at $g\sim22.5$.
The full stack is used to extract a catalog of \ptf\ sources from each of the coadded \ptf\ images.
The light-curves (flux as a function of time) for all of these \ptf\ sources are measured.

\section{Quasar target classes}
\label{sec:qts}

As only a limited number of fibers are available in the \eboss\ experiment, each target
class is assigned a different target density to optimize scientific return. \eboss\ will
attempt to make the first 2\% measurement of the BAO scale at a redshift near $z\sim1.5$, and
the uniqueness of this measurement led to statistically selected $0.9 < z < 2.2$ quasars being
prioritized at a density of 90\perdegsq\ fibers. As noted in \S\ref{sec:SRD} because objects targeted
by past \sdss\ projects do not need to be reobserved, this fiber allocation effectively corresponds to a density
of 115\perdegsq\ targets.
\eboss\ will also attempt to augment \boss\ measurements of clustering in the \LyA\ Forest, improving BAO constraints
from near 2\% to closer to 1.5\%. This program is assigned the remaining available \eboss\
fibers once other target classes have been accounted for, typically resulting in $\sim 20$\perdegsq\ targets.
The combined cosmological constraints that can be achieved by this overall program design are detailed in
G.\ Zhao et al.\ (2016, in preparation).

As further discussed in \S\ref{sec:goals}; there are therefore two distinct target classes in \eboss: CORE quasars and
\LyA\ quasars. The CORE quasars are targeted in a
statistically reproducible fashion, with the intention of using them to measure clustering over redshifts of $0.9 < z < 2.2$. 
The \LyA\ quasars are targeted to lie at $z > 2.1$ to augment the BAO signal detected by \boss. These
two categories of quasars are not mutually exclusive, in that the CORE quasars are not
constrained to lie at $z < 2.1$ and so the CORE selection algorithm can also identify 
\LyA\ quasars. In the rest of this section, we discuss each of the \eboss\ target classes in detail.
The full targeting algorithm is also depicted by a flow-chart in Fig.\ \ref{fig:flowchart}.

\subsection{Broad overview of the CORE quasar sample}
\label{sec:core}

The \eboss\ CORE sample is designed to provide a statistically selected sample of
$115$\perdegsq\ targets that, after \eboss\ spectroscopy of the $90$\perdegsq\ targets that
do not have existing good \sdss\ spectra, comprises $> 58$\perdegsq\ total quasars with accurate redshifts
in the range $0.9 < z < 2.2$ (see \S\ref{sec:goals}). This $> 58$\perdegsq\ quasars will consist of
both new quasars from \eboss\ spectroscopy and previously known quasars from the sample of
$25$\perdegsq\ targets that have existing \sdss\ spectroscopy. To achieve this goal \eboss\ uses 
two complementary methods; an optical selection using the \xdqsoz\ method of \citet{Bov12}, and a mid-IR-optical color cut using \wise\ imaging. The specifics of these two methods are detailed in the next few sections.

The starting sample for CORE targeting is all {\em point sources} in \sdss\ imaging that are {\tt PRIMARY}, have (de-extincted) PSF magnitudes of $g < 22$ {\tt OR} $r < 22$ and a {\tt FIBER2MAG}\footnote{{\tt FIBER2MAG} corresponds to the flux through a fiber with a 2\arcsec\ diameter,
appropriate to \boss. Surveys with the \sdss\ spectrographs instead used {\tt FIBERMAG}, appropriate
to a 3\arcsec\ fiber diameter.} of $i > 17$, and that have good {\tt IMAGE\_STATUS}.\footnote{in fact, all target classes detailed in this paper undergo these cuts with the exception of the variability-selected sample discussed in \S\ref{sec:variability}} These basic initial cuts are discussed further in \S\ref{sec:flags}. 

Point sources in the \sdss\ are denoted by the flag {\tt objc\_type == 6}, corresponding
to a magnitude cut based on star-like or galaxy-like profile fits of ${\tt psfMag -  modelMag} \leq 0.145$ \citep{EDR}.
A concern might be that a selection to $r \sim 22$ might suffer incompleteness to quasars at $r~\geqsim\,21$
where star-galaxy separation in \sdss\ imaging was initially argued to break down due to errors on profile fits \citep[e.g.][]{EDR,Scr02}. In general,
though, at the limit of the \sdss\ imaging the trend is to classify faint, ambiguous sources as point-like. The expectation is then that a selection
approaching $r\sim 22$ will become increasingly contaminated by galaxies that are classified as unresolved, rather than miss quasars
that are classified as resolved
\citep[see also the discussion in \S4.5.1 of][]{Ric09KDE}. Further, requiring {\tt objc\_type == 6} 
{\em and} applying \xdqsoz\ reduces galaxy contamination to $\leqsim\,10\%$ even at $i\sim22$ \citep[see Figure 11 of][]{Bov12},
 so we expect our selection to remain robust even to $r\sim22$ (which, on average, corresponds to $i\sim 21.85$ for $0.9 < z < 2.2$ quasars).

From the initial sample of magnitude-limited {\tt PRIMARY} point sources, objects are targeted if they have an \xdqsoz\ 
probability of being a quasar at $z > 0.9$ of more than 20\%, i.e., PQSO($z > 0.9) > 0.2$.
{\em It is important to note the subtle distinction between the specific goal of the CORE sample
and the sample it produces}. The {\em goal} of the CORE is to uniformly target $> 58$\perdegsq\ quasars 
in the redshift range $0.9 < z < 2.2$ but {\em no attempt is made to restrict the upper redshift range of the CORE quasar sample}.
The CORE is left free to recover quasars at $z > 2.2$ because, although such quasars are outside the preferred CORE redshift range, they remain useful as tracers of the Lyman-$\alpha$ Forest. To this moderate-probability \xdqsoz\ sample, a \wise-optical color cut is applied to further reduce the target density by filtering out obvious stars based on optical-mid-IR colors. Finally, objects are {\em not} targeted if they have existing good spectroscopy from earlier iterations of the \sdss\ unless a visual inspection as part of \boss\ produced an ambiguous classification. The resulting set of objects comprises the \eboss\ CORE quasar sample.

\subsubsection{\xdqsoz}
\label{sec:xdqsozcuts}

\xdqso\ \citep{Bov11a}  is a method of classifying quasars in flux-space using {\em extreme deconvolution}
\citep[\xd;][]{Bov11b} to estimate the density distribution of quasars as compared to non-quasars. Effectively,
\xdqso\ takes any test point in flux-space, together with its flux errors, and convolves that error
envelope with deconvolved distributions of the quasar and of the non-quasar loci. By weighting this convolution
with a prior representing the expected numbers of quasars and non-quasars, the test point is
assigned a probability of being
a quasar. \xdqso\
inherits many desiderata from \xd, including the rigorous incorporation of (and extrapolation from)
errors on fluxes, and the ability to distinguish the effect on quasar probabilities of data that are 
completely missing from data that are {\em merely of low significance}.
This feature is a boon for quasar classification near the limits of imaging
data where flux errors are large. For \eboss\ targeting, we adopt the \xdqsoz\ method \citep{Bov12} which
extends the \xdqso\ schema to provide probabilistic classifications for quasars in any specified range of redshift.

In pursuit of the \eboss\ CORE goal of $> 58$\perdegsq\ $0.9 < z < 2.2$ quasars, a test spectroscopic survey in the 
W3 field of the CFHT Legacy
Survey\footnote{\url{http://terapix.iap.fr/cplt/oldSite/Descart/summarycfhtlswide \\ .html}} was conducted. 
This CFHTLS-W3 test survey was
deemed necessary as no iteration of the \sdssi/\project{II}/\project{III} specifically targeted quasars to as faint as $r\sim22$ over the redshift range $0.9 < z < 2.2$. Although the CFHTLS-W3 test survey informed the initial quasar target
selection for \eboss, and so will be used to describe the broad ideas behind that target selection, 
it only contained $\sim 1{,}600$ quasars and was easily supplanted by the \sequels\ survey described in \S\ref{sec:results}, which
comprised $\sim 21{,}700$ quasars. Readers interested in an up-to-date description and depiction of the properties of \eboss\ quasars as compared
to \sdssi/\project{II}/\project{III}, should therefore consult \S\ref{sec:properties} and, in particular, Fig.\ \ref{fig:sequelsnz} and Fig.\ \ref{fig:sequelsMz}.

The CFHTLS-W3 test survey is detailed
in the appendix of \citet{DR12}. Broadly, an optical selection was applied to \sdss\dre\ imaging, 
restricting to {\tt PRIMARY} point sources in the (PSF, unextincted) magnitude 
range $17 < r < 22$. From this initial sample, objects were targeted for follow-up spectroscopy if they had an \xdqsoz\ probability 
of greater than 0.2 of being a quasar at {\em any} redshift (i.e., PQSO($z > 0.0) > 0.2$). 

\begin{table}[t]
\label{tab:cfhts}
\centering
\caption{\small
Efficiency of Quasar Target Selection in the CFHTLS-W3 test survey as a function of \xdqsoz\ probability cut}
\begin{tabular}{|c |r r| r r| r r|}
\hline
{\footnotesize ID}  & \multicolumn{6}{c|}{{\footnotesize PQSO}} \\
  \hline 
{\footnotesize (rows 1--4)} & & & & & \multicolumn{2}{|c|}{{\footnotesize ($z > 0.0) > 0.2$}}  \\
{\footnotesize $z_{\rm spec}$ range} & \multicolumn{2}{c|}{{\footnotesize ($z > 0.0)$}} & \multicolumn{2}{|c|}{{\footnotesize ($z > 0.9)$}} & \multicolumn{2}{|c|}{{\footnotesize \&\&}}  \\
{\footnotesize for quasars} & \multicolumn{2}{c|}{{$> 0.2$}} & \multicolumn{2}{|c|}{{\footnotesize $> 0.2$}} & \multicolumn{2}{|c|}{{\footnotesize ($z > 0.9) < 0.2$}}  \\
{\footnotesize (rows 5--7)} & {\footnotesize N} & {\footnotesize \%} & {\footnotesize N} & {\footnotesize \%} & {\footnotesize N} & {\footnotesize \%} \\ 
\hline
{\footnotesize Stars}  & 27.0 & 18.2\% & 23.3 & 16.8\% & 3.6 & 39.6\%  \\
{\footnotesize Galaxies} & 13.9 & 9.4\% & 12.3 & 8.8\% & 1.6 & 17.8\%  \\
{\footnotesize Unidentified}  & 2.4 & 1.6\% & 2.2 & 1.6\% & 0.2 & 2.0\%  \\
{\footnotesize Quasars} & 105.0 & 70.8\% & 101.3 & 72.8\% & 3.7 & 40.6\%  \\
\hline
{\footnotesize $z < 0.9$} & 13.2 & 8.9\% & 10.9 & 7.9\% & 2.3 & 24.8\%  \\
{\footnotesize $0.9 < z < 2.2$} & 70.9 & 47.8\% & 69.7 & 50.1\% & 1.2 & 12.9\%  \\
{\footnotesize $z > 2.2$} & 20.9 & 14.1\% & 20.7 & 14.9\% & 0.3 & 3.0\%  \\
\hline
{\footnotesize Total} & 148.3 & 100\% & 139.1 & 100\% & 9.2 & 100\%  \\
\hline
\end{tabular}
\tablecomments{The total survey area was 11.0\degsq\ and ``N,'' the number of spectroscopically confirmed targets, is always expressed in \perdegsq\ over this area.}
\end{table}

\begin{figure}[t]
\centering
\includegraphics[width=0.48\textwidth,height=0.3\textwidth]{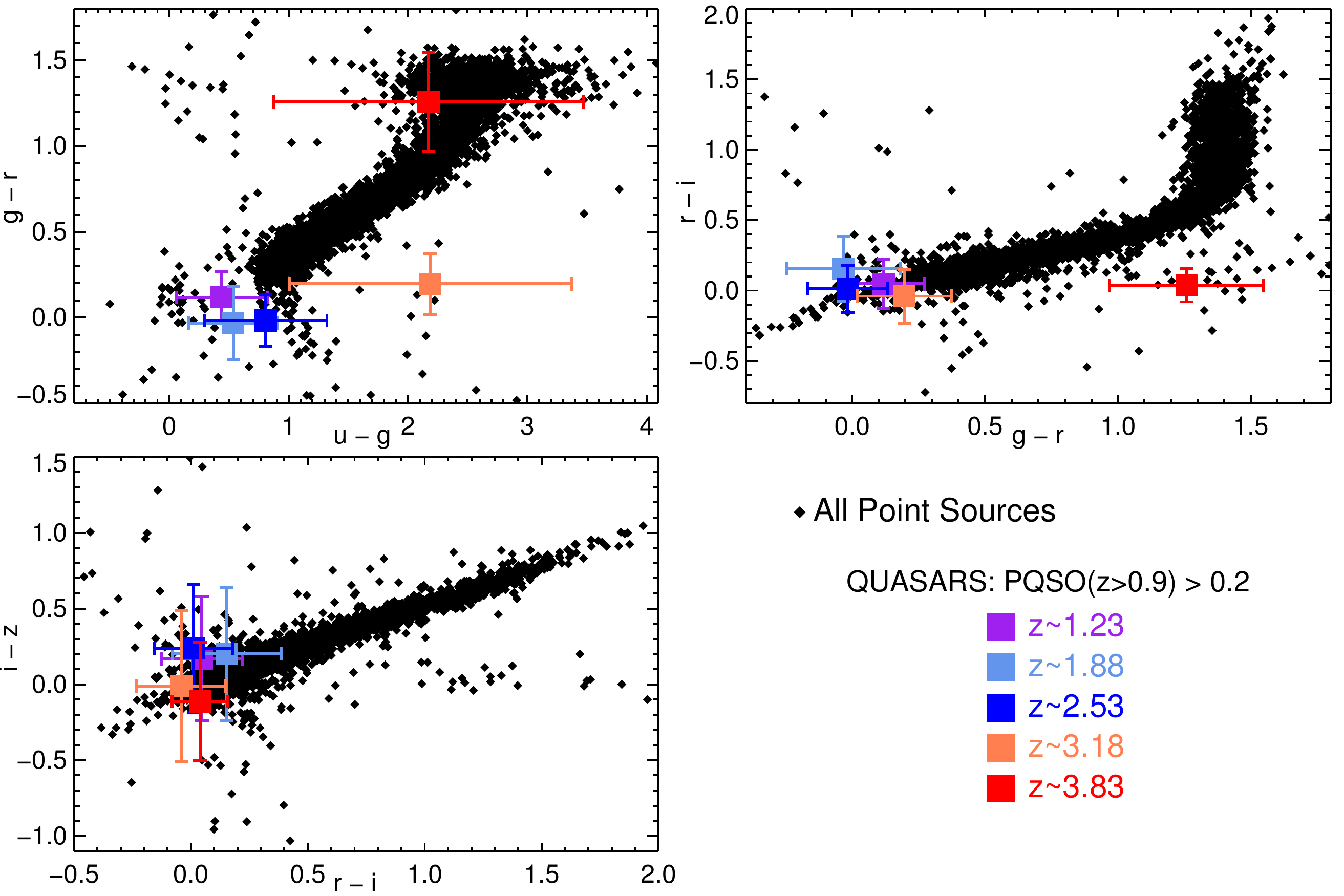}
\caption{\small 
The position of \xdqsoz-selected PQSO($z > 0.9) > 0.2$ quasars in $ugriz$ optical color space (using PSF magnitudes).
Black points depict $r < 19$ {\tt PRIMARY} point sources from a randomly chosen \sdss\ imaging run (5225). The $r < 19$ limit
is chosen in order to illustrate the position of the stellar locus in \sdss\ filters; at fainter limits the locus widens considerably \citep[see, e.g., Figures 5 and 6 of][]{Bov11a}. 
Spectroscopically confirmed 
PQSO($z > 0.9) > 0.2$ quasars from \boss\ (\drt; squares) are plotted as a function of redshift, from $z=0.9$ to $z=4.15$ in bins of $\Delta z=0.65$. 
The error bars indicate the 1$\sigma$ scatter. 
}
\label{fig:opticalcut}
\end{figure}

As the CFHT W3 test survey targeted objects regardless of their redshift probability density 
(all objects with PQSO($z > 0.0) > 0.2$) the results of the survey could be optimized to better recover quasars
in the \eboss\ CORE redshift range of $0.9 < z < 2.2$.
One initial outcome of the CFHT W3 test survey, then, was that objects with PQSO($z > 0.0) > 0.2$ but 
PQSO($z > 0.9) < 0.2$ were rarely quasars in the \eboss\ redshift range of interest, 
as demonstrated in Table~1.
Further, restricting the redshift range of
\eboss\ quasar targets to $z > 0.9$ is desirable to mitigate losses of, e.g., \eboss\ Luminous Red Galaxies targeted at 
$z < 0.9$ \citep[c.f.][]{Prakash15inprep???} due to fiber collisions between neighboring targets. Therefore, it was decided to focus only on targets with 
PQSO($z > 0.9) > 0.2$ for \eboss\ targeting; we will subsequently restrict our discussion to such targets.

Fig.\ \ref{fig:opticalcut} shows the typical positions of \xdqsoz\ PQSO($z > 0.9) > 0.2$ quasars in 
\sdss\ colors. To demonstrate the position of \xdqsoz-selected quasars in optical color space, we use the large spectroscopically
confirmed quasar sample from the \drt\ quasar catalog of \citet{Par14}.
In general, \xdqsoz\ selects similar regions of color space to \sdss\ targets from earlier surveys \citep[e.g.,][]{Ric01}, with the majority of the quasar-star 
separation occuring in the $ugr$ filters. 

\begin{figure}[t]
\centering
\includegraphics[width=0.48\textwidth,height=0.3\textwidth]{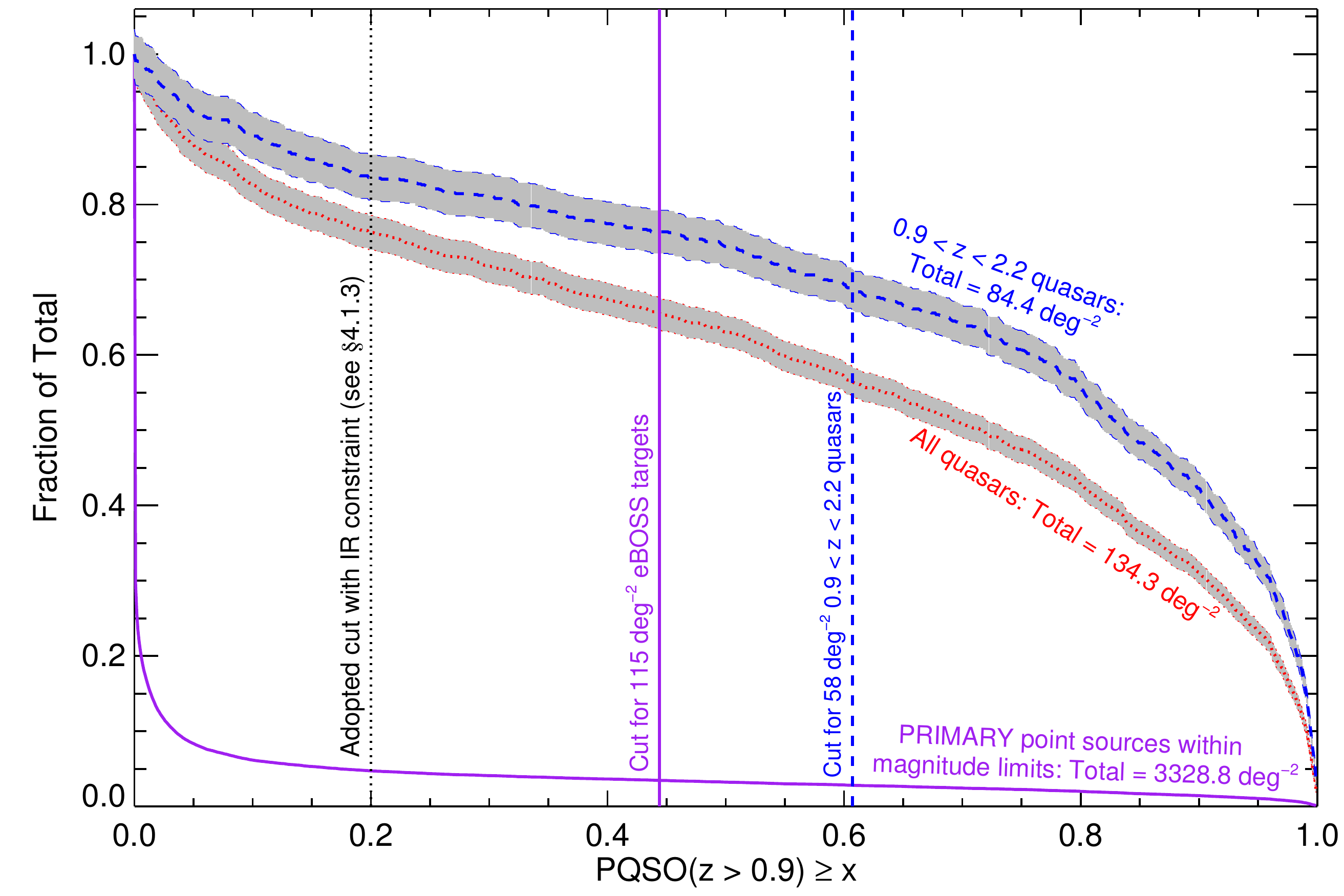}
\caption{\small 
The cumulative sky density of quasars and targets as a function of $z > 0.9$ \xdqsoz\ probability.
The upper curves represent all quasars (red), and $0.9 < z < 2.2$ quasars (blue), from the CFHTLS-W3 test program.
These curves yield an estimate of the {\em completeness} of \eboss\ to quasars
for various PQSO($z > 0.9)$ constraints. Grey contours illustrate the (Poisson) errors.
The lowest curve represents all sources from \sdss\ imaging in the CFHTLS-W3 test 
region (magenta). 
This curve yields an estimate of the necessary {\em fiber budget} for \eboss. 
A quantitative example of how to use the curves 
to predict quasar and target densities is provided in \S\ref{sec:xdqsozcuts}. The vertical lines depict the adopted cut
for \eboss\ (after also applying an optical-IR color cut; see \S\ref{sec:combcuts}), 
the cut for the \eboss\ requirement of 58\perdegsq\ $0.9 < z < 2.2$ quasars, and the cut to assign $< 115$ \perdegsq\eboss\ fibers (the maximum assignable;
see \S\ref{sec:SRD}).
All samples depicted have been limited to \sdss\ {\tt PRIMARY} point sources with {\tt FIBER2MAG} of $i > 17$ and de-extincted PSF magnitudes of 
$g < 22$ OR $r < 22$ (the initial cuts for the \eboss\ CORE). 
}
\label{fig:opticalcomp}
\end{figure}

Whether an \xdqsoz\ PQSO($z > 0.9)$ selection alone is sufficient to
meet the \eboss\ targeting goal of 58\perdegsq\ quasars is investigated in Fig.\ \ref{fig:opticalcomp}, where the sky density of \xdqsoz-selected
targets as a function of probability threshold is compared to that of confirmed quasars in the requisite CORE 
redshift range ($0.9 < z < 2.2$; see \S\ref{sec:SRD}). Fig.\ \ref{fig:opticalcomp} displays three curves that correspond to source densities in
the CFHTLS-W3 test program,
which can be used to
estimate the ``true'' densities of quasars and targets expected in \eboss. The lowest (magenta) curve represents  
all sources in \sdss\ imaging in the CFHTLS-W3 field that meet the basic CORE cuts (i.e., {\em PRIMARY} point sources within the
CORE magnitude limits); as a fraction of the total density of $\sim3330$\perdegsq\ such sources. The central (red) curve represents all quasars that
were spectroscopically confirmed as part of the CFHTLS-W3 program as a fraction of the total density of $\sim135$\perdegsq\ such sources. 
The upper (blue) curve represents all quasars in the specific CORE redshift range of $0.9 < z < 2.2$ that
were spectroscopically confirmed as part of the CFHTLS-W3 program as a fraction of the total density of $\sim85$\perdegsq\ such sources. 
As the CFHTLS-W3 program was limited to PQSO($z > 0.0) > 0.2$, the test sample is partially incomplete to quasars that have
PQSO($z > 0.9) < 0.2$; such quasars only appear in the CFHTLS-W3 test data due to targeting approaches that did not use \xdqsoz-selection.  
Fig.\ \ref{fig:opticalcomp} therefore provides best estimates only for PQSO($z > 0.9) > 0.2$.

Fig.\ \ref{fig:opticalcomp} can be used to estimate the total density of quasars and targets that might be expected in \eboss\ for
different PQSO($z > 0.9$) constraints. For example, to estimate the sky density of all quasars at PQSO($z > 0.9) > 0.6$, one would
find the corresponding {\em Fraction of Total}  ($\sim0.57$) and multiply by the total for all quasars (134.3\perdegsq) to obtain $\sim77$\perdegsq. 
The vertical lines in Fig.\ \ref{fig:opticalcomp} depict the necessary constraints to achieve the requisite \eboss\ CORE density of
58\perdegsq\ $0.9 < z < 2.2$ quasars and the requisite \eboss\ target density of 115\perdegsq\ (see \S\ref{sec:SRD}).
The maximum target density of 115\perdegsq\ is achieved at PQSO($z > 0.9) > 0.45$, which would result in 
64.9\perdegsq\ CORE quasars. In actuality, a more relaxed constraint of PQSO($z > 0.9) > 0.2$ is adopted for \eboss
\footnote{Note that this
parameter space extends well beyond the effective equivalent cut of PQSO($2.2 < z < 3.5) > 0.424$ that was adopted for \boss.}, which
further improves quasar targeting. This relaxed constraint, which is 
labeled ``Adopted cut with IR constraint (see \S\ref{sec:combcuts})'' in Fig.\ \ref{fig:opticalcomp}, was achieved through an 
additional constraint on mid-IR-optical color (see also \S\ref{sec:wisecuts}).

\begin{figure}[t]
\centering
\includegraphics[width=0.48\textwidth,height=0.31\textwidth,clip=true,trim=20 0 0 25]{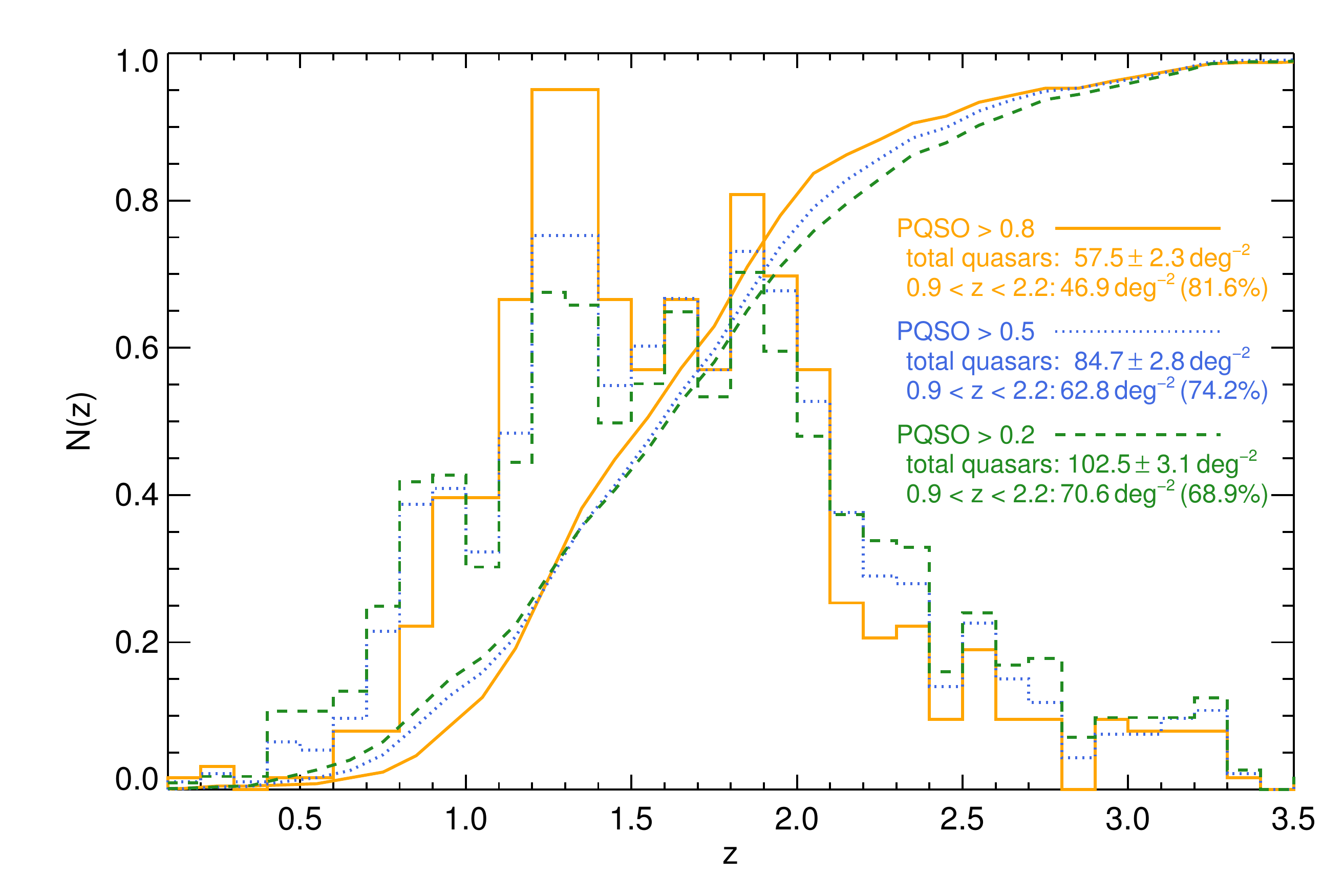}
\caption{\small 
The redshift distribution of spectroscopically confirmed quasars from the CFHTLS-W3 test program. The distributions that peak in the $0.9 < z < 2.2$ range are the 
redshift Probability Density Functions (PDFs). The distributions that climb to 1 near $z\sim3.5$ are cumulative. The distributions for three 
different cuts on the $z > 0.9$ \xdqsoz\ probability are depicted; PQSO($z > 0.9) > 0.8$ (orange, solid),
PQSO($z > 0.9) > 0.5$ (blue, dotted), and PQSO($z > 0.9) > 0.2$ (green, dashed).
}
\label{fig:opticalNz}
\end{figure}

Fig.\  \ref{fig:opticalNz} depicts how relaxing constraints on PQSO($z > 0.9$) to thresholds as low
as our adopted PQSO($z > 0.9) > 0.2$ affects the redshift distribution of targeted quasars.
The resulting $N(z)$ distributions are
broadly similar, but the PQSO($z > 0.9) > 0.2$ selection has a tail to $z < 0.9$ and contains a smaller fraction of
quasars in the CORE target range of $0.9 < z < 2.2$. This drop is more than offset by the PQSO($z > 0.9) > 0.2$
selection containing more total quasars (c.f., Fig.\ \ref{fig:opticalcomp}). The peak near $z\sim1.3$ is
likely an artifact of the small sample size in the CFHTLS-W3 test program (c.f., Fig.\ \ref{fig:sequelsnz}).
Fig.\  \ref{fig:opticalNz} demonstrates that the majority of quasars 
selected at PQSO($z > 0.9) > 0.2$ remain useful for \eboss\ by being in the CORE redshift 
range of $0.9 < z < 2.2$. In fact, there is an additional advantage to relaxing the \xdqsoz\ probability; 
doing so tends to introduce new quasars at $z > 2.1$ while retaining the quasars in the
CORE redshift range. Quasars at $z > 2.1$ remain useful for the purposes of \eboss\ as part of 
the \LyA\ sample (see \S\ref{sec:LyA}).

\subsubsection{Mid-IR-optical color cuts}
\label{sec:wisecuts}

Starlight tends to greatly diminish at wavelengths
redwards of 1--$2\mu$m, making galaxies, and in particular stars, dim in the mid-IR, whereas Active Galactic Nuclei 
(AGN) have considerable IR emission. Photometric selection 
techniques based on \wise\ data can therefore be used to target active galaxies, and such techniques uncover both
unobscured and obscured quasars over a range of luminosities \citep[e.g.][]{Ste12,Ass13,Yan13}.

\begin{figure}[t]
\centering
\includegraphics[width=0.48\textwidth,height=0.3\textwidth]{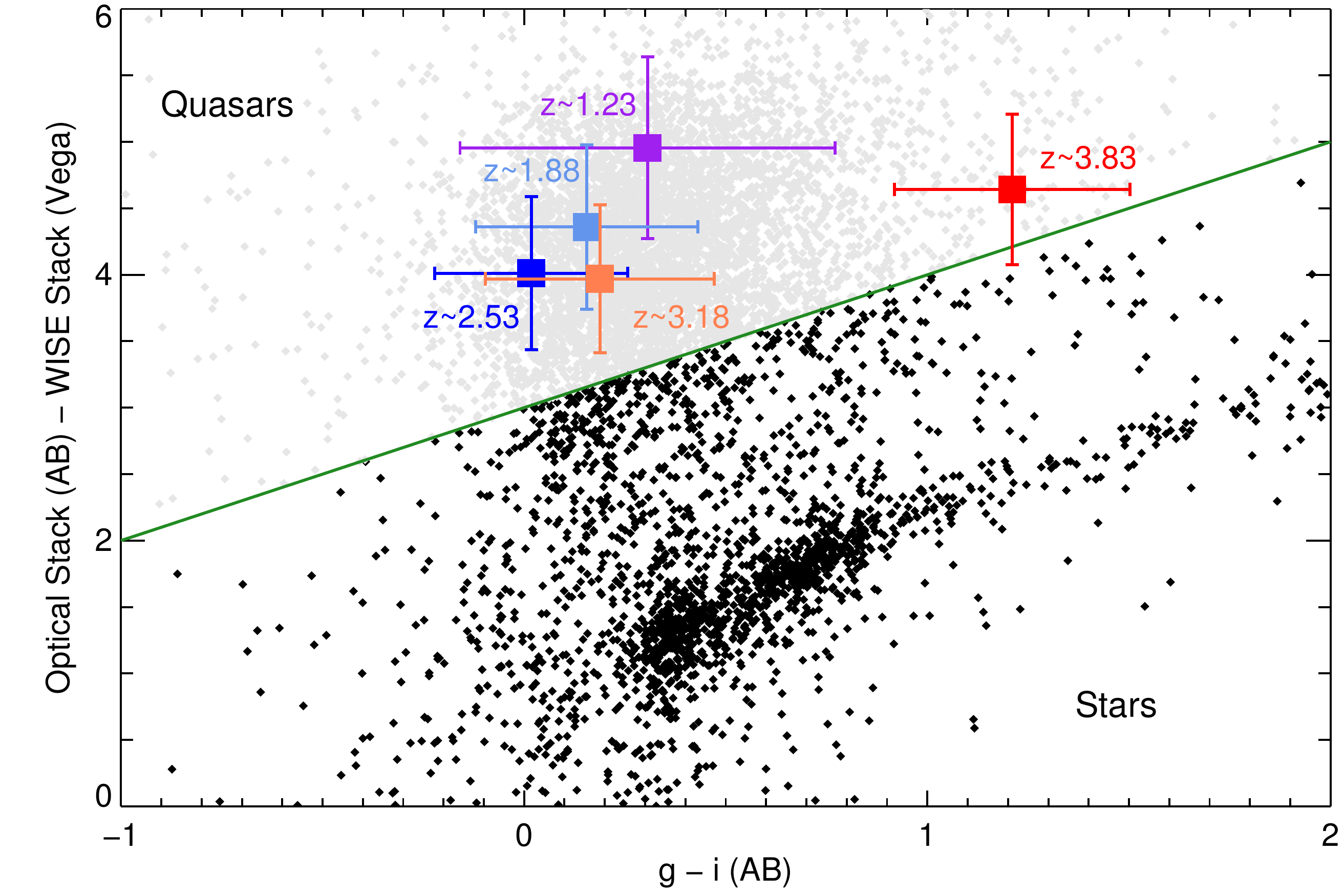}
\caption{\small 
The optical-IR cut (applied to PSF magnitudes) used to define \eboss\ CORE quasar targets. The green line depicts the color cut in
\sdss\ ($f_{g} + 0.8f_{r}+0.6f_{i})/2.4$ and \wise\ ($f_{W1} + 0.5f_{W2})/1.5$ stacks versus $g-i$ that was
used to target quasars as part of the CFHTLS-W3 test program. Quasars of 
interest to \eboss\ ($z~\leqsim\,3.5$) generally occupy the region above this line; the stellar locus is a dense 
region in the lower part of the plot. Black points depict objects with {\em any} \eboss\ targeting bit set (see \S\ref{sec:bits}) 
from a randomly chosen \sdss\ imaging run (5225) limited to $g < 22$. 
Spectroscopically confirmed quasars from \boss\ (\drt; squares) are plotted as a function of 
redshift, from $z=0.9$ to $z=4.15$ in bins of $\Delta z=0.65$. The error bar indicates the 1$\sigma$ scatter. 
}
\label{fig:wisecut}
\end{figure}

Significantly more than half of the objects targeted using mid-IR selection are low-luminosity unobscured
AGN at $z < 1$ or obscured quasars \citep[e.g.,][]{Lac13,Hai14}. This makes a pure \wise\ selection approach imperfect for \eboss\ targeting, as 
objects without an optical spectrum and/or AGN at $z < 0.9$ will not typically have utility for the
\eboss\ CORE goal of targeting $> 58$\perdegsq\ $0.9 < z < 2.2$ quasars. \wise\ remains ideal, however, for {\em removing
contaminating stars} from \eboss\ quasar selection. Fig.\ \ref{fig:wisecut} demonstrates the utility of a
\wise-optical color cut in selecting against stars. This color cut is based on
stacking optical and \wise\ fluxes to attain as great a depth as possible. A stack is created from \sdss\ PSF fluxes 
according to

\begin{equation}
{\rm Optical~Stack} = f_{\rm opt} = (f_{g} + 0.8f_{r}+0.6f_{i})/2.4 ~,
\label{eqn:optstack}
\end{equation}

\noindent and from fluxes in the bluest (and also deepest) \wise\ bands according to

\begin{equation}
{\rm \wise\ Stack} = f_{\rm WISE} = (f_{W1} + 0.5f_{W2})/1.5  ~.
\label{eqn:wisestack}
\end{equation}

\noindent where the weights are chosen to roughly yield the highest combined S/N for a typical $z < 2$ quasar. The sample depicted by black points in Fig.\ \ref{fig:wisecut} represents objects with {\em any} \eboss\ quasar targeting bit set 
(see \S\ref{sec:bits}). This sample has been limited to $r > 21$ and $g < 22$ to illustrate the scatter at the faint end of \eboss, demonstrating the
power of the \wise\ data in filtering 
stars that other methods target due to these stars' resemblance to quasars in optical colors.

As part of the the CFHTLS-W3 test survey introduced in \S\ref{sec:xdqsozcuts} 
\wise\ was photometered at the positions of 
\sdss\ {\tt PRIMARY} sources (see \S\ref{sec:WISE}) in the CFHT Legacy survey W3 field.
A \wise-\sdss\ selected sample was created by applying the cut depicted in Fig.\ \ref{fig:wisecut} 
to these W3-test-field sources;

\begin{equation}
m_{\rm opt} - m_{\rm WISE} \geq (g-i) + 3 ~,
\end{equation}

\noindent where $m_{\rm opt}$ and $m_{\rm WISE}$ are as defined in Eqn.\ \ref{eqn:optstack} and Eqn.\ \ref{eqn:wisestack} 
after converting the stacked fluxes to magnitudes\footnote{This 
cut was also eventually used for \eboss\ CORE quasar target selection}. An inclusive star-galaxy separation
of {\tt objc\_type == 6} OR $m_{\rm opt} - m_{\rm model} < 0.1$, where $m_{\rm model}$ is the equivalent of
Eqn.\ \ref{eqn:optstack} but for \sdss\ model magnitudes, was adopted. This is inclusive in the sense that
{\tt objc\_type == 6} corresponds to a star-galaxy separation of ${\tt psfMag -  modelMag} \leq 0.145$ (as also discussed further 
in \S\ref{sec:qts})
but based on \sdss\ fluxes in {\em all} bands, not just the bands stacked in $m_{\rm opt}$. 
In addition, magnitude limits of $17 < m_{\rm opt} < 22$ were enforced. Finally, an optical color cut of $g-i < 1.5$ was applied in
an attempt to excise the highest redshift quasars (this cut is not obvious in Fig.\ \ref{fig:wisecut} as other programs 
in the CFHTLS-W3 test program repopulated this parameter space). The squares
with error bars in Fig.\ \ref{fig:wisecut} depict the typical range of colors of 
spectroscopically confirmed quasars in different redshift bins. The separation of these
points from the green line suggests that \wise\ is robust for quasar selection across the CORE redshift range of $0.9 < z < 2.2$.

\begin{figure}[t]
\centering
\includegraphics[width=0.48\textwidth,height=0.3\textwidth]{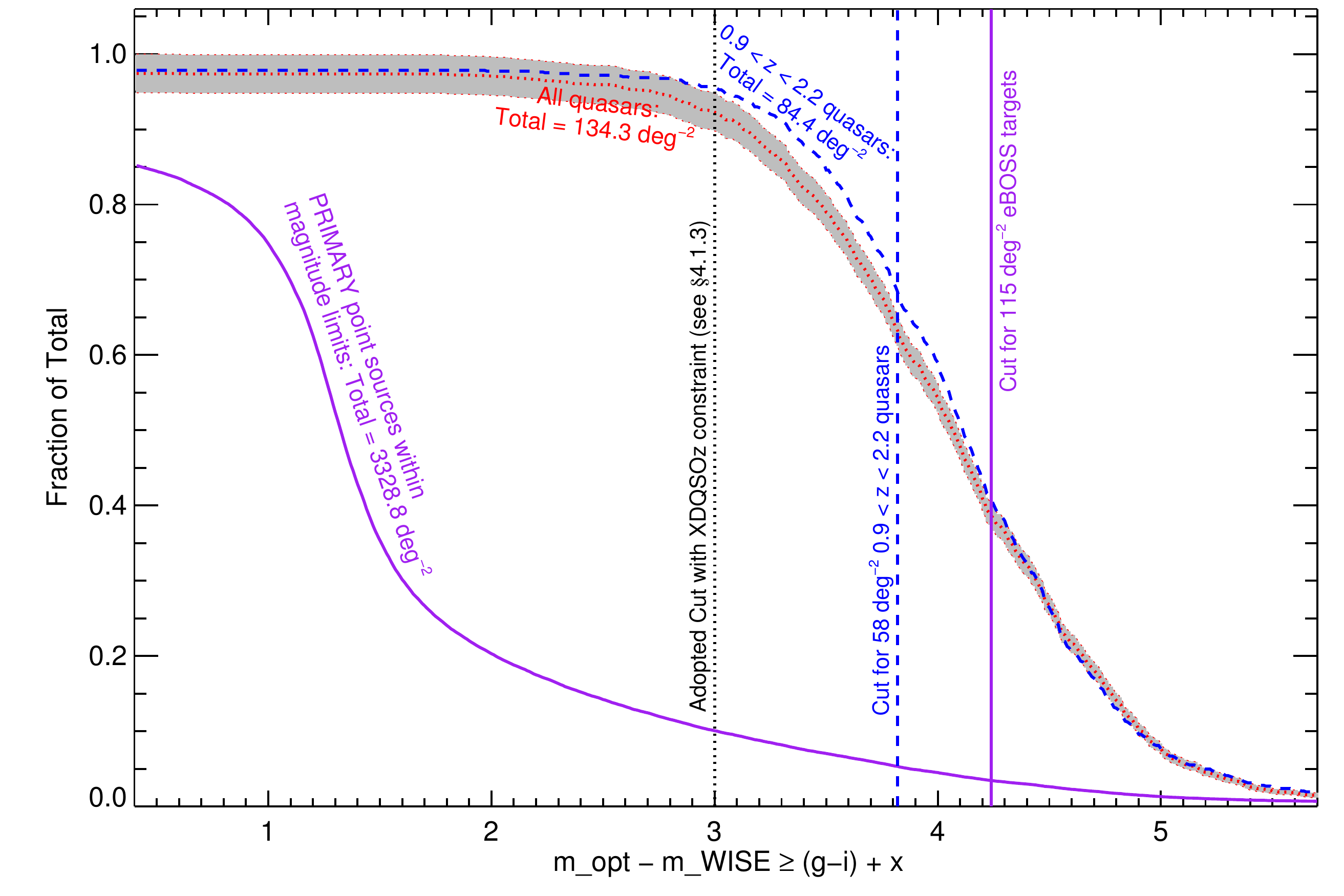}
\caption{\small 
As for Fig.\ \ref{fig:opticalcomp} but for the adopted \wise-optical cut. The x-axis depicts the number
of sources for a cut of $\geq x$ where $x$ is defined by ($m_{\rm opt} - m_{\rm WISE}) = (g-i) + x$
and $m_{\rm opt}$ and  $m_{\rm WISE}$ are the magnitudes from the optical and \wise\ stacks. 
The grey (Poisson) error contours have been omitted from the blue curve for visual clarity, but are comparable to the errors on the red curve.
All samples depicted have been limited to \sdss\ {\tt PRIMARY} point sources with {\tt FIBER2MAG} of $i > 17$ and de-extincted PSF magnitudes of 
$g < 22$ OR $r < 22$ (the initial cuts for the \eboss\ CORE). As the CFHTLS-W3 program was limited to $(m_{\rm opt} - m_{\rm WISE}) > (g-i) + 3$ the test sample is partially 
incomplete to quasars for $x < 3$. This figure can be used to estimate target densities in a similar manner to
Fig.\ \ref{fig:opticalcomp}.
}
\label{fig:wisecomp}
\end{figure}

Fig.\ \ref{fig:wisecomp} demonstrates whether a \wise-optical cut of 
$m_{\rm opt} - m_{\rm WISE} \geq (g-i) + x$ is sufficient, in isolation, to
meet the \eboss\ targeting goal of 58\perdegsq\ $0.9 < z < 2.2$ quasars, 
(modulo our additional restrictive cuts to the W3-test-field targets, such as $g - i < 1.5$). 
Fig.\ \ref{fig:wisecomp} is an exact analog of Fig.\ \ref{fig:opticalcomp}, and a detailed
description of how these figures can be interpreted is provided in \S\ref{sec:xdqsozcuts}.
Fig.\ \ref{fig:wisecomp} implies that a 
cut of about $m_{\rm opt} - m_{\rm WISE} \geq (g-i) + 4.25$ is necessary to meet the requisite \eboss\ target density of 115\perdegsq\
and that, therefore, only 34.1\perdegsq\ CORE quasars could be obtained with a \wise-optical 
selection alone. As discussed further in \S\ref{sec:combcuts}, by combining \xdqsoz\ selection with \wise\ \eboss\ could
use the ``Adopted cut...'' plotted in Fig.\ \ref{fig:wisecomp}. This relaxed cut
{\em does} achieve \eboss\ targeting goals.

\begin{figure}[t]
\centering
\includegraphics[width=0.48\textwidth,height=0.305\textwidth,clip=true,trim=20 0 0 25]{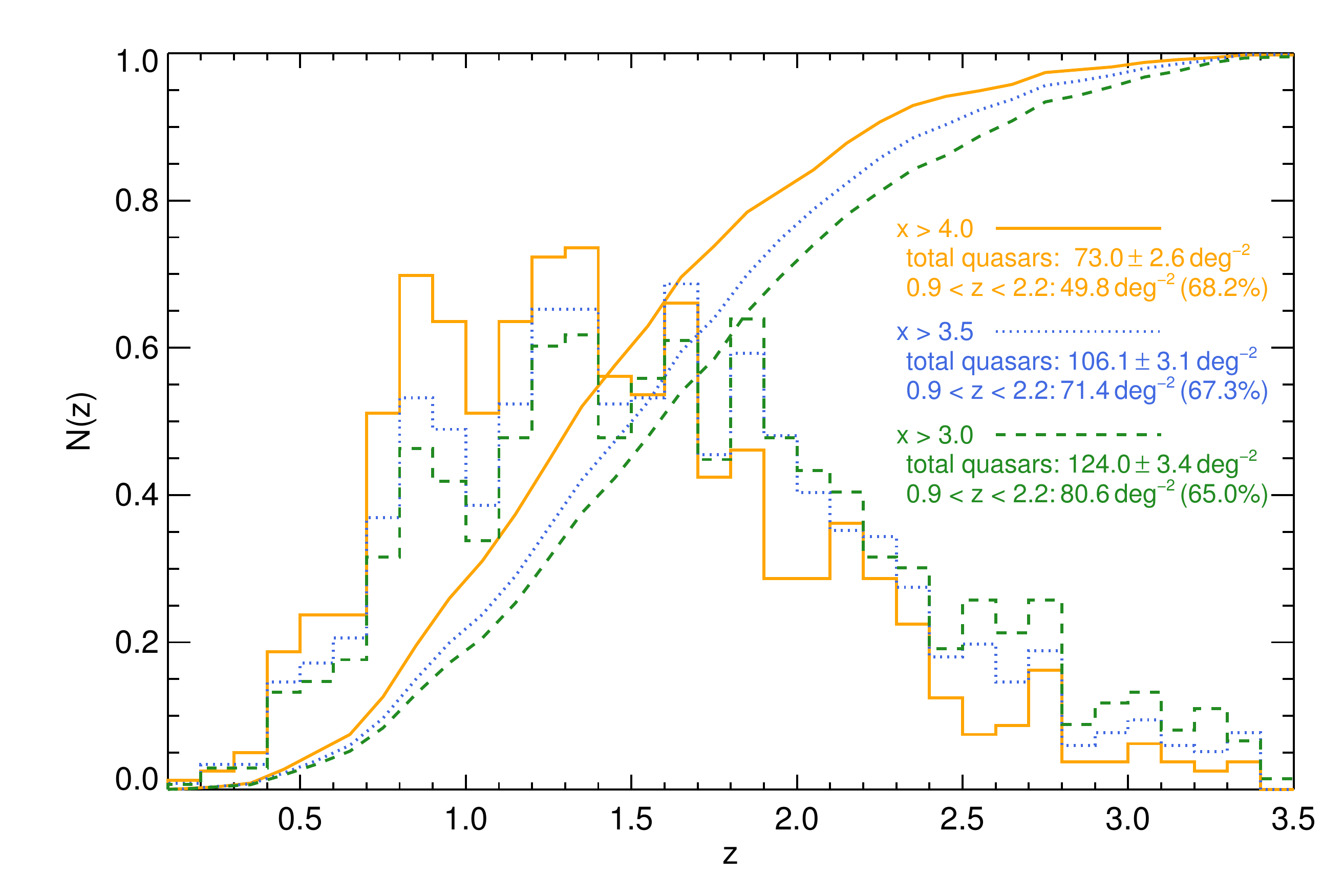}
\caption{\small 
As for Fig.\ \ref{fig:opticalNz} but for the adopted \wise-optical cut. The distributions for three 
different cuts on $x$ are depicted, where $x$ is defined by ($m_{\rm opt} - m_{\rm WISE}) = (g-i) + x$
and $m_{\rm opt}$ and  $m_{\rm WISE}$ are the magnitudes from the optical and \wise\ stacks. These
cuts are $x > 4.0$ (orange, solid), $x > 3.5$ (blue, dotted), and $x > 3.0$ (green, dashed).
}
\label{fig:wiseNz}
\end{figure}

Fig.\  \ref{fig:wiseNz} demonstrates that relaxing cuts on $x$ in the function 
$m_{\rm opt} - m_{\rm WISE} \geq (g-i) + x$ does not strongly affect the redshift distribution of 
targeted quasars. This figure shows that 65--70\% of quasars selected by this \wise-\sdss\ cut
are in the CORE redshift range regardless of the value of $x$. Overall, there is 
less variation in the \eboss\ CORE $0.9 < z < 2.2$ redshift distribution with $x$ as compared to the variation in 
Fig.\ \ref{fig:opticalNz}, because the \wise-optical cut has less power to discriminate redshift
as compared to $ugriz$ over most of the CORE range (c.f.\ Fig.\ \ref{fig:wisecut}). Instead of augmenting the
CORE quasar range, relaxing $x$ tends to expand the fraction of quasars at about $z > 2$.
This outcome is desirable, given that $z > 2.1$ quasars
can be used as part of the \eboss\ \LyA\ sample (see \S\ref{sec:LyA}).

\begin{figure}[t]
\centering
\includegraphics[width=0.48\textwidth,height=0.3\textwidth]{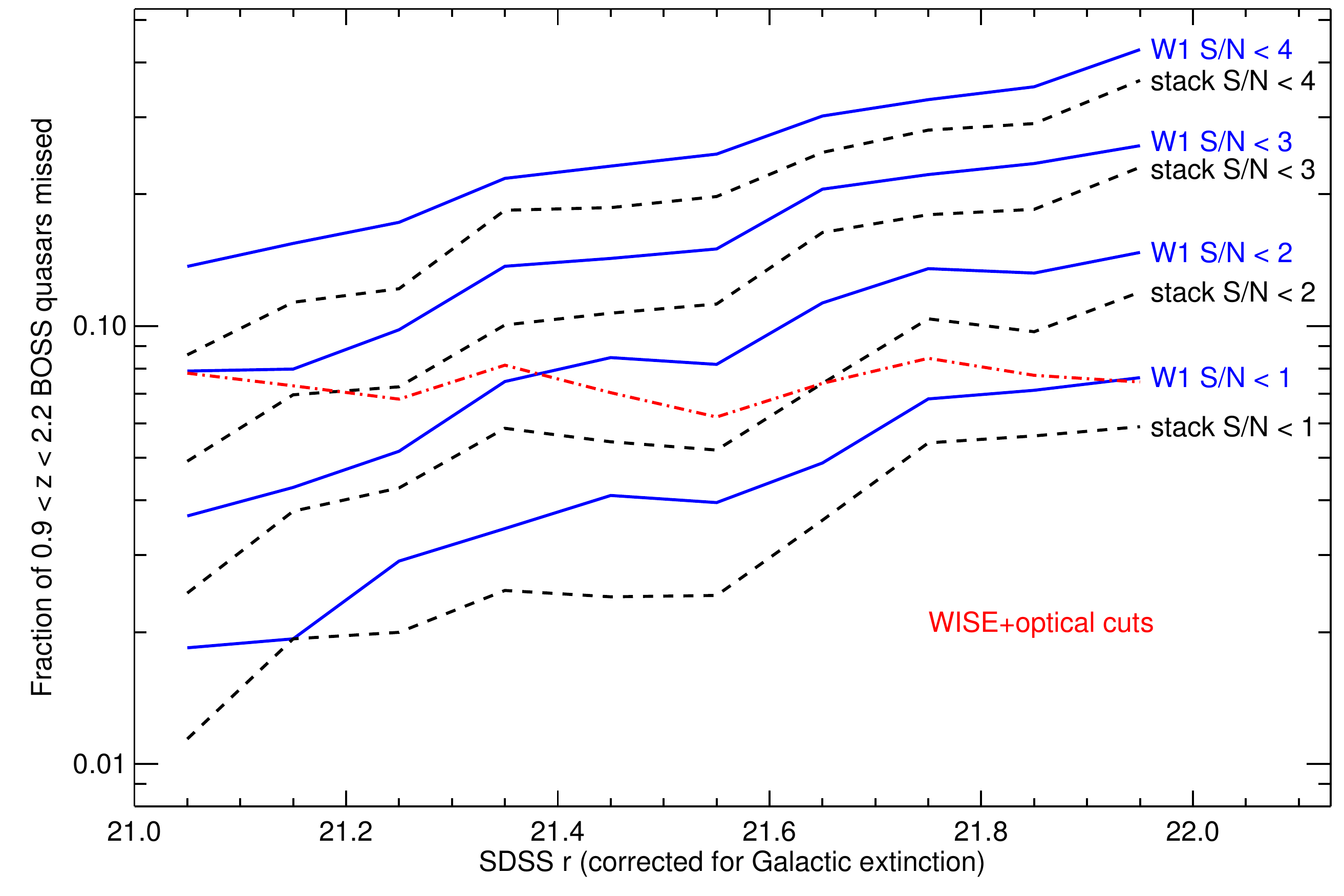}
\caption{\small 
The fraction of $0.9 < z < 2.2$ (DR10) \boss\ quasars that are missed as a function of \wise\ signal-to-noise ratio in the $W1$ band (blue solid line) 
and in the stack of ($f_{W1} + 0.5f_{W2})/1.5$ that is actually used in \eboss\ CORE quasar selection (black dashed line). The red (dot-dashed) line displays the fraction of such 
quasars {\em missed by the overall \eboss\ CORE quasar target selection}.
}
\label{fig:wisedepth}
\end{figure}

By redshifts of $z\sim6$, about half of quasars aren't detected in the \wise\ $W1$ and $W2$ bands \citep{Bla13}. 
In addition, a $10\sigma$ detection in \wise\ $W2$ is equivalent to $i\sim19.8$ \citep{Ste12}, which may not 
detect all quasars to the effective \eboss\ limits of $r\sim22$. Thus it is worth investigating 
whether the \wise\ data photometered for \eboss\ targeting (see \S\ref{sec:WISE}) are sufficiently 
{\em deep} for our purposes. Fig.\ \ref{fig:wisedepth} addresses this issue by
plotting known \drt\ quasars as a function of signal-to-noise in our \wise\ stack ($m_{\rm WISE}$). 
The stack depth is 
sufficient to identify 90\% of $0.9 < z < 2.2$ \boss\ quasars at a S/N ratio of 2 in the stack to $r < 21.9$.
Although the depth of \wise\ becomes limiting near $r\sim22$ for \eboss\ CORE quasars, 
about 93\% of
$0.9 < z < 2.2$ \boss\ quasars would be selected by our \wise-optical cut; this is because of the combined effect that few quasars are {\em both} blue in $g-i$ {\em and} faint in \wise.

\subsubsection{Combined mid-IR and optical selection}
\label{sec:combcuts}

After analyzing our CFHTLS-W3 test data (as outlined in \S\ref{sec:xdqsozcuts} and \S\ref{sec:wisecuts}) it
became clear that the overall number of CORE quasars targeted at the \eboss\ fiber density
could be increased by combining an
\xdqsoz\ probability limit {\em with} a \wise-optical cut.
It was possible to only partially study the \xdqsoz\ probability and \wise-optical cut 
beyond the limits to which they had been tested in the CFHTLS-W3 
program---using those \xdqsoz-selected quasars that failed the \wise-optical cut and vice versa. 
As the combination of the two original test cuts exceeded \eboss\ goals, however, it 
was decided to proceed with an \eboss\ CORE quasar target selection corresponding to {\em both} of

\begin{eqnarray}
{\rm PQSO}(z > 0.9) > 0.2 ~~~~~~~~~~~~{\rm \&\&} \nonumber \\
m_{\rm opt} - m_{\rm WISE} \geq (g-i) + 3 ~~~~~~.
\end{eqnarray}

The ``Adopted cut...'' lines in \noindent Fig.\ \ref{fig:opticalcomp} and Fig.\ \ref{fig:wisecomp} demonstrate that in combination these constraints easily achieve the 
 \eboss\ CORE goal of 58\perdegsq\ $0.9 < z < 2.2$ quasars. It turns out that the combined \xdqsoz-and-\wise-optical constraints that
 correspond to these adopted cuts require close to the
maximum \eboss\ quasar target density of 115\perdegsq\ (see \S\ref{sec:SRD})
and achieve an overall density of $\sim70$\perdegsq\ $0.9 < z < 2.2$ quasars. 
The expected \eboss\ CORE quasar density arising from these constraints is explored in more detail in \S\ref{sec:sequels}.

\subsection{Broad overview of the Lyman-$\alpha$ quasar sample}
\label{sec:LyA}

The goal of  \eboss\ Lyman-$\alpha$ quasar targeting is to compile as large a sample of new $z > 2.1$ quasars as
possible using the remaining available fibers that were not allocated to other \eboss\ targets.
The \eboss\ \LyA\ sample is not required to 
be homogeneously selected; it is therefore targeted using several different
selection algorithms and sources of imaging---even imaging that only partially 
covers the \eboss\ footprint. 

The majority of new \eboss\ \LyA\ quasars are targeted using
two techniques. First, the CORE sample described in \S\ref{sec:core} is a source of
new \LyA\ quasars, since its selection contains no requirement to intentionally remove $z>2.1$ quasars.
Second, a variability selection is used to target additional \LyA\ quasars. The CORE and the variability-selected samples each
select $\sim5$\,\perdegsq\ new Lyman-$\alpha$ quasars, with only $\sim1.5$\,\perdegsq\ in common
(see also Table~\ref{tab:stats} in \S\ref{sec:targeff}). The variability-selected targets undergo a different 
set of initial flag and flux cuts as compared to other target classes (see \S\ref{sec:variability}).

\eboss\ uses two additional techniques to target more \LyA\ quasars and to
acquire more signal in the \LyA\ Forest. First, all previously unidentified sources within
 1\arcsec\ of a radio detection in the  FIRST survey \citep{FIRST,Hel15} are targeted. Finally, quasars
 that had low signal-to-noise ratio spectra in \boss\ are re-targeted.
The target categories specific to Lyman-$\alpha$ selection are detailed below, and are summarized in \S\ref{sec:bits}. 

\begin{figure}[t]
\centering
\includegraphics[width=.42\textwidth,height=0.28\textwidth,clip=true,trim=0 0 40 20]{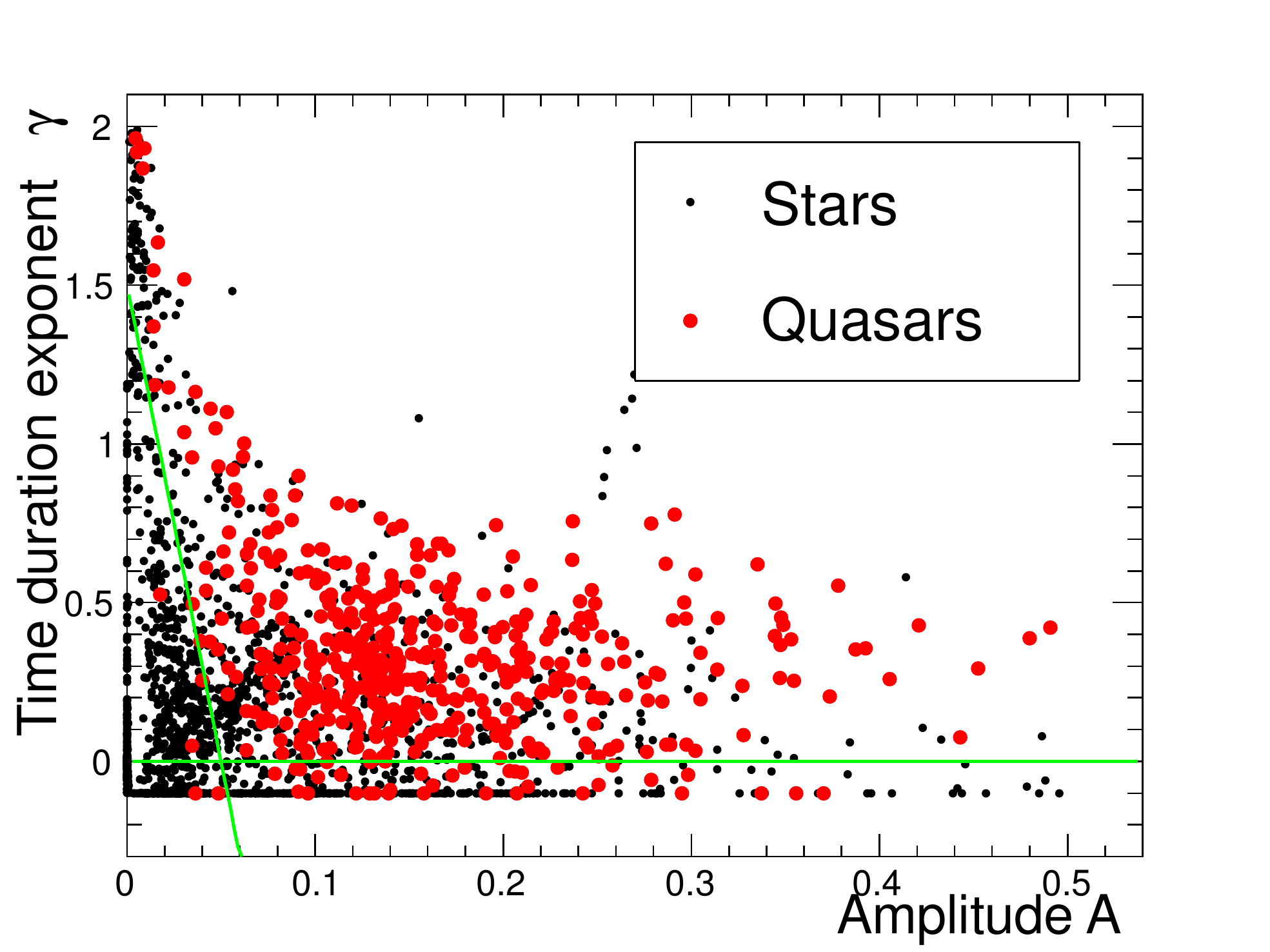}
\caption{\small Structure function parameters for 6-epoch $R$-band light curves from \ptf. Quasars (red) and stars (black), 
whether variable or non-variable, populate distinct regions of the $\gamma-A$ plane. 
Stars are a subsample of 1500 random point-like objects delimited in Equatorial Coordinates by $52^\circ<\delta_{\rm J2000}<54^\circ$ and $211^\circ<\alpha_{\rm J2000}<216^\circ$. Quasars are the previously identified quasars (mostly from \boss) in the same field.
}
\label{fig:PTF}
\end{figure}

\subsubsection{Variability selection}
\label{sec:variability}
Time-domain photometric measurements can exploit 
quasars' intrinsic variability in order to distinguish
them from stars of similar colors \citep[e.g.,][]{van73,Haw83,Cim93,Ren04,Ren04b,Cla06,Ses07,Koz10,Mac10,Sch10,Pal11,Pal13qlf,Pal15}. 
The time-variability of astronomical sources can be described using the ``structure function,'' a measure of the amplitude of the observed variability as a function of the time delay between two 
observations \citep[e.g.,][]{Cri96,Giv99,Van04,Ren06}.  This function can be modeled as a power law parameterized in terms of $A$, the mean amplitude of the variation on a one-year timescale (in the observer's reference frame), and $\gamma$, the logarithmic slope of the variation amplitude with respect to time \citep{Sch10}. With $\Delta m_{ij}$ defined as the difference between the magnitudes of the source at time $t_i$ and $t_j$, and assuming an underlying Gaussian distribution of $\Delta m$ values, the model predicts an evolution of the variance  $\sigma^2(\Delta m)$ with time according to
\begin{equation}
\sigma^2(\Delta m) = \left[ A (\Delta t_{ij})^\gamma \right]^2 + (\sigma_i^2 + \sigma_j^2) \, ,
\end{equation}
where $\sigma_i$ and $\sigma_j$ are the imaging errors at time $t_i$ and $t_j$.
Quasars should lie at high $A$ and $\gamma$, non-variable stars near $A=\gamma=0$ and variable stars should 
have $\gamma$ near 0 even if $A$ is large. In addition, variable sources (whether stars or quasars) are expected to deviate greatly from a model with constant flux. This deviation is quantified by computing the $\chi^2$ of the fit of the light curve compared to a constant-flux model.

\begin{figure}[t]
\centering
\includegraphics[width=.41\textwidth,height=0.28\textwidth,clip=true,trim=10 0 50 23]{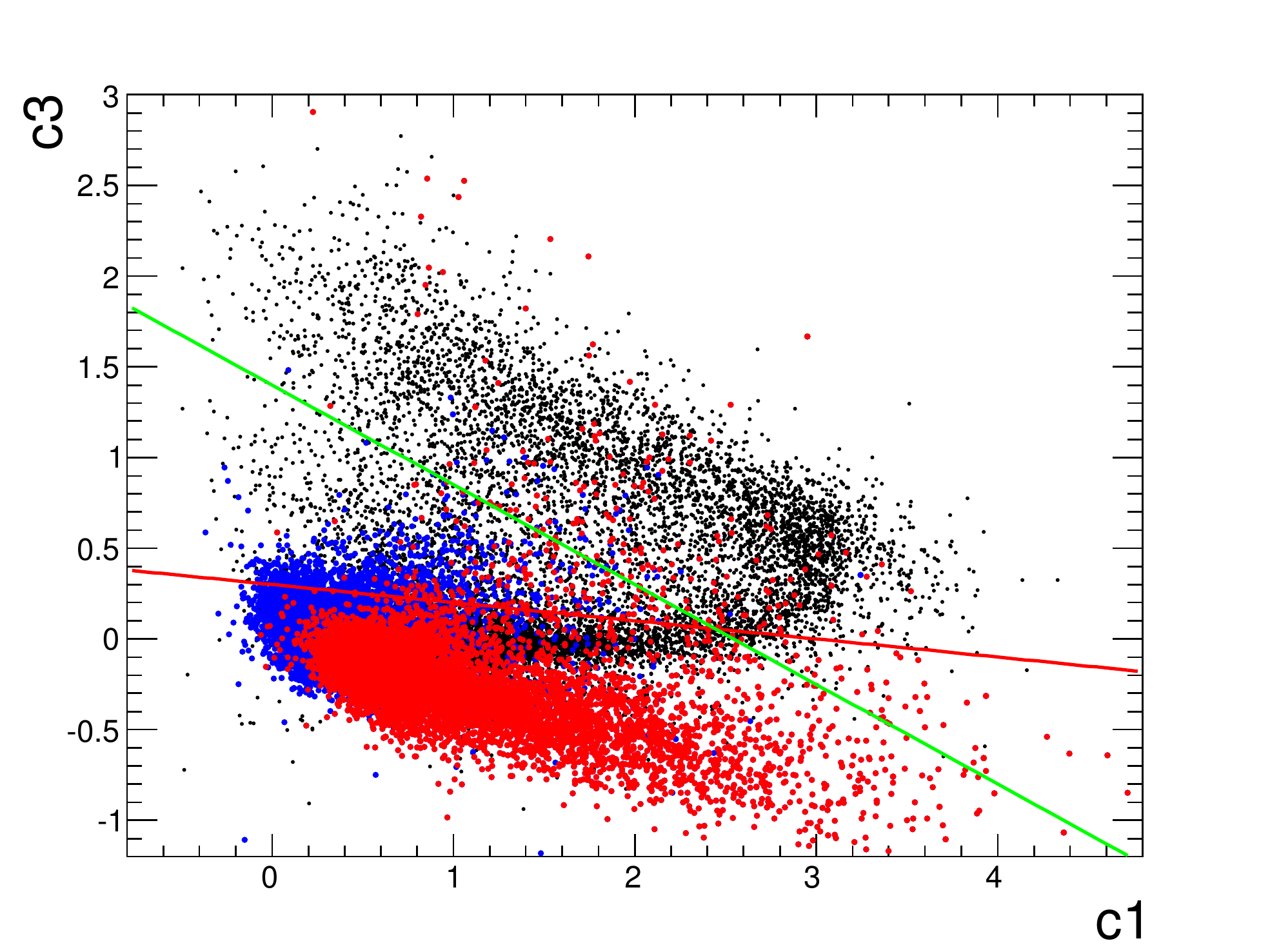}
\caption{\small The adopted loose color-cut designed to reject stars. Black, blue and red points represent stars, $z<2.1$ and $z>2.1$ quasars, respectively. The colors of each set of objects are taken from the \sdss\ {\em Catalog Archive Server}. Stars are obtained from a 7.5\degsq\ region delimited by $357^\circ< \alpha_{\rm J2000} <360^\circ$ and $-1.25^\circ< \delta_{\rm J2000} <1.25^\circ$ (i.e.\ they represent a random sample of point-like objects). Quasars are a subsample of spectroscopically confirmed sources from the \sdss\ surveys.
}
\label{fig:lya_colorcut}
\end{figure}

Using customized \ptf\ $R$-band stacks (see section \S\ref{sec:PTF}), light curves are built for all of the \ptf\ sources. 
The \ptf\ sources are matched to \sdss\ imaging catalogs, and the selection is restricted to \sdss\
{\tt PRIMARY} point sources. With the \ptf\ light curves in hand, all additional cuts are
then applied using \sdss\ imaging information. \sdss\ cuts of $g < 22.5$ and $r>19$ are then applied. 
When \sdss\ $r$-band data are available, the $R$-band \ptf\ light curve, adjusted to \sdss\ $r$, is extended to include the \sdss\ fluxes. 
These PTF+SDSS light-curves typically contain 3 to 4 PTF ``coadded epochs,'' where each PTF ``coadded epoch'' is obtained by coadding the exposures within a given PTF observational season. 
The number of exposures in each season varies from $\sim 10$ to a few dozen for typical fields.

Because the density of \ptf\ images varies across the sky, so does the efficiency of the variability-based selection. To account
for this, the thresholds of the variability cuts are adapted as a function of position in order to reach an average target 
density of $\sim 20$\degsq\ across the \eboss\ footprint. Constraints of 
$5.0<\chi^2<200.0$ for combined \ptf $+$\sdss\ measurements are typically necessary; smaller 
$\chi^2$ values are obtained for non-variable sources, while larger values often signify artifacts. 
The parameters of the variability structure function are
forced to lie in the parameter space bounded by $\gamma>0$ and $\gamma>-30A+1.5$, as illustrated by the green lines in Fig.\ \ref{fig:PTF}. Tighter $\chi^2$ cuts are applied to light curves for which the variability parameters $A$ and $\gamma$ cannot be computed reliably, such as light curves with fewer than 3 \ptf\ epochs. 

To maximize the efficiency of quasar selection, the variability selection is complemented by loose 
color cuts designed to reject stars. 
Cuts of $c_3<1.4-0.55\times c_1$ and $c_3<0.3-0.1\times c_1$ are imposed, where 

\begin{align}
c_1 &~~=& 0.95(u-g)+0.31(g-r)+0.11(r-i) \nonumber \\
c_3 &~~=& -0.39(u-g)+0.79(g-r)+0.47(r-i) ,
\end{align} 

\noindent as defined in \cite{Fan99}. In these equations, $ugri$ are PSF magnitudes measured in the \sdss\ imaging. 
This color cut is illustrated in Fig.\ \ref{fig:lya_colorcut}, where the regions above the red and green lines are rejected.

\begin{figure}[t]
\centering
\includegraphics[width=.52\textwidth,height=.6\textwidth,clip=true,trim=10 0 0 20]{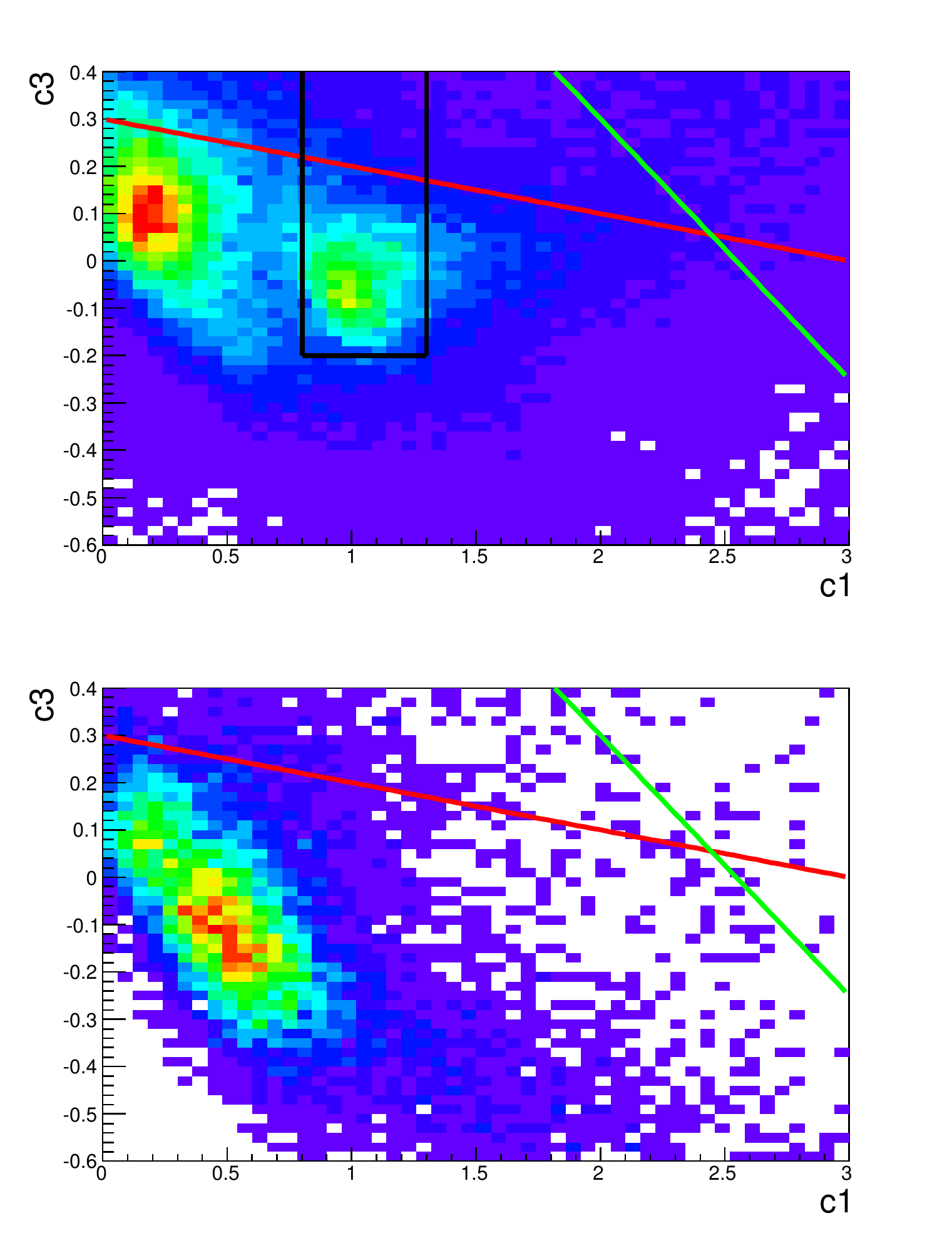}
\caption{\small $c_1 - c_3$ color plots for sources passing the variability criteria defined in \S\ref{sec:variability}. The upper panel depicts 
all objects: the two peaks correspond to quasars (left-most density peak) and bright variable stars (right-most density peak). The lower 
panel shows previously known quasars only (mostly $z>2.1$ quasars from \boss). The contaminating population in the top plot is 
variable stars that are removed with a dedicated set of color cuts illustrated by the black box (see \S\ref{sec:variability} for more details). 
}
\label{fig:lya_var}
\end{figure}

Finally, a region in color-space mostly populated by bright variable stars, that passes both the color and the variability cuts, is removed. These stars are apparent in the top panel of
 Fig.\ \ref{fig:lya_var}---but are clearly absent in the lower panel, which depicts known quasars. These contaminating variable stars are removed by rejecting 
 sources that lie in the color box $0.85<c_1<1.35$ and $c_3>-0.2$ if they are brighter than $r=20.5$. This cut is not applied to fainter sources. 

\subsubsection{Reobservation of \boss\ quasars}
\label{sec:reobs}
The mean density of \LyA\ quasars in \boss\ (once Broad Absorption Line quasars 
are removed) is $\sim 15$\perdegsq. Roughly $60 \%$ of these quasars
have a signal-to-noise ratio (SNR) $<3$, thus reducing their utility 
for tracing large-scale structure. Here, SNR is defined as the mean S/N
per \LyA\ Forest pixel measured over the rest-frame wavelength range of 
1040\,\AA$\,< \lambda < 1200$\,\AA.
With the exception of \boss\ spectra that have
SNR\,pixel$^{-1}=0$ (signifying an observational error) quasars with 
$0\leq {\rm SNR\,pixel}^{-1}<0.75$ do not contribute as much to the Forest signal as 
placing a fiber on a new quasar target, so such quasars are
not worth reobserving. Within \eboss, \boss\ quasars are therefore targeted if they lie in the \eboss\ footprint and have  
$0.75\leq {\rm SNR\,pixel^{-1}} <3$ OR ${\rm SNR\,pixel^{-1}}=0$. The density of these targets varies over the \eboss\ footprint from $\sim 6$\perdegsq\ to $\sim 10$\perdegsq, 
depending upon the underlying density of \boss\ \LyA\ quasars.

\subsubsection{Radio selection}
\label{sec:radio}
\eboss\ also targets all \sdss\ point sources that are within 1\arcsec\ of a radio detection in the 13 June 05 version\footnote{\url{http://sundog.stsci.edu/first/catalogs/readme\_13jun05.html}}
 of the \first\ point source catalog \citep{FIRST,Hel15}. The density of such sources (that are not already included
 in another target class) is low ($< 1$\perdegsq), and these additional targets
 are expected to identify some previously unknown high redshift quasars.

\subsection{Additional Cuts}
\label{sec:flags}

\sdss\ imaging includes a great deal of meta-data\footnote{e.g.\ see Tables 5, 6, 8 and 9 of \citet{EDR}},
and, notably, contains flags (in the form of bitmasks) that can be used to characterize photometric quality\footnote{see Table 9 of \citet{EDR}}. 
Initially, \eboss\ adopts a set of obvious and necessary cuts on \sdss\ imaging parameters. The target selection is restricted to {\tt PRIMARY} sources in the \sdss\ to avoid
duplicate sources. Targets are cut on (deextincted) {\tt PSFMAG} to near the limits of \sdss\ imaging, in part driven by the necessary exposure
times to obtain spectra of reasonable signal-to-noise ratio. These limits are $g < 22$ OR $r < 22$ for CORE quasars and $g < 22.5$ for the
\LyA\ quasar sample---which can be more speculative and inhomogeneous in its selection. A bright limit of
{\tt FIBER2MAG} $i > 17$ 
is adopted for all \eboss\ targets to prevent light leaking between adjacent fibers \citep[see][]{ebosspaper???}.
Quasars selected by variability and intended purely for \LyA\ studies have a more restrictive bright-end cut of $r > 19$, as there 
are few high-redshift quasars brighter than $r=19$. Finally, 
the restriction that quasar targets must be unresolved in imaging ({\tt objc\_type==6}) is imposed. 
This is necessary as at fainter magnitudes, extended sources
begin to dominate \sdss\ imaging, and at $r > 21.2$ there are three times as many  {\tt objc\_type==3} (extended)
sources as {\tt objc\_type==6} (point-like) sources. Targeting
extended sources would greatly increase the \eboss\ fiber budget, while recovering few $z > 0.9$ quasars.

Our CFHTLS-W3 test program (outlined in \S\ref{sec:xdqsozcuts})
had relaxed limits on star-galaxy separation and magnitude, meaning that it
is possible to show that our basic flag cuts for \eboss\ quasar targeting represent sensible choices. 
Adopting the selection outlined in \S\ref{sec:combcuts}, 
a cut on {\tt objc\_type==6} discards only 4.6\% of quasars but requires $3.5\times$ fewer fibers. Enforcing faint limits
of $g < 22$ OR $r < 22$  discards 5.8\% of quasars but requires $11.5\times$ fewer fibers.

\begin{figure}[t]
\centering
\includegraphics[width=0.48\textwidth,height=0.3\textwidth,clip=true,trim=0 0 0 -15]{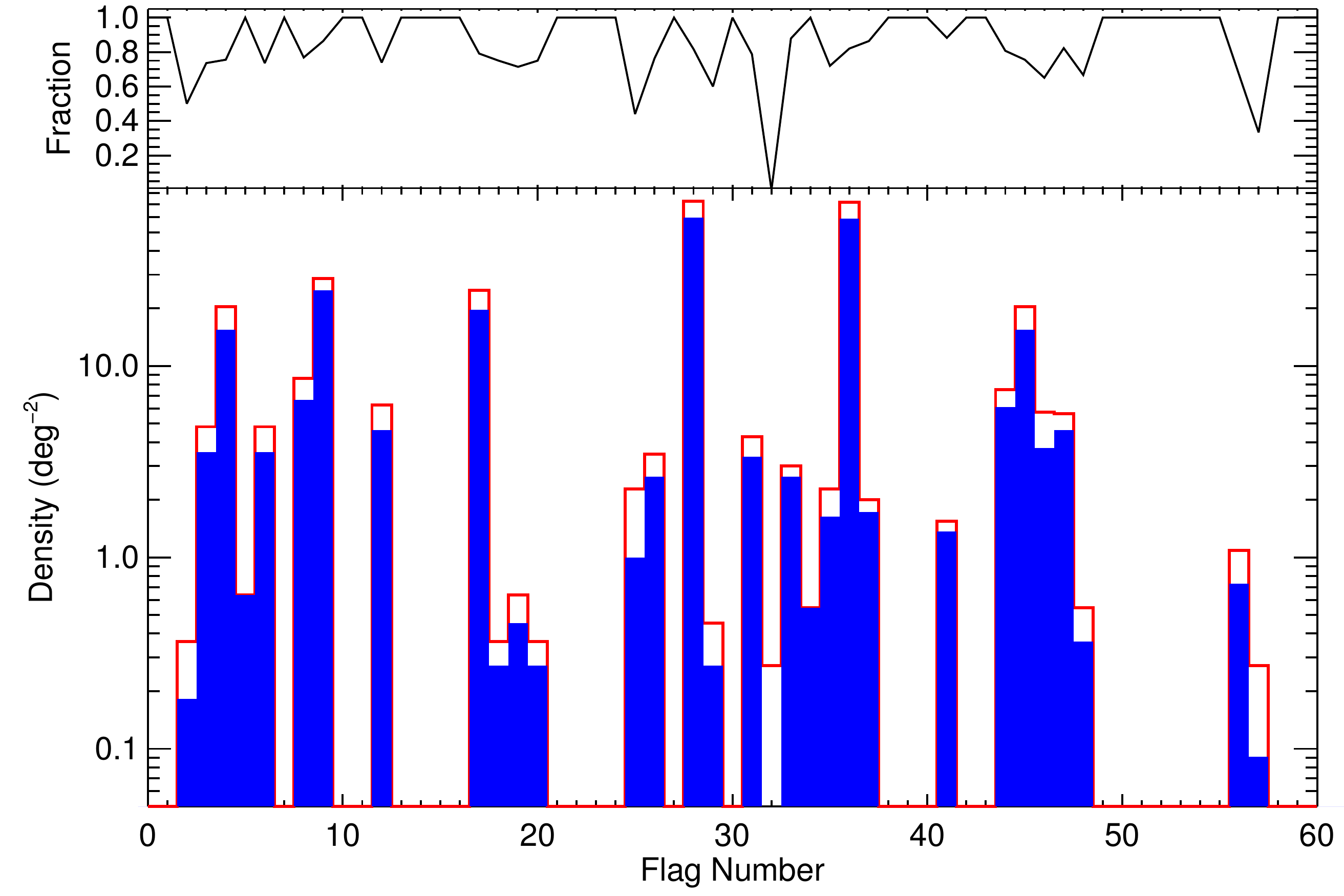}
\caption{\small 
Sky density of quasars and targets {\em removed} by a specific \sdss\ flag cut. 
Flag numbers 0--31 correspond to the 32 bits in the \sdss\ {\tt objc\_flags} bitmask and 
flag numbers 32--63 are the 32 bits in the \sdss\ {\tt objc\_flags2} bitmask. The final three bits
in {\tt objc\_flags2} 
do not correspond to an imaging flag.
The red (empty) 
histogram is the density of targets discarded from the CFHTLS-W3 test data and the blue (filled) histogram
is the density of genuine $z > 0.9$ quasars discarded by the same flag cut. In the upper panel we display
the ratio of the two histograms, which is the fraction of targets discarded that would be useful quasars for \eboss.
}
\label{fig:checkflags}
\end{figure}

Typically, previous \sdss\ quasar targeting algorithms \citep{Ric02,Ros12} have employed additional constraints on image quality to 
reduce spurious targets. Given that the CFHTLS-W3 test program did not adopt strict flag cuts,
it could be used to assess which flag cuts might be worthwhile for \eboss\ targeting (see Fig.\ \ref{fig:checkflags}).
A range of individual \sdss\ flag cuts are plotted in Fig.\ \ref{fig:checkflags}, which demonstrates 
that there are essentially no \sdss\ flags that discard targets
without also discarding useful $z > 0.9$ quasars. The one exception is the {\tt DEBLENDED\_AS\_MOVING} flag (number 32), which does not
obviously discard quasars, but which only saves 0.3\perdegsq\ targets. In addition to the results in Fig.\ \ref{fig:checkflags}, we
also tested numerous standard combinations of flags used by other \sdss\ quasar targeting algorithms, 
such as the {\tt INTERP\_PROBLEMS} and {\tt DEBLEND\_PROBLEMS}
combinations outlined in the appendices of \citet{Bov11a} and \citet{Ros12}. In no case did we find a flag combination that
removed significant numbers of targets {\em without also discarding useful quasars}. We do not study why the \sdss\ image quality flags
have limited utility for \eboss\ targeting---speculatively the flags may become less meaningful near the faint limits of \sdss\ imaging and/or
our incorporation of \wise\ data may ameliorate \sdss\ artifacts. In any case, based on this analysis and the fact that the basic
\eboss\ selection already achieves the requisite target density, we make no additional \sdss\ flag cuts.

It is likely that certain regions of the \sdss\ imaging will have to be masked further for quasar clustering analyses, due to, e.g.,
areas around bright stars (both in \wise\ and \sdss\ imaging), or bad imaging fields \citep[e.g.\ see][and \S\ref{sec:homo}]{Ros11}. For instance,
due to how the \sdss\ geometry was initially defined for  ``uber-calibration,'' small overlap regions ($\sim 1$\degsq) in \sdss\ run 752 are
mis-aligned between \sdss\ and our \wise\ photometering. Such
regions do not have a major impact on target homogeneity, however, and may differ for different \eboss\ target classes, 
so such geographic areas will be masked post-facto depending on a specific science purpose. One set of regions that was masked
{\em a priori} for \boss\ quasar targeting corresponded to bad $u$-columns \citep[e.g.\ see Fig.\ 1 of][]{Whi12}. Specifically testing
target density in areas with bad \sdss\ $u$-columns did not suggest they have greatly different \eboss\ CORE target densities ($\sim
116$--118\perdegsq\ versus the average of $\sim115$\perdegsq\ for the typical survey area), so bad $u$-columns are
not specifically masked {\em a priori} for \eboss\ targeting.

In general, the only large geographic areas that should certainly {\em not} be photometric in \sdss\ imaging are regions with
catastrophic values of {\tt IMAGE\_STATUS}\footnote{\url{http://www.sdss3.org/dr8/algorithms/bitmask\_image\_status \\ .php}}. 
For \eboss\ CORE quasar targeting, we avoid all areas with {\tt IMAGE\_STATUS} set to any of 
{\tt BAD\_ROTATOR, BAD\_ASTROM, BAD\_FOCUS, SHUTTERS, FF\_PETALS, DEAD\_CCD} or {\tt NOISY\_CCD} 
in any filter. Quasars targeted on the basis of their variability in \ptf\ for \LyA\ studies do not undergo cuts on {\tt IMAGE\_STATUS} as
there is no requirement for \LyA\ quasars to be selected homogeneously.
The full set of flag cuts eventually adopted is outlined succinctly in Fig.\ \ref{fig:flowchart}.

\begin{table}[t]
\centering
\caption{\small
\eboss\ quasar targeting bits and their numerical equivalents}
\begin{tabular}{c l c l}
\hline\hline
Bit & Name & Bit & Name  \\ \hline
${0}$ & {\tt DO\_NOT\_OBSERVE} \\
${10}$ & {\tt QSO\_EBOSS\_CORE} &
${15}$ & {\tt QSO\_BAD\_BOSS} \\
${11}$ & {\tt QSO\_PTF} &
${16}$ & {\tt QSO\_BOSS\_TARGET} \\
${12}$ & {\tt QSO\_REOBS} &
${17}$ & {\tt QSO\_SDSS\_TARGET} \\
${13}$ & {\tt QSO\_EBOSS\_KDE} &
${18}$ & {\tt QSO\_KNOWN} \\
${14}$ & {\tt QSO\_EBOSS\_FIRST}  &
${19}$ & {\tt DR9\_CALIB\_TARGET} \\
\hline\hline
\end{tabular}
\label{tab:bits}
\end{table}

\subsection{Targeting bits}
\label{sec:bits}

The tests summarized in \S\ref{sec:qts}--\ref{sec:flags} provide sufficient information to justify the choices made to target quasars in \eboss.
This section provides an outline of how the \eboss\ targeting bits directly correspond to the specified choices. A visual
representation of the overall targeting algorithm is also provided in Fig.\ \ref{fig:flowchart}. Unless otherwise
specified, each target class is derived from the imaging outlined in \S\ref{sec:imaging} and
undergoes the basic flag cuts outlined in \S\ref{sec:flags} ({\tt PRIMARY}, {\tt objc\_type==6}, magnitude cuts, and good
{\tt IMAGE\_STATUS}). The numerical value of each of the \eboss\ quasar targeting bits is listed in Table~\ref{tab:bits}.
The density and success rate of each class of target is described further in \S\ref{sec:results}.

\subsubsection{\tt QSO\_EBOSS\_CORE}

Quasars that comprise the main \eboss\ CORE sample are assigned the {\tt QSO\_EBOSS\_CORE} bit. The main goal of the CORE sample is
to obtain $> 58$\perdegsq\ $0.9 < z < 2.2$ quasars (assuming an exactly 7500\degsq\ footprint for \eboss).
We make no attempt to limit the upper end of the CORE redshift range, meaning that the CORE also selects
$z > 2.1$ quasars that have utility for \LyA\ Forest studies. Quasars in the CORE are selected by \xdqsoz\ and
\wise\ as described in \S\ref{sec:combcuts}

\subsubsection{\tt QSO\_PTF}

Quasars intended for \LyA\ Forest studies typically do not have to be selected in a uniform manner. This freedom allows
variability selection to be applied to inhomogeneous imaging in order to target additional $z > 2.1$ quasars for \eboss. The
{\tt QSO\_PTF} bit indicates such quasars, which have been selected using multi-epoch imaging 
from the Palomar Transient Factory. \ptf\ targets undergo slightly different initial cuts to other quasar
target classes; they are limited in magnitude to $r > 19$ and $g < 22.5$ and they {\em are} observed in areas
with bad {\tt IMAGE\_STATUS}. These choices are justified in \S\ref{sec:flags}.
\ptf\ quasars are selected as described in \S\ref{sec:variability}.

\subsubsection{\tt QSO\_REOBS}

Quasars previously confirmed in \boss\ that are of reduced (but not prohibitively low) signal-to-noise ratio have
decreased utility for  \LyA\ Forest studies. In addition, high probability \boss\ quasar targets that have
zero spectral signal-to-noise ratio in \boss\ are likely to have been spectroscopic glitches.
The {\tt QSO\_REOBS} bit signifies quasars that were measured to have $0.75\leq {\rm SNR\,pixel}^{-1} <3$ or
SNR\,pixel$^{-1}= 0$ in \boss. Quasars are selected for reobservation as described in \S\ref{sec:reobs}.

\subsubsection{\tt QSO\_EBOSS\_KDE}

The {\tt QSO\_EBOSS\_KDE} bit has been discontinued for \eboss\ but formed part of the targeting for \sequels\ (see \S\ref{sec:sequels}).
Targets that had the {\tt QSO\_EBOSS\_KDE} bit set in \sequels\ were drawn from the Kernel Density Estimation catalog of 
\citet{Ric09KDE} and had {\tt uvxts==1} set within that catalog. As the {\tt QSO\_EBOSS\_KDE} bit is discontinued, the origin of this target class is 
not described further in this paper.

\subsubsection{\tt QSO\_EBOSS\_FIRST}

Powerful radio-selected quasars can be detected by FIRST at $z > 2.1$ and can therefore have utility for \LyA\ Forest studies. The {\tt QSO\_EBOSS\_FIRST}
bit indicates quasars that are targeted because they have a match in the FIRST radio catalog, as described in \S\ref{sec:radio}.

\subsubsection{\tt QSO\_BAD\_BOSS} 

Some likely quasars with spectroscopy obtained as part of \boss\ have uncertain classifications or redshifts upon visual
inspection. Such objects are designated as {\tt QSO?} or {\tt QSO\_Z?} in \drtwq\ \citep[c.f.][]{Par14}. The {\tt QSO\_BAD\_BOSS} 
bit signifies such objects, to ensure that ambiguous \boss\ quasars are {\em always reobserved}, regardless of which other targeting
bits are set. Prior to 4 November, 2014 \citep[effectively prior to the {\tt eboss6} tiling; see][]{ebosspaper???}
a close-to-final but preliminary version of \drtwq\ was used to define this sample, but as of {\tt eboss6} the final sample of
\drtwq\ was used to define the {\tt QSO\_BAD\_BOSS}  bit.
This change effectively means that a small number of quasars with ambiguous \boss\ spectra may not have been reobserved prior
to {\tt eboss6}.

\subsubsection{\tt QSO\_BOSS\_TARGET}
\label{sec:qsobosstarget}

In an attempt to reduce the overall target density, \eboss\ quasar targeting does not retarget any objects with good spectra from \boss\ 
unless otherwise specified. The {\tt QSO\_BOSS\_TARGET} bit is set to indicate such objects.
We define an object as having good \boss\ spectroscopy if it appears in the file of all spectra that have been observed by
\boss\footnote{Specifically the combination of v5\_7\_0 and v5\_7\_1 of the \boss\ SpAll file
(\url{http://data.sdss3.org/datamodel/files/BOSS\_SPECTR \\ O\_REDUX/RUN2D/spAll.html}) circa May 30, 2014} and
if it does {\em not} have either {\tt LITTLE\_COVERAGE} or {\tt UNPLUGGED} set in the {\tt ZWARNING} 
bitmask \citep[see Table 3 of][]{Bol12}. 

\subsubsection{\tt QSO\_SDSS\_TARGET}                                

\eboss\ quasar targeting will not retarget objects with good pre-\boss\ spectra from the \sdss\ 
(i.e., spectra from prior to \dre). The {\tt QSO\_SDSS\_TARGET} bit is set to indicate such objects.
A ``good'' spectrum is defined using {\tt LITTLE\_COVERAGE} and {\tt UNPLUGGED} as for the {\tt QSO\_BOSS\_TARGET} bit.
\sdss\ spectral information is obtained from
the final \dre\ spectroscopy files\footnote{Specifically the (line-by-line) parallel spectroscopy and imaging catalogs at
\url{http://data.sdss3.org/sas/dr8/sdss/spectro/redux/ \\ photoPosPlate-dr8.fits} and
\url{http://data.sdss3.org/sas/dr8/sdss/sp \\ ectro/redux/specObj-dr8.fits}}.

\subsubsection{\tt QSO\_KNOWN}

\eboss\ quasar targeting will not reobserve objects with previous good spectra (defined by the {\tt QSO\_BOSS\_TARGET} 
and {\tt QSO\_SDSS\_TARGET} bits). The purpose of the {\tt QSO\_KNOWN} bit is to track which previously known objects have a reliable, visually inspected
(or otherwise highly confident) redshift and classification from prior spectroscopy. Objects classified as having excellent prior
spectroscopy are those that are of \sdss\ provenance and match the sample used to define known objects in \boss\ \citep[see][]{Ros12}, or those that match
the final \boss\ quasar catalog \citep[\drtwq; c.f.][]{Par14}. The {\tt QSO\_KNOWN} bit is intended to represent that subset of objects
deliberately not observed that have a {\em reliable} spectrum---because objects without such a reliable spectrum are almost certainly {\em not} quasars. The main
utility of this bit is to populate catalogs for scientific analyses with reliable previous redshifts and classifications. The version of the 
\drtwq\ catalog used to set {\tt QSO\_KNOWN} changed at the time of the {\tt eboss6} tiling in the same manner as
described for the {\tt QSO\_BAD\_BOSS} bit.


\begin{figure}[t]
\centering
\includegraphics[width=0.48\textwidth,height=0.3\textwidth]{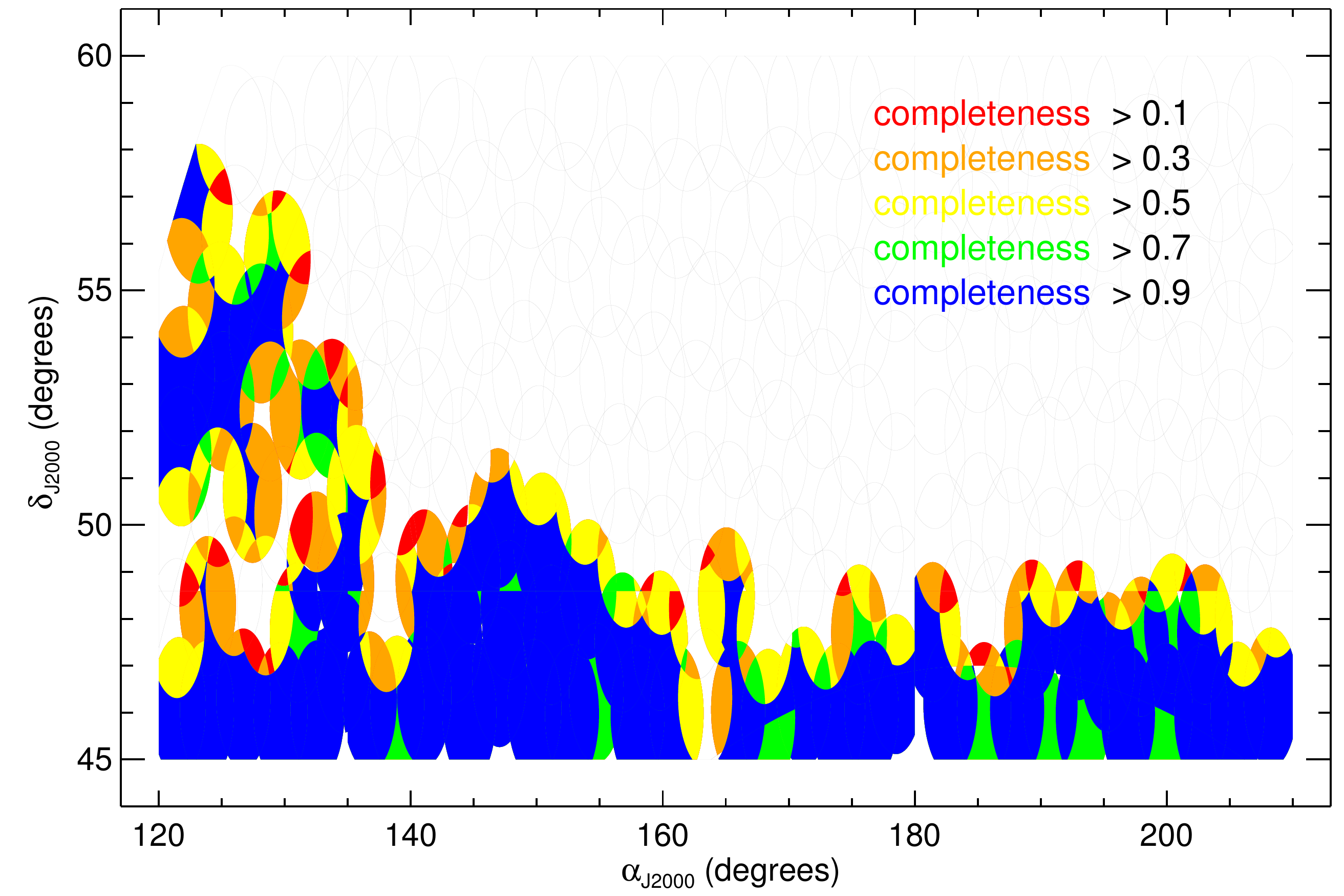}
\caption{\small 
The targeting completeness of CORE quasars as a function of position across the first 66 plates of \sequels. Blue corresponds to a completeness of greater than 90\%, red of only greater than 10\%. Gray lines depict sectors of \sequels\ that have yet to be observed. The structure of the overlapping plates in defining complete areas is apparent, and the quasar density is a function of that completeness. Overall, the depicted \sequels\ plates with completenesses above zero comprise 299.3\,deg$^2$ of area, but an effective area (area $\times$ targeting completeness) of only 236.3\,deg$^2$.
}
\label{fig:sequelscomp}
\end{figure}

\subsubsection{{\tt DO\_NOT\_OBSERVE}: Which previously known quasars are targeted?}
\label{sec:donotobs}

The parameter space for \eboss\ quasar targeting overlaps that of earlier iterations of the \sdss. The bits 
{\tt QSO\_BAD\_BOSS, QSO\_BOSS\_TARGET, QSO\_SDSS\_TARGET}, and {\tt  QSO\_KNOWN} work together to determine a
sample of objects for which \eboss\ does not need to obtain an additional spectrum because a good
classification and redshift should already exist (if the object {\em is} a quasar). Targets are not observed if any of
{\tt QSO\_BOSS\_TARGET, QSO\_SDSS\_TARGET} or {\tt  QSO\_KNOWN} are set {\em unless} {\tt QSO\_BAD\_BOSS} is set.
In addition, {\tt QSO\_REOBS} always forces a reobservation of an earlier \boss\ quasar.
In Boolean notation, {\tt DO\_NOT\_OBSERVE} is then set according to quasar target bits if:

\begin{equation}
\begin{split}
{\tt  (QSO\_KNOWN~ ||~QSO\_BOSS\_TARGET~||~QSO\_SDSS\_TARGET)} \\ {\tt \&\&~!(QSO\_BAD\_BOSS~||~QSO\_REOBS)} ~.
\end{split}
\end{equation}

\noindent The reduction in target density from implementing this schema is significant. Broadly, the total density of \eboss\ CORE quasar targets 
that have to be allocated a fiber drops
from $\sim115$\perdegsq\ to close to $\sim90$\perdegsq\ with effectively no loss of useful quasars (see \S\ref{sec:results}). This filtering allows \eboss\ to target
a larger number of \LyA\ quasars using the {\tt QSO\_PTF} method, and 
may ultimately result
in a larger total area for \eboss.

\subsubsection{{\tt DR9\_CALIB\_TARGET}: Which version of the \sdss\ imaging was used?}

\eboss\ quasar targeting always uses the updated imaging described in \S\ref{sec:v5b}. In 
\S\ref{sec:results} we describe a preliminary survey called \sequels\ that bridged the \sdss-\project{III} and
\sdss-\project{IV} surveys. \sequels\ targeted quasars selected in both the \drn\ imaging used for \boss\ {\em and}
the updated imaging used in \eboss. The {\tt DR9\_CALIB\_TARGET} bit signifies quasars that were selected for
\sequels\ using the \drn\ imaging calibrations.

\section{Results from a large pilot survey}
\label{sec:results}

The approaches discussed so far for \eboss\ quasar targeting were mostly based upon an $\sim 11$\degsq\ test 
survey, which is further described in the appendix of \citet{DR12}, that was conducted in the CFHT Legacy Survey W3 field
(e.g., see \S\ref{sec:xdqsozcuts} and \S\ref{sec:wisecuts}). This test field alone was sufficient to define a mature \eboss\ quasar targeting process, which
is outlined in \S\ref{sec:bits}. To determine whether the targeting approaches detailed so far in this paper truly met \eboss\ goals,
and to provide a sample for initial scientific analyses, a larger pilot survey was conceived as part of \sdss-\project{III}. This
section describes the targeting results from this survey, the {\em Sloan Extended QUasar, ELG and LRG Survey} (\sequels), in the context
of whether they meet the goals outlined in \S\ref{sec:SRD}.

\begin{table}[t]
\centering
\caption{\small
Redshift and classification efficiency from \sequels\ for CORE quasars upon visual inspection}
\begin{tabular}{c c c c c c c}
\hline\hline
$r <$ & $f_{\rm conf}$ & $f_{\rm z}$ & $f_{\rm qsoconf}$ & $f_{\rm qsoz}$ & $f_{\rm coreconf}$ & $f_{\rm corez}$ \\ 
(1) & (2) & (3) & (4) & (5) & (6) & (7) \\ \hline
21.0 & 0.981 & 0.960 & 0.996 & 0.970 & 0.997 & 0.973 \\
21.1 & 0.980 & 0.960 & 0.995 & 0.970 & 0.996 & 0.973 \\
21.2 & 0.978 & 0.958 & 0.994 & 0.970 & 0.996 & 0.972 \\
21.3 & 0.977 & 0.958 & 0.993 & 0.970 & 0.995 & 0.972 \\
21.4 & 0.977 & 0.957 & 0.993 & 0.970 & 0.995 & 0.972 \\
21.5 & 0.975 & 0.956 & 0.992 & 0.969 & 0.995 & 0.972 \\
21.6 & 0.971 & 0.953 & 0.991 & 0.968 & 0.993 & 0.971 \\
21.7 & 0.968 & 0.950 & 0.989 & 0.967 & 0.992 & 0.970 \\
21.8 & 0.964 & 0.947 & 0.987 & 0.966 & 0.990 & 0.970 \\
21.9 & 0.960 & 0.944 & 0.986 & 0.966 & 0.989 & 0.969 \\
22.0 & 0.957 & 0.941 & 0.984 & 0.965 & 0.987 & 0.968 \\
\hline\hline
\end{tabular}
\tablecomments{(1) The $r$ limit for which the efficiencies are derived; (2) The fraction of all quasar targets with a highly confident classification and redshift;
(3) The fraction of all quasar targets for which the \sdss\ spectroscopic pipeline redshift is accurate; (4--5) As for columns (2--3) but for targets classified as quasars on visual inspection;
(6--7) As for columns (2--3) but for quasar targets classified as $0.9 < z < 2.2$ (i.e.\ ``CORE'') quasars on visual inspection.
}
\label{tab:zeff}
\end{table}

\subsection{Details of the \sequels\ survey}
\label{sec:sequels}

\sequels\ comprises two chunks of 
\boss\footnote{designated {\tt boss214} and {\tt boss217}; see \url{http://www.sdss3.org/ \\ dr10/algorithms/boss\_tiling.php\#chunks} for a description of \boss\ chunks} 
covering $\sim810$\degsq\ in total area. \sequels\ approximates the region bounded by the \sdss\ Legacy imaging footprint and 
$120^\circ \leq \alpha_{\rm J2000} <  210^\circ$ and $+45^\circ \leq \delta_{\rm J2000} <  +60^\circ$.
Targets are selected as described thus far for \eboss\ with five slight differences:

\begin{figure}[t]
\centering
\includegraphics[width=0.48\textwidth,height=0.3\textwidth]{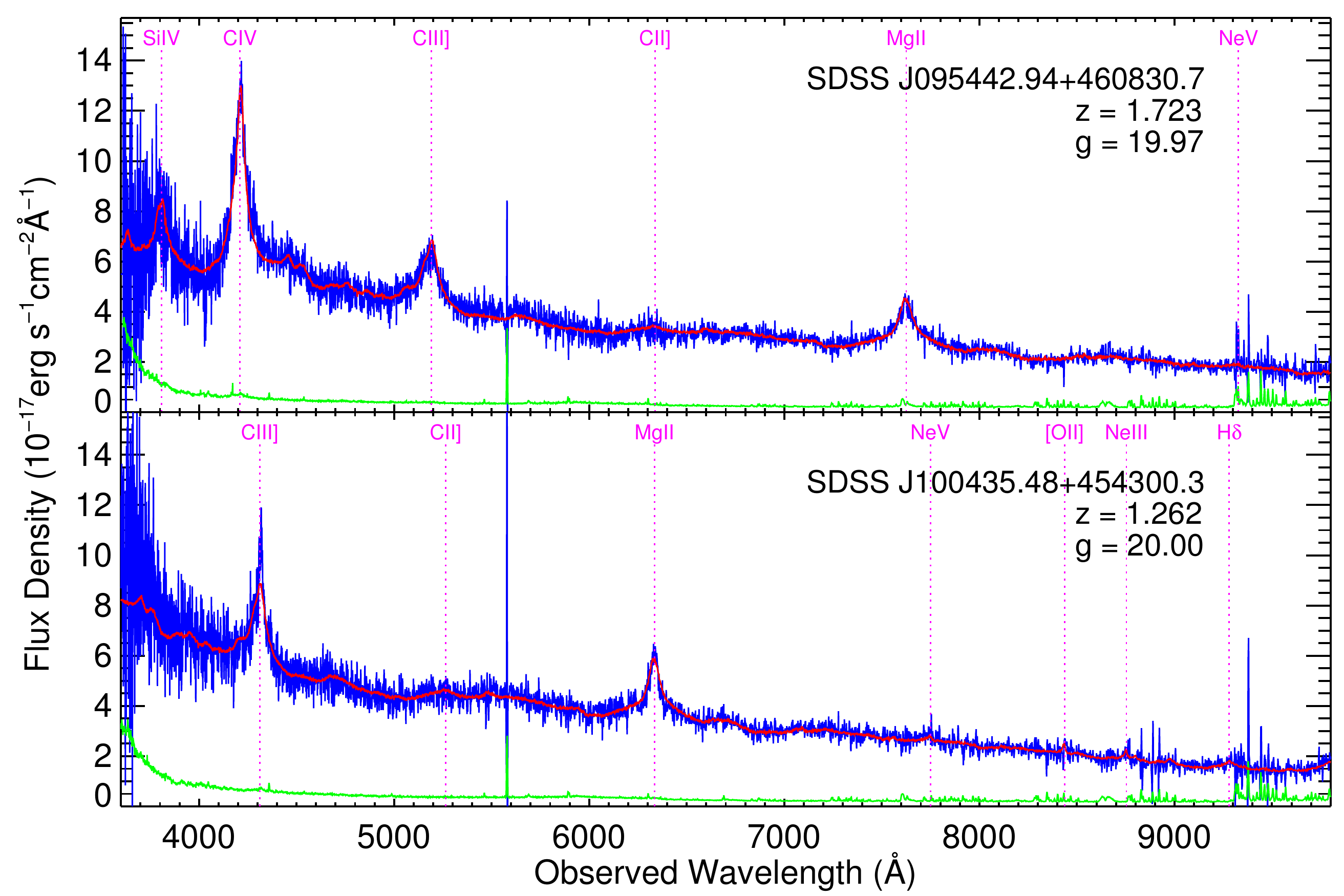}
\caption{\small 
Two representative spectra of $g\sim20$ quasars from \sdss\ plate 7284 (part of \sequels). Plate 7284 had a total exposure time of 75 minutes. The spectra 
have not been smoothed or otherwise enhanced. The dotted lines and associated labels mark the positions of some typical quasar emission lines
with rest-frame wavelengths taken from \citet{Van01}. Emission lines that are close to the edges of the covered wavelength range are not marked. Other labels are the object name, redshift, and (observed, not de-extincted) $g$-band target magnitude. The blue solid line depicts the flux density ($f_\lambda$), the green depicts the $1\sigma$ error on $f_\lambda$, and the red depicts the best-fit template output by the \sdss\ pipeline.}
\label{fig:g20spec}
\end{figure}

\begin{enumerate}

\item The bright-end cut enforced on all target classes in \sequels\ was $i > 17$ on {\tt FIBERMAG} {\em rather than} on {\tt FIBER2MAG}. This choice makes
a tiny difference to the selected targets, of order 0.2\%;

\item {\tt IMAGE\_STATUS} flags were not applied in \sequels. More than 97\% of the \sequels\ area has good {\tt IMAGE\_STATUS} according to our
definition from \S\ref{sec:flags}. The remaining $\sim3$\% of area, however, would not have been observed in \eboss\ proper;

\item The {\tt QSO\_EBOSS\_KDE} target class (see \S\ref{sec:bits}) was observed in \sequels\ but discontinued for \eboss;

\item CORE quasar targets in \eboss\ are all selected from the updated imaging described in \S\ref{sec:v5b}. In \sequels\ the superset arising from
{\em both} the updated and \drn\ imaging was targeted, because the updated imaging calibrations were considered to be preliminary.
As we shall outline in this section, the updated imaging is sufficient to meet \eboss\ goals, so targeting using \drn\ imaging was
discontinued after \sequels. In this section of the paper, we only discuss the results arising from the use of the updated imaging;

\item For \sequels\ the {\tt QSO\_PTF} target density was set at $\sim 35$\,deg$^{-2}$, which is higher than the typical \eboss\ density of
this target class of $\sim 20$\,deg$^{-2}$.

\end{enumerate}

\begin{figure}[t]
\centering
\includegraphics[width=0.48\textwidth,height=0.3\textwidth]{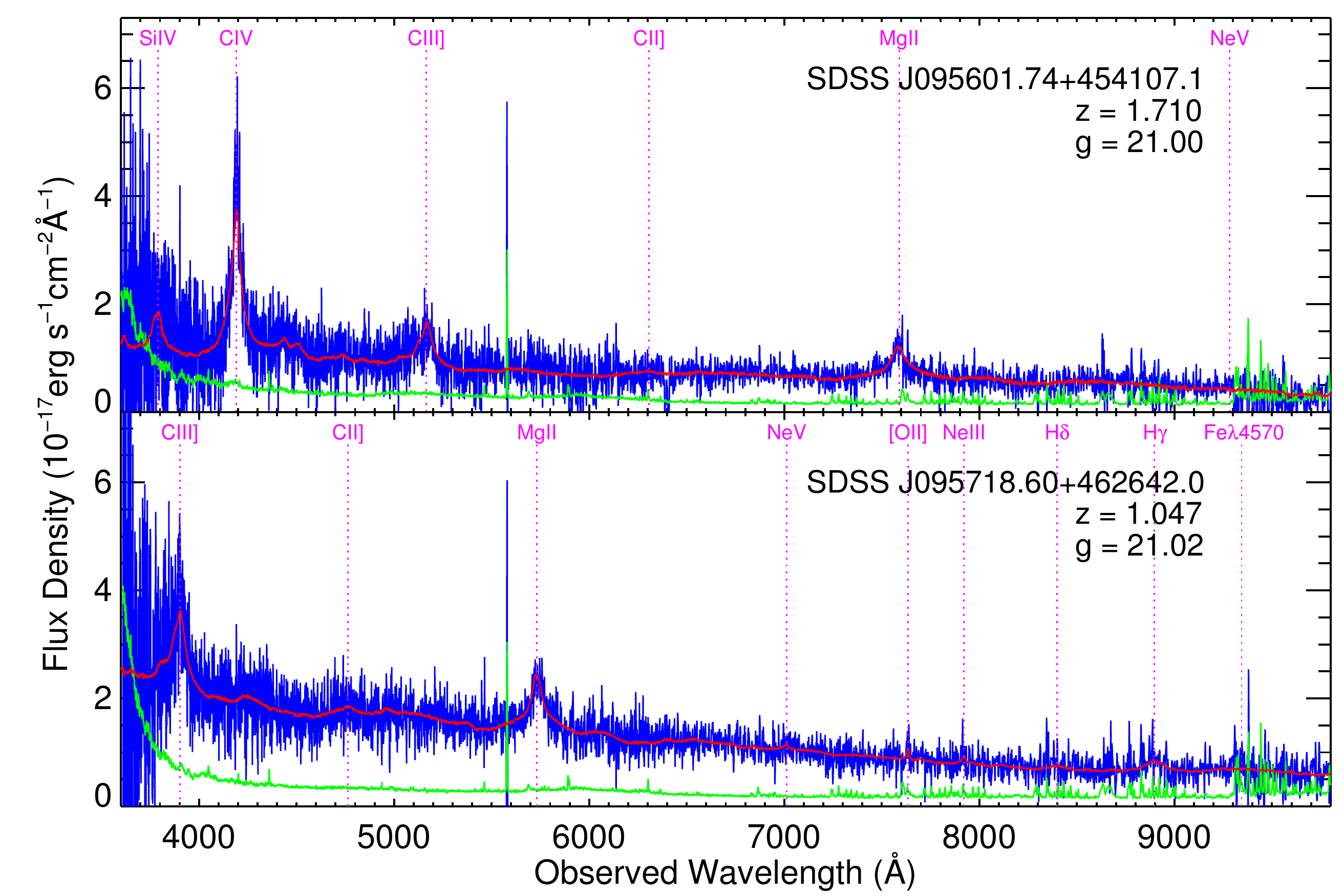}
\caption{\small 
As for Fig.\ \ref{fig:g20spec} but for $g\sim21$ quasars.
}
\label{fig:g21spec}
\end{figure}

\begin{figure}[t]
\centering
\includegraphics[width=0.48\textwidth,height=0.3\textwidth]{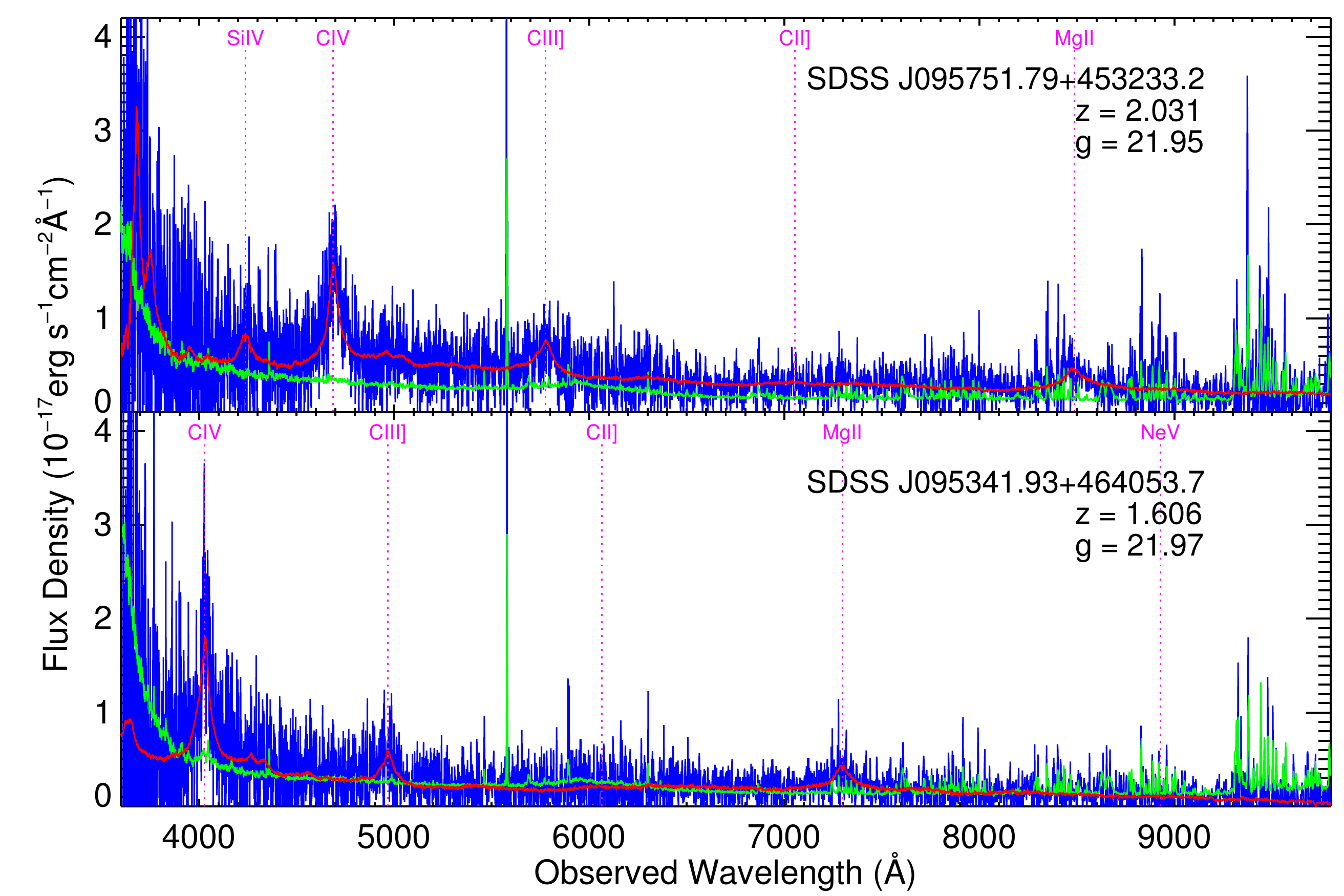}
\caption{\small 
As for Fig.\ \ref{fig:g20spec} but for $g\sim22$ quasars.
}
\label{fig:g22spec}
\end{figure}

\begin{table*}[t]
\centering
\caption{\small
Density of \sequels\ quasar targets that are confidently a quasar upon visual inspection}
\begin{tabular}{|c c c | c c c | c c c | c c c |}
\hline
{\footnotesize Comp.} & {\footnotesize Total}  & {\footnotesize Eff.}  & \multicolumn{3}{|c|}{{\footnotesize $0.9<z<2.2$ from CORE}} & \multicolumn{3}{|c|}{{\footnotesize ALL $z$ from CORE}} & \multicolumn{3}{|c|}{{\footnotesize New $z > 2.1$ from...}}  \\
{\footnotesize $>$} & {\footnotesize Area} & {\footnotesize Area} & {\footnotesize New} & {\footnotesize Known} & {\footnotesize Total} & {\footnotesize New} & {\footnotesize Known} & {\footnotesize Total} & {\footnotesize CORE} & {\footnotesize PTF} & {\footnotesize Total} \\ 
(1) & (2) & (3) & (4) & (5) & (6) & (7) & (8) & (9) & (10) & (11) & (12) \\ \hline
0.00 & 298.5 & 237.1 & 57.9 & 13.1 & 71.1 & 69.3 & 28.7 & 98.0 & 6.6 & 3.7 & 10.3 \\
0.80 & 189.9 & 183.5 & 58.3 & 13.4 & 71.6 & 69.7 & 29.0 & 98.7 & 6.6 & 4.6 & 11.2 \\ 
0.85 & 187.6 & 181.6 & 58.3 & 13.3 & 71.6 & 69.7 & 29.0 & 98.7 & 6.6 & 4.4 & 11.0 \\ 
0.90 & 174.5 & 170.0 & 58.4 & 13.4 & 71.8 & 69.8 & 29.2 & 99.0 & 6.6 & 4.4 & 11.0 \\ 
0.95 & 125.9 & 124.7 & 59.2 & 12.8 & 72.0 & 71.0 & 27.9 & 98.9 & 7.0 & 4.1 & 11.1 \\ 
\hline
\end{tabular}
\tablecomments{(1) Targeting completeness (fraction of CORE targets which received a fiber) limit of the sectors used for a given row of the table 
(see also Fig.\ \ref{fig:sequelscomp}). \eboss\ should be $>$ 95\% complete;
(2) Total \sequels\ area above this completeness (\degsq); (3) The effective area (area in \degsq\ weighted by per-sector completeness);
(4) Completeness-weighted total density of {\em new} (i.e.\ previously unconfirmed) $0.9 < z < 2.2$ quasars (\perdegsq) targeted by the CORE (i.e.\ having the {\tt QSO\_EBOSS\_CORE} bit set). We
define a quasar as an object classified {\tt QSO} or {\tt QSO\_Z?} as in Table~2 of \citet{Par14};  (5) The total density of previously confirmed $0.9 < z < 2.2$ quasars from 
earlier \sdss\ surveys (\perdegsq) targeted by the CORE;
(6) Total density (completeness-weighted) of $0.9 < z < 2.2$ quasars that would comprise the CORE clustering sample (\perdegsq). We only include objects classified as a quasar---a 
further 1.5--2\perdegsq\ of CORE targets are galaxies (or unidentifiable objects) at $0.9 < z < 2.2$; (7--9) As for columns (4--6) but for {\em all} quasars selected
by the CORE (not just those that are at $0.9 < z < 2.2$ on visual inspection); (10) New quasars selected by the CORE as for columns (4) and (7) but specifically
at $z > 2.1$ (the \LyA\ quasar redshift range); (11)
New quasars (heterogeneously) selected by {\em only} \ptf\ (i.e., having the {\tt QSO\_PTF} bit set), this column is 
{\em not} completeness-weighted; (12) Total density of {\em new} $z > 2.1$ quasars that would comprise 
the \eboss\ sample of \LyA\ quasars.
}
\label{tab:stats}
\end{table*}

Spectroscopic observations for \sequels\ were conducted in the same fashion as general \boss\ plates \citep[see][]{Daw13} with average exposure 
times of 75 minutes.
The \sequels\ observations contained in \drtw\ consist of 66 plates over an effective area
of 236.3\,deg$^2$.  The coverage is depicted in Fig.\ \ref{fig:sequelscomp}. The targeting completeness,
defined as the fraction of all targets that have received a fiber in each overlapping sector of the survey\footnote{see
\citet{Bla03} for the definition of a sector in the context of \sdss\ tiling}, is plotted. Sectors are derived using the MANGLE
software package \citep[e.g.][]{mangle}.

Every object targeted as a quasar or identified as a likely quasar by the automated pipeline \citep{Bol12}
was visually inspected following the procedures presented in \citet{Par14}.
The final classifications are described in \drtwq.
A summary of the results is reported in Table~\ref{tab:zeff}.
Fig.\ \ref{fig:g20spec}--\ref{fig:g22spec} display typical \sequels\ spectra as a function of $g$-band magnitude.
It is apparent that 
even the faintest quasars observed in \sequels\ (Fig.\ \ref{fig:g22spec}) can be identified
and assigned a redshift on visual inspection, even with no smoothing or other enhancements to the spectrum. A caveat
is that \sequels\ was conducted during particularly good observing conditions, and there is therefore no guarantee that the quality of
\sequels\ spectra will be representative of the full \eboss\ survey.

Based on Table~\ref{tab:zeff}, we expect of order 96\% of {\em all} quasar targets in \eboss\ will be confidently classified to $r < 22$, and
$\simeq99$\% of CORE quasars should be confidently identified. There are 
reasons to believe that \sequels\ may slightly overestimate our ability to classify quasars in {\em every} area of the \eboss\
survey, for a number of reasons. First, the \sequels\ area contains relatively good imaging when compared to several
\eboss\ areas in the \sdss\ SGC region (see \S\ref{sec:homo}). Second, as \sequels\  occurred concurrently with \boss\
observations, some $z > 2$ \boss\ quasars that would {\em not} be reobserved in \eboss\ were tagged as \sequels\
targets---and, in general, $z > 2$ quasars are easier to classify as the
strong \LyA\ line and the \LyA\ Forest are redshifted into the \boss\ spectrograph bandpass at about $z > 2$. More comprehensive details of the \eboss\ pipeline and
spectral classification procedures---and, in particular, whether the pipeline meets the requirements discussed in 
\S\ref{sec:SRD}---are provided in our companion overview paper \citep{ebosspaper???}.

\subsection{Projected \eboss\ Targeting efficiency}
\label{sec:targeff}

Perhaps the most critical aspect of \eboss\ quasar targeting is that a sufficiently high density of quasars is obtained to make
meaningful and/or improved measurements of the BAO distance scale. Contingent on the {\em effective} area of \sequels\
(as depicted in Fig.\ \ref{fig:sequelscomp}) we can estimate the quasar density expected for \eboss. 
Making this estimate is relatively straightforward---it is obtained by dividing the total number of spectroscopically confirmed quasars in \sequels\
by the completeness-weighted area of the survey as a function of targeting approach and of redshift. For this 
purpose, ``completeness'' means {\em targeting} completeness to the {\em statistically selected 
quasar sample}, which is defined, here, to be {\em the fraction of CORE
quasar targets that received a fiber for spectroscopic observation}. Targeting incompleteness
occurs in \sequels\ for two main reasons: First, due to collisions, a fiber cannot always be placed on neighboring targets, causing
general incompleteness on a plate; and, second, certain
plates in \sequels\ are yet to be observed, causing significant incompleteness in areas where yet-to-be-observed plates overlap
completed plates. Table~\ref{tab:stats}
presents estimates of the \eboss\ quasar density. In addition to weighting the
CORE quasar counts by completeness on a sector-by-sector basis, Table~\ref{tab:stats} details results as a function of
completeness. Ultimately, \eboss\ is expected to be have a targeting completeness of 0.95 (due to collisions, fibers will only be placed on 95\%
of quasar targets), so it is worth noting that the statistics in Table~\ref{tab:stats} are somewhat dependent on completeness.

The results in Table~\ref{tab:stats} have been produced in a manner that should reflect the eventual targeting schema for \eboss. One subtlety is that 
most, but not all, \boss\ observations had been completed in the depicted area in Fig.\ \ref{fig:sequelscomp} by the time of \sequels\ observations. 
To better mimic \eboss, estimates in Table~\ref{tab:stats} are produced by substituting non-\sequels\ (\boss) identifications from 
\drtwq\ over \sequels\ targets, where they exist, and such objects are treated as 
previously observed, known quasars---i.e., when such objects have a good spectrum from \drtwq,
they are treated 
as if they had a known redshift from \boss\ and as if the {\tt DO\_NOT\_OBSERVE} bit had been set (see \S\ref{sec:donotobs}).
At the outset of \sequels, 8921 potential \sequels\ targets had the {\tt DO\_NOT\_OBSERVE} bit set due to a 
prior good spectrum in \sdss-\project{I, II} or \project{III}. Based on our substitution process, only an additional
267 ($\sim 3$\%) quasars would have had the {\tt DO\_NOT\_OBSERVE} bit set due to yet-to-be-completed
\boss\ observations, and only 92 ($\sim 1$\%) of these additional quasars
would have been in the redshift range $0.9 < z < 2.2$.

It is critical for users of \eboss\ data to be able to accurately track previously known quasars from earlier versions of the \sdss. 
Table~\ref{tab:stats} implies that of order $\sim13$\perdegsq\ $0.9 < z < 2.2$ quasars will be included in
\eboss\ as a prior confirmation. This number of $\sim13$\perdegsq\ previously identified CORE quasars is
as might be expected. The \sdss\-\project{I/II} quasar catalog of \citet{DR7QSO} contains
$\sim75{,}000$ $0.9 < z < 2.2$ quasars spread over 9400\degsq\ ($\sim 8$\perdegsq). The \boss\ quasar catalog of
\drtwq\ contains $\sim65{,}000$ $0.9 < z < 2.2$ quasars spread over 10{,}700\degsq\
 ($\sim 6$\perdegsq). These
catalogs also contain $\sim 1$\,\perdegsq\ mutual $0.9 < z < 2.2$ quasars. 
Depending on \sequels\ sector, the number of known quasars in the CORE redshift 
range can vary widely from as few as 5\perdegsq\ to as many as 25\perdegsq\ due 
to the complex set of ancillary programs that were conducted as part of \boss\ \citep[see, e.g.,][]{Daw13}.

The main purpose of this section is to investigate whether the \eboss\ target selection as applied to \sequels\ meets the
requirements discussed in \S\ref{sec:SRD}, which amount to a success rate of
$> 58$\perdegsq\ $0.9 < z < 2.2$ quasars over 7500\degsq. Whether the {\em area} requirements
of \S\ref{sec:SRD} will be met are discussed in \citet{ebosspaper???}. The results from 
the \sequels\ area suggest that \eboss\ will meet its quasar targeting requirements in terms of number densities. For a targeting completeness
reflective of \eboss\ ($\sim95$\%), a completeness-weighted density of 72.0\perdegsq\ $0.9 < z < 2.2$ quasars were identified in \sequels. 
This suggests that the \eboss\ CORE quasar selection will identify ($0.95 \times 72.0 =$) 68.4\perdegsq\ $0.9 < z < 2.2$ quasars.

The
\sdss\ imaging in the \sequels\ area may be of above-average quality, which could inflate these expectations
(see \S\ref{sec:homo}). There are also reasons to believe, however, that the \eboss\ quasar density may
be higher than \sequels\ expectations. For instance, \sequels\ data were reduced using the \sdss-\project{III} spectroscopic
pipeline, which, with augmentations, might improve on the $\sim1$\% loss due to unidentifiable quasars listed in 
Table~\ref{tab:zeff}. Also, there are 1.5--2\perdegsq\ additional objects in the 
CORE redshift range in \sequels\ that are not included in Table~\ref{tab:stats} because they are classified
as ``unknown'' or as galaxies upon visual inspection. In theory these objects 
can also be used for \eboss\ clustering analyses (although such objects have a median redshift of $\sim1.1$).

\begin{figure}[t]
\centering
\includegraphics[width=0.48\textwidth,height=0.3\textwidth]{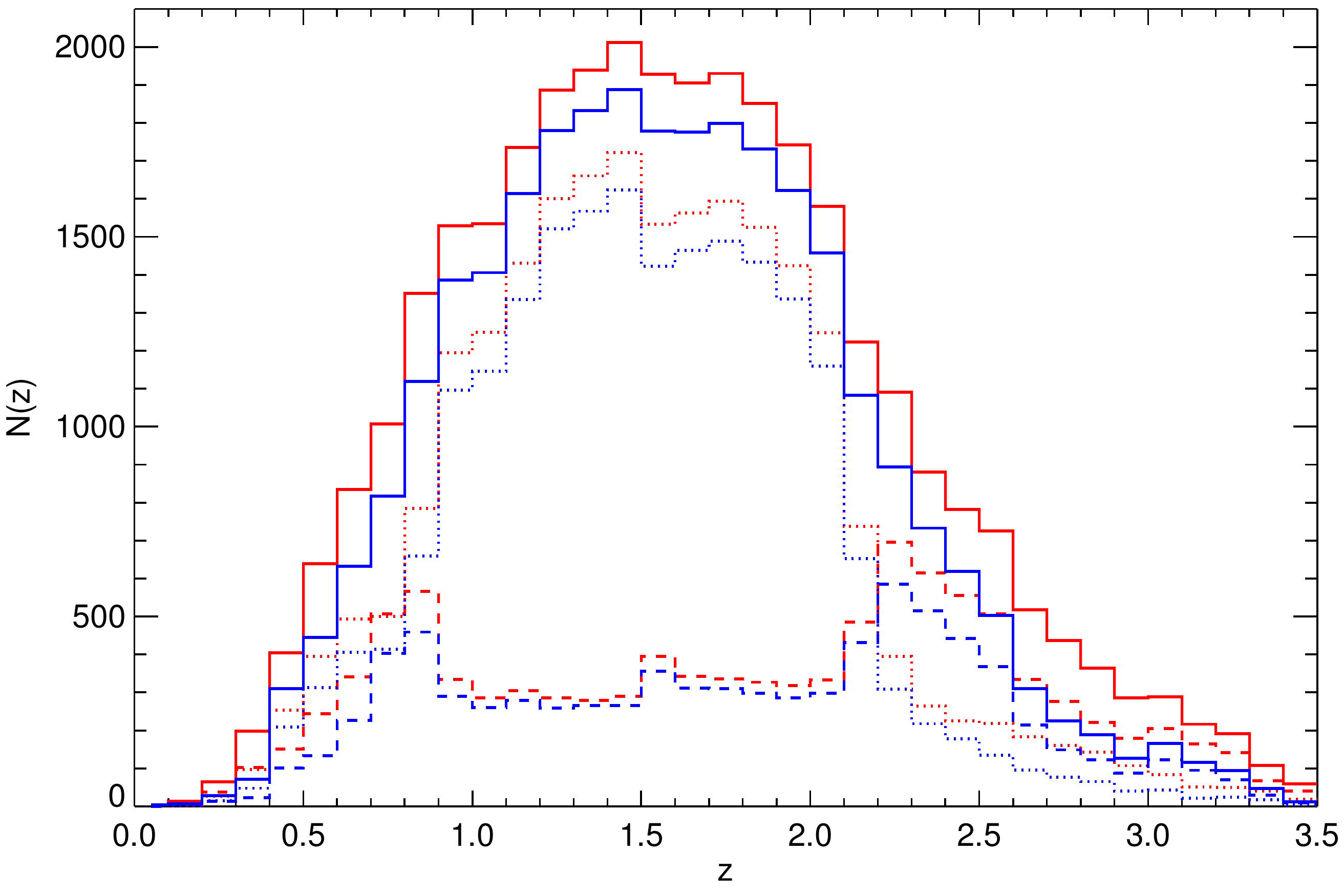}
\caption{\small 
The redshift distribution of quasars from \sequels. Red lines represent all quasars identified in \sequels, blue lines 
represent quasars targeted just by the CORE algorithm, and solid lines represent all quasars that would have been assigned 
a fiber by the \sequels\ targeting algorithm (i.e., including known \sdss\ or \boss\ quasars that do not need to be 
reobserved as they have the {\tt DO\_NOT\_OBSERVE} bit set). Dashed (dotted) lines represent quasars that were (were not) 
previously spectroscopically confirmed in the \sdss\ or \boss. The solid lines, which are the sum of the dotted and dashed lines, are quantified in 
columns 3 and 6 of Table~\ref{tab:sequelsnz} and have been completeness-corrected as described in that table.
}
\label{fig:sequelsnz}
\end{figure}

\begin{figure}[t]
\centering
\includegraphics[width=0.48\textwidth,height=0.3\textwidth]{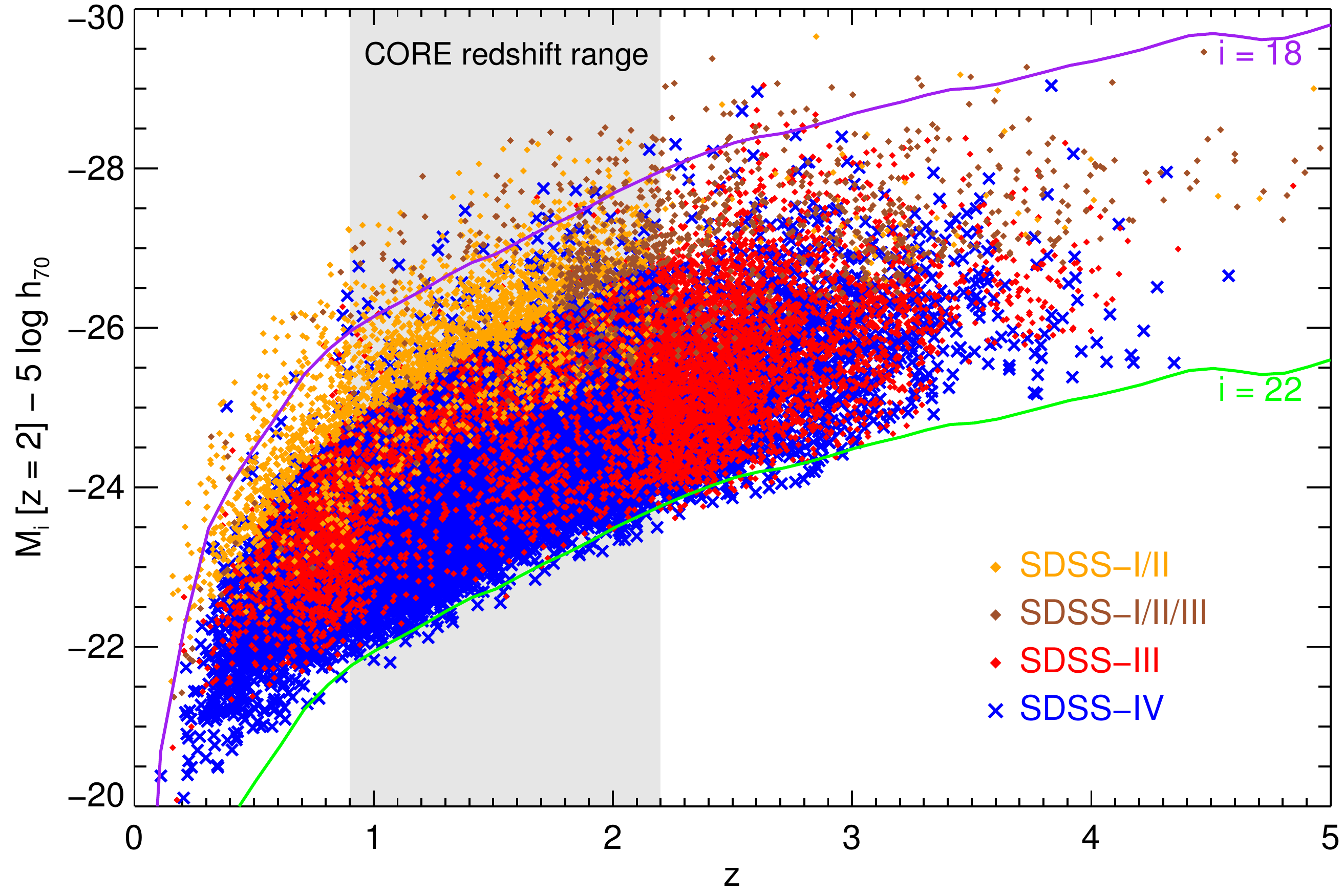}
\caption{\small 
The ($i$-band) absolute-magnitude-redshift plane for quasars targeted in \sequels. The blue crosses
depict new quasars that would be observed as part of \sdss-\project{IV}/\eboss. The other points represent
quasars that would be targeted by \eboss\ but that would not receive a fiber due to 
being previously observed in \sdss-\project{I/II} (orange)  \sdss-\project{III} (red)
or in both (brown; mostly 
ancillary targets or {\tt QSO\_KNOWN\_SUPPZ} targets; see \citealt{Daw13}). The lines track
quasars representative of the extremes of \sdss\ target selection between $i=18$ (purple)
and $i=22$ (green). The grey box illustrates the power of \eboss\ for detecting
new quasars in the CORE redshift range. All magnitudes are based on PSF fluxes, and have
been de-extincted. Absolute magnitudes have been $K$-corrected to $z=2$ using Table 4 of \citep{Ric06} and assume $H_0 = 70$\kms\perMpc.
}
\label{fig:sequelsMz}
\end{figure}

Fibers not allocated to other eBOSS target classes are assigned to finding new \LyA\ quasars ($z > 2.1$).
In Table~\ref{tab:stats} we show that \sequels\
contains ($7.0\times0.95$) = $\sim 6.7$\perdegsq\ new \LyA\ quasars acquired by the CORE selection and 
 ($4.1\times0.95$) = $\sim 3.9$\perdegsq\ new \LyA\ quasars acquired by other selections (mainly objects with
the {\tt QSO\_PTF} bit set). These results are likely robust for CORE targets (given the caveats discussed
in the previous paragraph). \LyA\ quasar target density may fluctuate across the survey with the
availability of \ptf\ imaging (see \S\ref{sec:variability}), so \sequels\ is a reasonable but imperfect
estimate of the success rate for new {\tt QSO\_PTF} \LyA\ quasars in \eboss. In particular, the 
target density of {\tt QSO\_PTF} sources was 35\perdegsq\ in \sequels\ but is expected to be close to 20\perdegsq\ across
the entire \eboss\ footprint (see \S\ref{sec:sequels}). The expected density of new $z > 2.1$ quasars from the \eboss\ {\tt QSO\_PTF} program 
is therefore quoted as 3--4\perdegsq\ in the abstract of this paper. There
are also reasons to believe, however that results from \sequels\ may {\em underestimate} the success of \eboss. Most
notably, our companion surveys such as \tdss\ \citep{Mor15} will target some
\LyA\ quasars in addition to those targeted by the {\tt QSO\_EBOSS\_CORE} and {\tt QSO\_PTF} approaches (see, 
e.g., J.\ Ruan et al.\ 2016, in preparation)



\subsection{Overall characteristics of \eboss\ quasars}
\label{sec:properties}

Beyond the cosmological goals of \eboss, the quasar sample produced by \sdss-\project{IV} should
be unparalleled, exceeding the depth and numbers of any previous quasar sample. As there is
likely to be significant interest in the nature of \eboss\ for quasar science, quasars observed as part of
\sequels\ are broadly characterized in this section. Because \sequels\ observations were
conducted in tandem with \boss, some quasars that would not normally receive a fiber in \eboss\
because of existing \boss\ spectroscopy did receive a \sequels\ fiber. Throughout this section, we treat such 
objects as if they had the {\tt DO\_NOT\_OBSERVE} bit set by correctly 
incorporating (non-\sequels) redshifts and classifications
from the \drtw\ quasar catalog (I.\ P{\^a}ris et al.\ 2016, in preparation), as also described in 
the discussion of Table~\ref{tab:stats} in \S\ref{sec:targeff}.

\begin{table}[t]
\centering
\caption{\small
$N(z)$ for \sequels\ quasars upon visual inspection}
\begin{tabular}{| c | r r c | r r c |}
\hline
 & \multicolumn{3}{|c|}{{\footnotesize CORE quasars}} & \multicolumn{3}{|c|}{{\footnotesize All quasars}} \\
$z$ & $N_{\rm raw}$ & $N$ & $dN$ & $N_{\rm raw}$ & $N$ & $dN$  \\ 
(1) & (2) & (3) & (4) & (5) & (6) & (7) \\ \hline
0.05 &  3 &      3.8 &  0.001 &  4 &      4.8 &  0.001 \\
0.15 &  6 &      6.3 &  0.002 &  14 &     14.3 &  0.004 \\
0.25 &  25 &     28.1 &  0.010 &  62 &     65.1 &  0.019 \\
0.35 &  61 &     70.8 &  0.025 &  189 &    198.8 &  0.059 \\
0.45 &  267 &    310.0 &  0.108 &  361 &    404.0 &  0.120 \\
0.55 &  381 &    445.2 &  0.155 &  575 &    639.2 &  0.190 \\
0.65 &  549 &    632.4 &  0.221 &  751 &    834.4 &  0.249 \\
0.75 &  732 &    817.2 &  0.285 &  922 &   1007.2 &  0.300 \\
0.85 &  983 &   1118.7 &  0.390 &  1215 &   1350.7 &  0.402 \\
0.95 &  1161 &   1386.6 &  0.484 &  1303 &   1528.6 &  0.455 \\
\hline
1.05 &  1170 &   1405.7 &  0.490 &  1299 &   1534.7 &  0.457 \\
1.15 &  1339 &   1613.5 &  0.563 &  1461 &   1735.5 &  0.517 \\
1.25 &  1467 &   1779.9 &  0.621 &  1574 &   1886.9 &  0.562 \\
1.35 &  1510 &   1832.5 &  0.639 &  1617 &   1939.5 &  0.578 \\
1.45 &  1555 &   1887.9 &  0.659 &  1679 &   2011.9 &  0.599 \\
1.55 &  1485 &   1778.7 &  0.620 &  1634 &   1927.7 &  0.574 \\
1.65 &  1475 &   1776.2 &  0.620 &  1604 &   1905.2 &  0.568 \\
1.75 &  1493 &   1798.1 &  0.627 &  1625 &   1930.1 &  0.575 \\
1.85 &  1435 &   1730.7 &  0.604 &  1556 &   1851.7 &  0.552 \\
1.95 &  1347 &   1621.8 &  0.566 &  1467 &   1741.8 &  0.519 \\
\hline
2.05 &  1219 &   1457.5 &  0.508 &  1342 &   1580.5 &  0.471 \\
2.15 &  949 &   1083.2 &  0.378 &  1089 &   1223.2 &  0.364 \\
2.25 &  833 &    893.5 &  0.312 &  1031 &   1091.5 &  0.325 \\
2.35 &  685 &    732.8 &  0.256 &  832 &    879.8 &  0.262 \\
2.45 &  584 &    619.5 &  0.216 &  746 &    781.5 &  0.233 \\
2.55 &  474 &    502.7 &  0.175 &  697 &    725.7 &  0.216 \\
2.65 &  291 &    310.7 &  0.108 &  498 &    517.7 &  0.154 \\
2.75 &  211 &    225.8 &  0.079 &  423 &    436.8 &  0.130 \\
2.85 &  174 &    188.5 &  0.066 &  349 &    364.5 &  0.109 \\
2.95 &  120 &    127.3 &  0.044 &  280 &    286.3 &  0.085 \\
\hline
3.05 &  156 &    165.8 &  0.058 &  278 &    288.8 &  0.086 \\
3.15 &  112 &    116.4 &  0.041 &  212 &    216.4 &  0.064 \\
3.25 &  89 &     93.9 &  0.033 &  188 &    191.9 &  0.057 \\
3.35 &  44 &     47.6 &  0.017 &  103 &    107.6 &  0.032 \\
3.45 &  12 &     12.8 &  0.004 &  58 &     58.8 &  0.018 \\
3.55 &  9 &     10.8 &  0.004 &  58 &     59.8 &  0.018 \\
3.65 &  6 &      7.0 &  0.002 &  61 &     62.0 &  0.018 \\
3.75 &  8 &      9.6 &  0.003 &  51 &     50.6 &  0.015 \\
3.85 &  6 &      6.5 &  0.002 &  37 &     39.5 &  0.012 \\
3.95 &  4 &      4.5 &  0.002 &  27 &     27.5 &  0.008 \\
\hline
4.05 &  3 &      3.3 &  0.001 &  19 &     19.3 &  0.006 \\
4.15 &  2 &      2.5 &  0.001 &  10 &     10.5 &  0.003 \\
\hline
\end{tabular}
\tablecomments{
(1) Redshift; (2) Number of \sequels\ quasars selected by the CORE targeting algorithm; (3) As for column (2)
but completeness-corrected; (4) As for column (3) but normalized;
(5-7) As for columns (2-4) but for \sequels\ quasars selected by any targeting algorithm. Completeness
corrections are conducted by multiplying the counts of all {\em newly} identified CORE quasars by 298.5/237.1 (see the
first row of Table~\ref{tab:stats}). Counts of all other quasars in \sequels\ are
not completeness-corrected as they are dominated by quasars that were previously confirmed in the \sdss\ or \boss---such quasars
are effectively assigned a fiber 100\% of the time. 
A quasar is defined using {\tt QSO} or {\tt QSO\_Z?} as in Table~\ref{tab:stats}.
\label{tab:sequelsnz}
}
\end{table}

The redshift distribution of quasars in \sequels\ is plotted in Fig.\ \ref{fig:sequelsnz} and is 
similar to the expectation from Fig.\ \ref{fig:opticalNz}.
The measurements of the \sequels\ $N(z)$ are listed in Table\ \ref{tab:sequelsnz}. When combined
with the expected total \eboss\ quasar target density over all redshifts of $\sim99$\perdegsq\ (see Table~\ref{tab:stats}) 
and the expected $7500$\degsq\ area of \eboss, the \sequels\ $N(z)$ should be
sufficient to project science results using an \eboss-like sample. 
The redshift-absolute-magnitude distribution of \sequels\ is provided in Fig.\ \ref{fig:sequelsMz}.
This figure illustrates why \eboss\ will be {\em the} next-generation quasar survey, complementing
the (largely) $i < 19$ space of \sdss-\project{I/II} and the (largely) $z < 0.9$ and $z > 2.2$ space of \boss, by
filling in the $i > 19$ and $0.9 < z < 2.2$ quasar space in an unprecedented fashion.

The overall expected
quasar numbers for \eboss\ can be estimated from the \sequels\ $N(z)$ and number densities. Projecting from Table~\ref{tab:stats} and
assuming a minimum \eboss\ area of $7500$\degsq\ (\S\ref{sec:SRD}), \eboss\ should, conservatively,
comprise {\em at least} 500{,}000 spectroscopically confirmed $0.9 < z < 2.2$ quasars selected in a uniform manner 
with which to pursue quasar clustering studies such as the BAO scale, and at least 500{,}000 
total {\em new} quasars (at any redshift) that have never before been spectroscopically identified and characterized.
Overall, at the completion of \eboss, 
the \sdss\ surveys will have provided unique spectra of over 800{,}000 total quasars, including \sdss\ areas
outside of the \eboss\ footprint as well as new quasars observed by the \tdss\ and
\spiders\ surveys.

\begin{figure}[t]
\centering
\includegraphics[scale=0.344,clip=true,trim=0 0 0 -27]{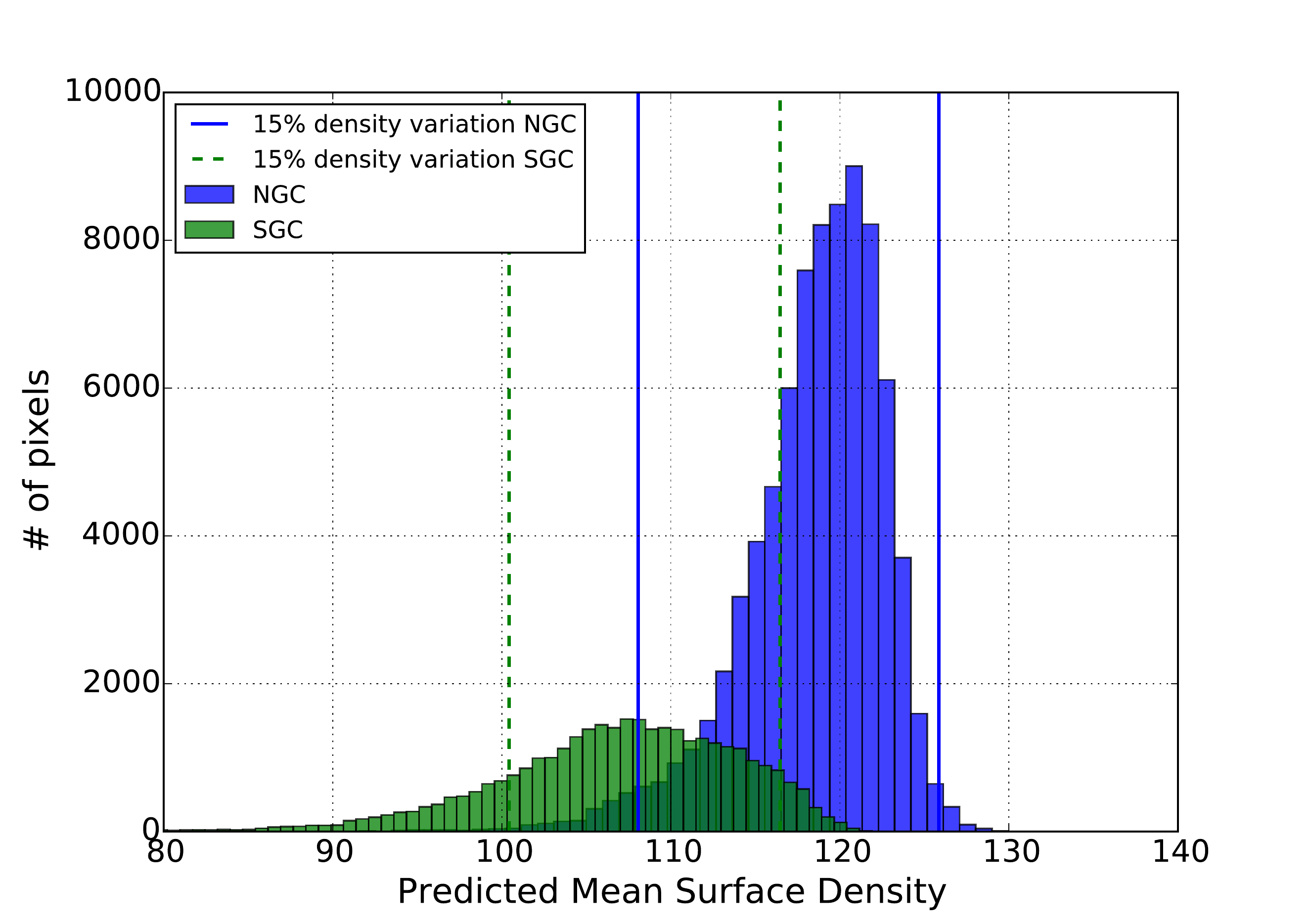}
	\caption{ 
	Histograms of the surface density of CORE quasar targets predicted by the regression models described in \S\ref{sec:homofluc} (the ``PSD''). The blue histogram represents the NGC, with solid blue lines depicting the window within which angular fluctuations in quasar target density meet the $\leq 15\%$ requirement of \S\ref{sec:SRD}. The green histogram and dotted green lines depict the same quantities for the SGC. 
	The histograms demonstrate that  $\sim97\%$ ($\sim77\%$) of the NGC (SGC) footprint meets the homogeneity requirements of \eboss\ (see \S\ref{sec:SRD}). The PSD and the fractional deviation from the mean PSD in each pixel are depicted as a sky map 
	in Fig.\ \ref{fig:qso_skymap}.
	}
\vspace{5pt}
\label{fig:qso_regress_pd}
\end{figure}

\begin{figure*}[t]
\centering
	\includegraphics[scale=0.468,clip=true,trim=-5 0 0 0]{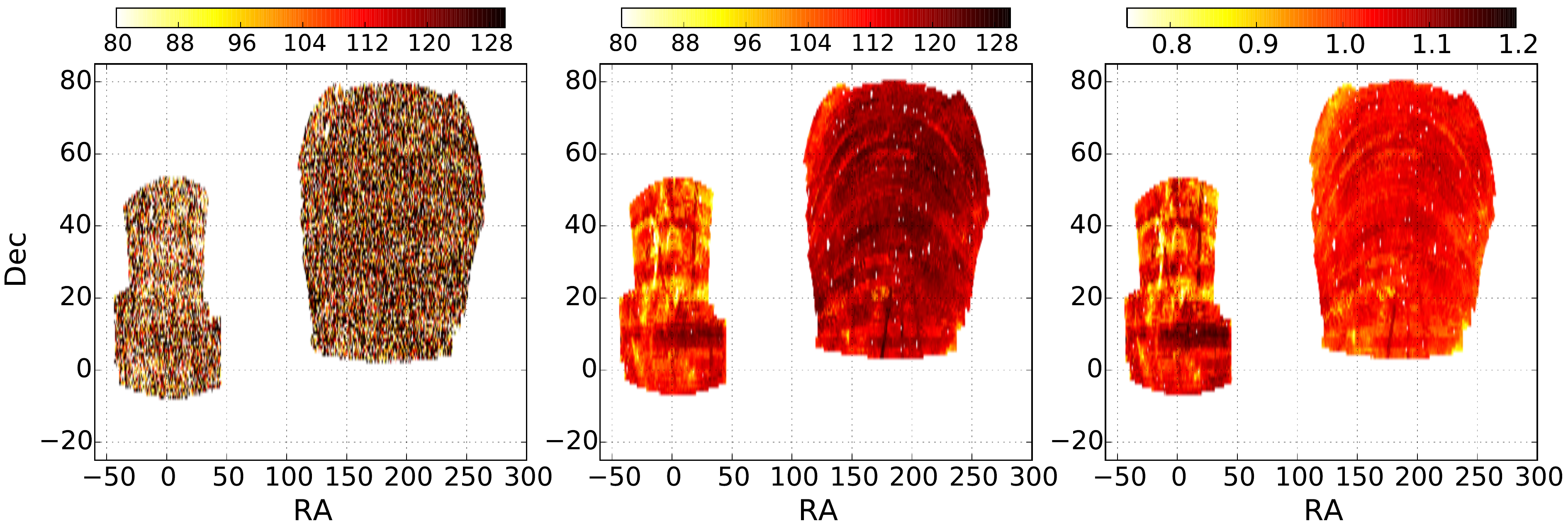}
	\caption{ Actual and theoretical maps of \eboss\ CORE quasar targets in J2000 Equatorial Coordinates (degrees).
	The left-hand panel shows the observed surface density sky-map of targets over the \boss\ footprint. \eboss\ will target quasars
	over a $\sim$7500\,\degsq\ subset of this area. As CORE quasar targets are relatively scarce ($\sim115$\,\perdegsq) fluctuations in this map
	are dominated by Poisson noise and sample variance. The central panel shows the theoretical map of CORE quasar target density predicted by the linear regression from imaging systematics (the PSD described in \S\ref{sec:homofluc}). The color bars above the left-hand and central panels represent target densities in \perdegsq. The right-hand panel rescales the map in the central panel so that it is expressed as a fractional deviation from the mean 
	(i.e.\  the color-bar above this panel represents the quantity PSD/$\langle {\rm PSD} \rangle$).}
\label{fig:qso_skymap}
\end{figure*}

\section{Tests of the homogeneity of the CORE quasar sample}
\label{sec:homo}

In order to perform clustering measurements to characterize the BAO scale, it is necessary
to mimic the angular distribution imposed by the target selection. This survey ``mask'' is often
expressed as a random catalog, or control sample, that mimics the characteristics of the 
targeted population but in the absence of any clustering. At its simplest, this process involves
uniformly distributing random points over the footprint of the target imaging. This simple 
approach, however, is rarely adequate because survey systematics such as seeing, sky brightness,
Galactic extinction etc.\ alter the target density in a complex manner. A related issue is that
zero-point calibrations in \sdss\ imaging can vary across the survey, also producing non-cosmological
variations in target density.

\subsection{Target density fluctuations due to systematics}
\label{sec:homofluc}

Previous studies of large-scale galaxy clustering over the
\sdss\ footprint \citep[e.g.,][]{Ros11} have demonstrated that systematics that produce target density 
variations at a level of $\sim$15\% or less can be controlled for by weighting the random catalog by a model of the
effect of that systematic. Beyond the 15\% level, systematics become more difficult to ``weight''
for, perhaps because some major systematics are covariant. When the effect of systematics exceeds
the 15\% level, that area of the survey may have to be excised from clustering analyses.

As part of \eboss\ target selection, a set of
regression tests have been devised to study how possible systematics in \sdss\ and \wise\ imaging may affect target
density---and whether such effects are below the $\sim15\%$ level that could be modeled 
with a suitable weighting scheme. The slate of systematics, which represents a reasonable (but not necessarily
exhaustive) list of quantities that could bias \eboss\ target density,
 is further detailed in a companion paper \citep{Prakash15inprep???}. Relevant to the \wise\ imaging; the systematics
include the median numbers of
exposures per pixel, the fraction of exposures contaminated by the Moon, and the total flux per pixel, 
all in the W1 band ({\tt W1covmedian, moon\_lev, W1median}). Relevant to the \sdss\ imaging;
the systematics include the FWHM and background sky-level in \sdss\ $z$-band, which are used
to track the quality of the seeing and the sky brightness. Additional systematics include Galactic latitude (to map the density
of possible contaminating stars) and Galactic dust (extinction in the $r$-band is used to represent this systematic).

The adopted regression technique is also detailed in \citet{Prakash15inprep???}. Briefly, 
the potential \eboss\ imaging footprint is deconstructed into equal-area pixels of 0.36\,\degsq. The 
\eboss\ CORE quasar target density and the mean value of each systematic is determined for each of these pixels. 
The observed surface density (${\rm SD}_{\rm obs}$) of \eboss\ CORE quasar targets in each pixel can be expressed as
a linear model of systematics

\begin{align}
      {\rm SD}_{\rm obs} = S_{0} + \sum\limits_{i=1}^7 S_{i}x_{i} + \epsilon,\label{eqn:sd_eqn}    
\end{align}

\noindent where $S_{0}$ is the mean target density across the pixels, $S_{i}$ is the weight accorded to fluctuations in target density ($x_{i}$) due to
systematic $i$, and $\epsilon$ is the combined effect of noise and variance, which is approximated
as a Gaussian. Multi-linear regression is used to determine $S_{0}$ and $S_{i}$ by minimizing the value of reduced $\chi^2$ across the pixels.
This regression is conducted separately in each Galactic hemisphere, such that different coefficients
are derived for the NGC and SGC regions of the \sdss\ imaging.

\begin{figure*}[t]
\centering
\includegraphics[scale=1.06,clip=true,trim=5 0 -20 0]{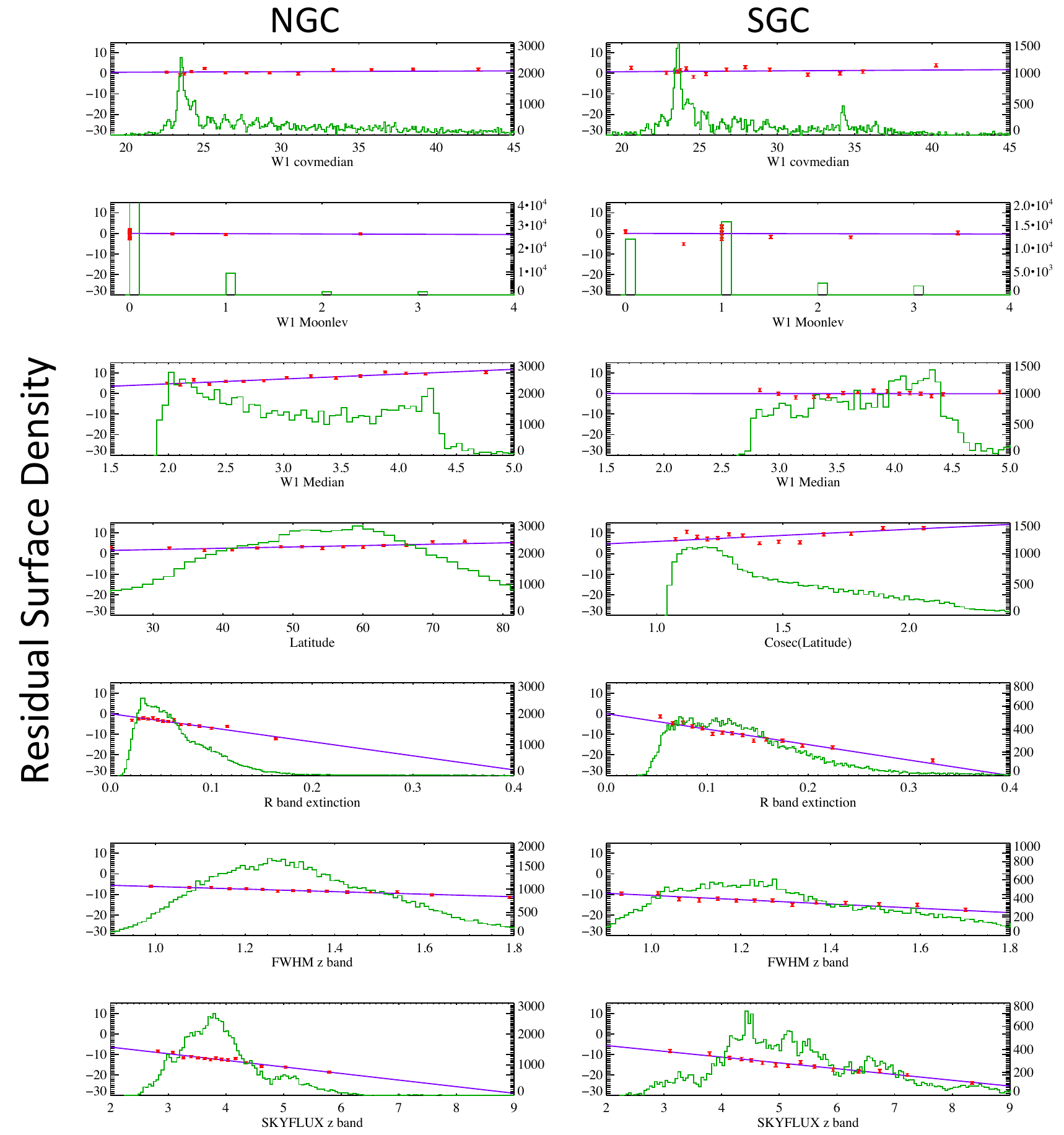}
\caption{
Systematics distributions and linear regression surface density models for \eboss\ CORE quasar targets. Each row of panels corresponds to
one of the systematics outlined in \S\ref{sec:homofluc} (``Latitude'' refers to Galactic latitude). 
The left-hand (right-hand) column of panels displays results for
these systematics for the NGC (SGC). The green histograms depict the 
distribution of pixels as a function of the mean value of each systematic in each pixel. The number of pixels is quantified on the right-hand axis of each
plot. The red data points and blue lines depict, instead, measures of the Residual\_SD (Eqn.\ \ref{eqn:rsd_eqn}), which is 
quantified on the left-hand axis of each plot. The points are the measured values of the Residual\_SD 
averaged over 4000 sky pixels in the NGC or 2000 pixels in the SGC. The error bars depict the standard error on the mean 
across the pixels. The lines show the best-fit regression models. A linear regression model appears to be an adequate description of how each displayed systematic affects \eboss\ CORE quasar
target density.
}
\label{fig:qso_regress_mp}
\end{figure*}

Once the coefficients of the linear regression model for systematics have been established,
a
statistic designated the {\em Predicted Surface Density}
or ``PSD'' is computed. The PSD is obtained by using $S_{0}$ and $S_{i}$ to calculate
what the \eboss\ CORE quasar density {\em should be} in a given pixel if the linear regression model is an
adequate description

\begin{equation}
      {\rm PSD} = S_{0} + \sum\limits_{i=1}^7 S_{i}x_{i}~~.\label{eqn:psd_eqn}\\
\end{equation}

\noindent Fig.\ \ref{fig:qso_regress_pd} presents
a histogram of the CORE quasar PSD as predicted from the derived linear regression model coefficients across all of the systematics. 
A total of 96.7\% of the \sdss\ imaging footprint in the NGC\footnote{only the area that could be useful for \eboss\ targeting,
due to scheduling constraints, is considered
\citep[see][and Fig.\ \ref{fig:qso_skymap}]{ebosspaper???}}
fluctuates in CORE quasar PSD at less than 15\%. The corresponding fraction is 76.7\% in the SGC footprint.

Fig.\ \ref{fig:qso_skymap} 
illustrates these deviations {\em on the sky} using a map of the PSD statistic, which serves to illustrate
the most problematic areas of the \sdss\ footprint for \eboss. The right-hand panel of Fig.\ \ref{fig:qso_skymap}
approximates the ``mask'' that will be necessary to ameliorate the effects of systematics on clustering measurements that use \eboss\
CORE quasars. The effective area or random catalog in each region of the \eboss\ footprint can be re-weighted
by the values displayed in the right-hand panel of Fig.\ \ref{fig:qso_skymap}, although
regions that deviate by more than 15\% from expectation 
may need to be excised from the survey in order to reach the target density variation requirement of \S\ref{sec:SRD}. 
The central panel of Fig.\ \ref{fig:qso_skymap} is a particularly clear illustration of why the PSD is regressed 
separately in the NGC and SGC regions---the NGC appears to be more robust to systematics than the SGC. 

To determine whether a linear regression adequately models the effect of systematics on the 
target density of \eboss\ CORE quasars, the statistics designated
the ${\rm Reduced\_PSD}_j$ and the ${\rm Residual\_PSD}_j$ in \citet{Prakash15inprep???} can be calculated.
The ${\rm Reduced\_PSD}_j$ is derived from the PSD by omitting the $j$'th systematic term when calculating the PSD---in order
to represent the deviation from the PSD caused by each systematic. 
The {\em difference} between the PSD and the
observed sky density of targets, called the {\em Residual Surface Density}, or ``Residual\_SD,'' is then calculated. 
If a linear model is an appropriate representation of the regression 
of a given systematic, then the ${\rm Residual\_PSD}$ should be well-represented by a model with a slope of 
$S_j$. Formally:

\begin{align}
      {\rm Reduced\_PSD}_j = {\rm PSD} - S_{j} \times x_{j}~~, \nonumber \\
      {\rm Residual\_SD}_j =  SD_{\rm obs} - {\rm Reduced\_PSD}_j ~~.\label{eqn:rsd_eqn}
\end{align}

Fig.\ \ref{fig:qso_regress_mp} shows how the CORE quasar Residual\_SD varies as a function of each of the
individual systematics, together with the underlying distributions of those systematics. In general, a linear regression
seems to be adequate for modeling variations in CORE quasar target density. Fig.\ \ref{fig:qso_regress_mp} 
suggests that sky brightness, and, in particular, Galactic extinction,
are the main culprits in causing variations in \eboss\ CORE quasar target density. The SGC has a 68\% range of $r$-band
extinction of 0.075 to 0.19 with a median of 0.12, whereas the NGC has a 68\% range of $r$-band 
extinction of 0.032 to 0.10, with a median of only 0.057. The corresponding numbers for $z$-band sky flux are
4.1 to 6.8 with a median of 5.1 in the SGC and 3.3 to 4.6 with a median of 3.8 in the NGC.
The higher median and wider range of values of these systematics in the SGC are likely responsible for both the suppressed density of SGC targets and the larger RMS in predicted surface density that can be seen in Fig.\ \ref{fig:qso_skymap}.
These systematics will act to reduce the effective depth of an exposure and hence to increase the error 
on the fluxes of a test object being assigned a quasar probability by the \xdqsoz\ method. In effect, as the flux errors for a test
object increase, the formal probability that the object is a quasar is reduced, and fewer objects are then assigned
PQSO($z > 0.9) > 0.2$ by \xdqsoz.

Overall, the \eboss\ quasar
targeting algorithm outlined in this paper is expected to produce quasar samples for clustering measurements that
are robust against systematics across essentially the entire NGC and across about three-quarters of the SGC. This statement may be
pessimistic, as \eboss\ does not attempt to restrict the CORE quasar redshift range 
to $0.9 < z < 2.2$ in advance of spectroscopic confirmation. Quasars at $z > 2.2$ are closer to the stellar locus in optical
color space, so the target density of quasars at $z > 2.2$ may fluctuate more due to systematics than at $z < 2.2$.
Weighting for systematics as a function of quasar redshift is a possible avenue 
for further improving \eboss\ clustering measurements once target redshifts have been
confirmed by spectroscopy. The final \eboss\ footprint is yet to be derived but in the worst-case scenario that the entire SGC
has to be observed, only $\sim86.7\%$ of \eboss\ will meet the requirements of \S\ref{sec:SRD}. This fraction of useful area is
almost exactly offset by the expected excess of \eboss\ CORE quasars. Table~\ref{tab:stats} implies
that \eboss\ will confirm ($0.95 \times 72.0 =$) 68.4\perdegsq\ $0.9 < z < 2.2$ quasars. Serendipitously, 
68.4\,\perdegsq$\times 0.867 = 59.3$\,\perdegsq, 
exceeding the requirement of 58\,\perdegsq\ $0.9 < z < 2.2$ quasars noted in \S\ref{sec:SRD}.


\begin{table}[t]
\centering
\caption{\small
Results of how zero-point fluctuations affect target density}
\begin{tabular}{|c c r c |}
\hline
 & {\footnotesize $N^{-1}(\Delta N/\Delta {\rm m})$} & {\footnotesize zero-point error} & {\footnotesize fluctuation} \\ 
& (1) & \multicolumn{1}{c}{(2)} & (3) \\ \hline
$u$ & 0.544  & $13 \times 10^{-3}$ & 2.8\%  \\
$g$ & 0.856 & $9 \times 10^{-3}$ & 3.1\%  \\
$r$ & 0.514  & $7 \times 10^{-3}$ & 1.4\% \\
$i$ & 0.475 & $7 \times 10^{-3}$ & 1.3\% \\
$z$ & 0.061 & $8 \times 10^{-3}$ &  0.2\% \\
$W$ & 0.223 &$20 \times 10^{-3}$  & 1.8\% \\
\hline
\end{tabular}
\tablecomments{
(1) Fractional deviation in target density that results from a $\pm0.01$\,mag scatter in each band;
(2) Zero-point RMS error in each band in magnitudes. Values for the \sdss\ are taken from 
D.\ Finkbeiner et al.\ (2016, in preparation). Values for the \wise\ stack are estimated from 
\citet{Jar11}; (3) 95\% ($\pm2\sigma$) values in target density fluctuation corresponding
 to $100\% \times 4 \times {\rm [z}$ero-point erro${\rm r]} \times [N^{-1}(\Delta N/\Delta {\rm m})]$
}
\label{tab:zeropoint}
\end{table}

\subsection{Target density fluctuations due to zero-point variations}

A further requirement of \eboss\ is that fluctuations in target density due to shifting zero-point calibrations
across the \sdss\ imaging footprint are well-controlled. Similar to \S\ref{sec:homofluc}, such fluctuations
need to be kept below the 15\% level (see also \S\ref{sec:SRD})\footnote{This 15\% limit is
on the two-tailed distribution (i.e.\ between the peaks due to a positive and a negative fluctuation in zero-point)}.
To study how changes in zero-point affect
the density of \eboss\ CORE quasar targets, each of the bands used in the \eboss\ CORE
quasar selection is offset by $\pm0.01$\,mags (i.e.\ scaled by 1\% in flux)
and the resulting fractional changes in target density are determined after re-running the target selection pipeline. Each \sdss\
band is tested individually. As the \wise\ bands are only incorporated into \eboss\ CORE
quasar target selection in a stack (see Eqn.\ \ref{eqn:wisestack}), both W1 and W2 are 
simultaneously shifted by $\pm0.01$\,mags and the result is reported as a single band (henceforth denoted $W$). 

The resulting fractional fluctuations in target density from these offsets ($N^{-1}[\Delta N/\Delta {\rm m}]$)
can then be multiplied by the zero-point RMS error expected for the imaging calibrations
used by \eboss\ (see \S\ref{sec:imaging}) to determine the expected RMS variation in number density 
due to zero-point calibrations shifting across the \eboss\ footprint. We adopt the zero-point errors in
[$u,g,r,i,z$] of [13, 9, 7, 7, 8]\,mmag RMS from D.\ Finkbeiner et al.\ (2016, in preparation) and conservatively estimate a
zero-point error of 20\,mmag RMS for the $W$\,stack \citep[see][]{Jar11}. Assuming that the zero-point errors
can be modeled using a Gaussian distribution, 95\% of CORE quasar targets in \eboss\ will be within
$\pm2\sigma$ of the expected RMS variation. In other words, 95\% fractional variance in target density
can be interpreted as meaning that 95\% of the area of the sky is expected to be described by fluctuations 
of $\pm2\sigma$. Thus, the overall 95\% fractional variance in target density due to
zero-point errors can be expressed (as a percentage) as 
$100\% \times 4 \times {\rm [z}$ero-point erro${\rm r]} \times [N^{-1}(\Delta N/\Delta {\rm m})]$.
Table~\ref{tab:zeropoint} displays the results of this analysis, which indicate
that $g$-band is the least robust to zero-point variations when
selecting \eboss\ CORE quasars. Even $g$-band, however, causes a ($2\sigma$) variation of only 3\%, far less than
the 15\% limit outlined in \S\ref{sec:SRD}. \eboss\ CORE quasar target selection is thus completely
robust to zero-point errors.

\section{Conclusions and Summary}
\label{sec:con}

The fourth iteration of the {\em Sloan Digital Sky Survey} will include the
{\em extended Baryon Oscillation Spectroscopic Survey}, a project with the overarching goal of 
using galaxies and quasars to measure the BAO scale across a range of redshifts.
This paper details the construction of a sample of quasars 
that can provide the first 2\%
constraints on the BAO scale at redshifts $0.9 < z < 2.2$ through clustering measurements, referred to as the \eboss\ ``CORE'' sample.
The final \eboss\ CORE algorithm, which is designed to be a homogeneous and reproducible selection, is:

\begin{enumerate}

\item Take all targets in the D.\ Finkbeiner et al.\ (2016, in preparation) recalibrations of \sdss\ imaging, which are
stored in the {\tt calib\_obj} or ``Data Sweep'' format \citep{Bla05}

\item Select {\tt PRIMARY} point sources ({\tt objc\_type==6}) that have (de-extincted) PSF 
magnitudes of $g < 22$ OR $r < 22$, a {\tt FIBER2MAG} of $i > 17$, and good {\tt IMAGE\_STATUS}

\item Apply the \xdqsoz\ method of \citet{Bov12} to these sources and restrict to objects with
PQSO($z > 0.9) > 0.2$.

\item Force-photometer \wise\ imaging at the positions of the resulting sources using the \citet{unwise} approach, or, equivalently, match
to the force-photometered catalog of \citet{Lan14}.

\item Create band-weighted stacks from the fluxes of these sources using photometry from the \sdss\ 
$f_{\rm opt} = (f_{g} + 0.8f_{r}+0.6f_{i})/2.4$ and from \wise\
$f_{\rm WISE} = (f_{W1} + 0.5f_{W2})/1.5$

\item Convert these flux stacks to magnitudes and restrict to sources with
$m_{\rm opt} - m_{\rm WISE} \geq (g-i) + 3$

\end{enumerate}

\noindent The resulting set of sources comprise the \eboss\ CORE quasar sample. Not all
such sources, however, are targeted for spectroscopy in \eboss. The \eboss\ survey does not place a
fiber on any target that has an existing good spectrum from earlier iterations of the
\sdss\ (see \S\ref{sec:donotobs}).

This paper also describes a $z > 2.1$ quasar sample that can be used to refine the BAO scale measured from 
clustering in the \LyA\ Forest, referred to as the \eboss\ ``\LyA'' sample.
The various techniques used to target \LyA\ quasars for \eboss\ are not designed to be homogeneous and
reproducible, so are only discussed in full in the body of this paper (see, e.g., Fig.\ \ref{fig:flowchart}). 

The CORE and \LyA\ quasar targeting algorithms have been used to select targets for
a spectroscopic survey over a large area in the 
\sdss\ NGC region, in order to test whether these algorithms meet the requirements for
\eboss.
This $\sim810$\degsq\ survey is known as the
{\em Sloan Extended QUasar, ELG and LRG Survey} (\sequels). Observations over the first 
$\sim300$\degsq\ of \sequels\ have been completed and 
visual inspections of all \sequels\ targets are used to project outcomes for \eboss\ (see, e.g., Table~\ref{tab:stats}).

The algorithms developed in this paper 
meet all of the requirements of \eboss\ quasar targeting that can be projected from
\sequels. In particular, the
requisite number densities for \eboss\ are $>58$\perdegsq\ uniformly selected quasars
in the redshift range $0.9 < z < 2.2$, leaving as many fibers as possible to target new \LyA\ quasars.
Results from \sequels\ suggest that \eboss\ will recover  $\sim 70$\perdegsq\ $0.9 < z < 2.2$ quasars using
the CORE selection technique and $\sim 10$\perdegsq\ new $z > 2.1$ quasars from
various \LyA\ selection techniques\footnote{These \LyA\ quasar densities will be reduced slightly 
by the fact that \ptf\ imaging is only expected to be available over
$\sim$90\% of the \eboss\ footprint, as detailed in our companion 
overview paper \citep{ebosspaper???}}. In addition, the adopted \sdss\ and \wise\ imaging 
is sufficiently homogeneous for quasar targeting that 
the statistics projected from \sequels\ are expected to remain valid over
close to 90\% of the \eboss\ footprint.
The few \eboss\ quasar sample 
requirements or assumptions that are not discussed in this paper are
verified elsewhere. These include a survey area of at least 7500\degsq\ and
precise and accurate redshifts for quasars \citep[see][]{ebosspaper???}.

Ultimately, \eboss\ will uniformly target in 
excess of 500{,}000 quasars in the redshift range $0.9 < z < 2.2$, exceeding previous
such clustering samples by a factor of more than ten. 
Samples of 
{\em new} spectroscopically confirmed quasars across all redshifts in \eboss\ will exceed 
500{,}000 quasars, which will be at least three times larger than all previous samples across 
the \eboss\ footprint in combination.  At the conclusion of \eboss, in excess of
800{,}000 confirmed quasars should have spectra from some iteration of the \sdss.
In essence, \eboss\ will be {\em the} next-generation quasar survey, and, in the wake of 
20 years of observations from \sdss-\project{I, II, III} and \project{IV}, 
\eboss\ will usher in the era of million-fold spectroscopic quasar samples.

\acknowledgements 

We are grateful for insightful discussions about quasar selection statistics with 
Joe Hennawi, David Hogg and Gordon Richards. ADM acknowledges
a generous research fellowship from the
Alexander von Humboldt Foundation at the Max-Planck-Institut f\"{u}r Astronomie
and was supported in part by NASA-ADAP awards NNX12AI49G and 
NNX12AE38G and by NSF awards 1211112 and 1515404.
JPK acknowledges support from the ERC advanced grant LIDA.

This paper includes targets derived from the images of
the Wide-Field Infrared Survey Explorer, which is a
joint project of the University of California, Los Angeles,
and the Jet Propulsion Laboratory/California Institute
of Technology, funded by the National Aeronautics and
Space Administration.

This paper represents an effort by both the SDSS-III and SDSS-IV collaborations.
Funding for SDSS-III was provided by the Alfred
P. Sloan Foundation, the Participating Institutions, the
National Science Foundation, and the U.S. Department
of Energy Office of Science. Funding for the Sloan Digital Sky Survey IV has been provided by
the Alfred P. Sloan Foundation, the U.S. Department of Energy Office of
Science, and the Participating Institutions. SDSS-IV acknowledges
support and resources from the Center for High-Performance Computing at
the University of Utah. The SDSS web site is www.sdss.org.

SDSS-IV is managed by the Astrophysical Research Consortium for the
Participating Institutions of the SDSS Collaboration including the
Brazilian Participation Group, the Carnegie Institution for Science,
Carnegie Mellon University, the Chilean Participation Group,
the French Participation Group, Harvard-Smithsonian Center for Astrophysics,
Instituto de Astrof\'isica de Canarias, The Johns Hopkins University,
Kavli Institute for the Physics and Mathematics of the Universe (IPMU) /
University of Tokyo, Lawrence Berkeley National Laboratory,
Leibniz Institut f\"ur Astrophysik Potsdam (AIP),
Max-Planck-Institut f\"ur Astronomie (MPIA Heidelberg),
Max-Planck-Institut f\"ur Astrophysik (MPA Garching),
Max-Planck-Institut f\"ur Extraterrestrische Physik (MPE),
National Astronomical Observatory of China, New Mexico State University,
New York University, University of Notre Dame,
Observat\'ario Nacional / MCTI, The Ohio State University,
Pennsylvania State University, Shanghai Astronomical Observatory,
United Kingdom Participation Group,
Universidad Nacional Aut\'onoma de M\'exico, University of Arizona,
University of Colorado Boulder, University of Portsmouth,
University of Utah, University of Virginia, University of Washington,
University of Wisconsin,
Vanderbilt University, and Yale University.


\bibliography{qso_technical}
\bibliographystyle{apj}

\end{document}